\documentclass[fleqn,usenatbib]{mnras}

\usepackage{newtxtext,newtxmath}
\usepackage[T1]{fontenc}
\usepackage{gensymb}
\usepackage{breqn}
\usepackage{enumitem}
\usepackage{appendix}
\usepackage{graphicx}	% Including figure files
\usepackage{amsmath}	% Advanced maths commands
\usepackage{xcolor}
\usepackage{amsbsy}
\title[Grain Alignment at the Galactic Centre]{Alignment and Rotational Disruption of Dust Grains in the Galactic Centre Revealed by Polarized Dust Emission}

\author[Akshaya and Hoang]{
M. S. Akshaya$^{1}$\thanks{E-mail: akshayams@kasi.re.kr (MSA)}
and Thiem Hoang$^{1,2}$
\\
$^{1}$Korea Astronomy and Space Science Institute, Daejeon 34055, Republic of Korea\\
$^{2}$Korea University of Science and Technology, 217 Gajeong-ro, Yuseong-gu, Daejeon 34113, Republic of Korea
}

\date{Accepted XXX. Received YYY; in original form ZZZ}

\pubyear{2023}

\newcommand{\jyarcsec}{Jy arcsec$^{-2}$}
\newcommand{\cm}{cm$^{-2}$}
\newcommand{\mS}{$\mathcal{S}$}

\begin{document}
\label{firstpage}
\pagerange{\pageref{firstpage}--\pageref{lastpage}}
\maketitle

\begin{abstract}
We study the alignment and rotational disruption of dust grains at the centre of our Galaxy using polarized thermal dust emission observed by SOFIA/HAWC+ and JCMT/SCUPOL at 53, 216, and 850 \micron. We analyzed the relationship between the observed polarization degree with total emission intensity, dust temperature, gas column density, and polarization angle dispersion. Polarization degree from this region follows the predictions of the RAdiative Torque (RAT) alignment theory, except at high temperatures and long wavelengths where we found evidence for the rotational disruption of grains as predicted by the RAdiative Torque Disruption mechanism. The grain alignment and disruption sizes were found to be around 0.1 \micron\ and 1 \micron\, respectively. The maximum polarization degree observed was around $p\sim13$\% at 216 \micron\ and comes from a region of high dust temperature, low column density, and ordered magnetic field. Magnetically Enhanced RAT alignment (MRAT) was found to be important for grain alignment due to the presence of a strong magnetic field and can induce perfect alignment even when grains contain small iron clusters. We estimated the mass fraction of aligned grains using a parametric model for the fraction of the grains at high-$J$ attractors and found it to correlate weakly with the observed polarization degree. We observe a change in the polarization ratio, from $p_{216\mu m}/p_{850\mu m}<1$ to $p_{216\mu m}/p_{850\mu m}>1$ at $T_{\rm d}\gtrsim35$ K, which suggests a change in the grain model from a composite to a separate population of carbon and silicate grains as implied by previous numerical modeling.
\end{abstract}

\begin{keywords}
 dust, extinction -- Galaxy: centre -- infrared: ISM -- ISM: magnetic fields -- ISM: general -- polarization
\end{keywords}

%%%%%%%%%%%%%%%%%%%%%%%%%%%%%%%%%%%%%%%%%%%%%%%%%%

%%%%%%%%%%%%%%%%% BODY OF PAPER %%%%%%%%%%%%%%%%%%

\section{Introduction}
Magnetic fields and dust are ubiquitous in the universe and play an important role in various astrophysical processes, from star and planet formation, gas heating and cooling, galaxy evolution, to cosmic ray propagation. Starlight polarization as well as polarized thermal dust emission have demonstrated that dust grains in the the interstellar medium (ISM) are non-spherical and preferentially aligned with the interstellar magnetic field \citep{Hall1949,Hiltner1949,Crutcher2012}. Thus, polarization caused by the alignment of dust grains are widely used probes to study the orientation and strength of the magnetic fields \citep{Hildebrand1988,Lazarian2007JQSRT,PlanckXIX2015,PattleReview2019}. The degree of dust polarization is determined by a combination of the strength and structure of the magnetic field, dust grain alignment efficiency, and the physical and chemical properties of the grains. This makes dust polarization a powerful diagnostic in astrophysics on one hand, but on the other hand, the interpretation of dust polarization is complex, especially in dense dynamic environments. In this regard, measurements of dust polarization across multiple wavelengths are useful as each wavelength might trace a different population of dust \citep{Hildebrand1999,Santos2019,Michail2021,Fanciullo2022} and can give information about the physical properties of dust grains (e.g., shape, composition, and size), fundamental physics of grain alignment, and the magnetic field properties \citep{Chen.2019}. 

The alignment process of non-spherical grains along a preferred direction in space (e.g., the magnetic field) is a two-stage process, starting with the internal alignment which is the alignment of the axis of maximum moment of inertia of the grain along the angular momentum, followed by the external alignment which is the alignment of the angular momentum along a preferred direction in space, which could be the magnetic field, the anisotropic radiation field or the gas flow direction \citep{LazarianMET2007,LazarianRAT2007,Hoang2014,Hoang2016,Hoang2022ApJ}. For typical ISM conditions, the characteristic timescale for grains to align with the magnetic field is much smaller than the other two possible orientations. However, this can change if the grains are in very dense environments like dense molecular clouds, protostellar cores and disks \citep{Hoang2021,Hoang2022AJ}. In general, the external alignment of dust grains can be driven by paramagnetic relaxation \citep{Davis1951}, radiative torques \citep[RATs;][]{Dolginov1976,Draine1997,LazarianRAT2007} or mechanical torques \citep[METs;][]{LazarianMET2007,Hoang2018}, see \citealt{AnderssonReview2015,LazarianReview2015} for reviews. Latest studies demonstrate that RATs play a key role in grain alignment, whereas magnetic relaxation plays an additional role of its enhancement \citep{Hoang2008,Hoang2016}. RATs are experienced by irregular dust grains subject to anisotropic radiation due to the differential scattering and absorption of incident photons. In contrast, METs are due to the drift of irregular grains through ambient gas resulting in direct collisions with gaseous atoms. The dominant process that drives the grain rotation depends on the characteristics of the dust grains as well as the local conditions. 

Recent investigations favor the grain alignment driven by RATs, called the RAdiative Torque Alignment theory \citep[RAT-A;][]{Draine1997,LazarianRAT2007,Hoang2016,Lee2020,Hoang2021,Tram2021Doradus1}. RAT-A predicts an enhanced alignment in the regions of higher radiation field or dust temperature, leading to an increase in the observed polarization degree with dust temperature. If the grains subject to RAT-A have embedded iron inclusions, the alignment of these grains can be further enhanced by magnetic relaxation. This combined effect of RAT-A and magnetic relaxation is called Magnetically Enhanced RAT-A \citep[MRAT;][]{Lazarian2008,Hoang2016,Hoang2022AJ}. Another important mechanism in the context RATs is the RAdiative Torque Disruption (RAT-D) of rapidly spinning grains due to centrifugal stress \citep{Hoang2019NatA,Hoang2019}. Note that the idea of grain disruption by centrifugal stress was first discussed in \cite{Purcell.1979} and later in \cite{Li1997} in the context of grain spin-up by various grain surface processes \citep{Purcell.1979}. The RAT-D theory predicts the disruption of large grains in strong radiation environments when the centrifugal stress induced by fast grain rotation spun-up by RATs exceeds the tensile strength of the grain material. This results in a depletion of large grains and a consequent drop in the polarization at higher dust temperatures. The combined effect of RAT-A and RAT-D are modeled and observed in many environments by recent studies \citep{Lee2020,Tram2021Ophi,Tram2021Doradus1,Tram2021Orion,Ngoc2021,HoangM172022}.

In this paper, we study the alignment and disruption of dust grains in the context of the above discussed RAT paradigm at the centre of our Galaxy, around the supermassive black hole Sgr A$^*$ \citep[$D=8$ kpc;][]{Reid2004}. Early observations of the region around Sgr A$^*$ revealed that there is a stream of ionized gas spiraling towards it called the ``minispiral" or the Sgr A West \citep{Lo1983,Zhao2009,Irons2012} and a torus of warm gas and dust orbiting around the Sgr A$^*$ called the Circumnuclear Disk (CND) \citep{Genzel1985,Morris1996,Balick1974,Genzel1989}. The CND is the closest molecular reservoir to the Galactic centre. Magnetic fields in this region play an important role in the kinematics of the material and their interaction with gravity. Previous studies of the mid-infrared polarization in the CND suggest a strong magnetic field (about $2-10$ mG) which is the preferred direction of grain alignment and any variation in the degree of polarization in this region to be due to the change in the inclination of the magnetic field on the plane of the sky \citep{Aitken1986,Aitken1998,Hsieh2018,Roche2018}. A multi-wavelength study of dust polarization from the Galactic centre allows us to trace any changes in the grain alignment efficiency from different populations of grains along the line of sight and also gives us a deeper understanding of the properties of dust in this region. The ratio of polarization degree at different wavelengths can also yield information about the model of the dust grains, whether they are composite or separate populations of silicate and carbon grains \citep{Lee2020,Tram2021Ophi}. 

The Galactic centre is a complex environment consisting of ionized, neutral, and molecular gas mixed with dust and is an excellent candidate to probe the effects of extreme environmental conditions on the interaction between the ISM and the magnetic fields \citep{Genzel2010,Bryant2021}. We will use dust polarization data at 53, 216, and 850 \micron\ to understand the dominant process of the grain alignment in this complex environment and how the alignment of the dust varies with the wavelength on sub-parsec scales of $\sim0.2-0.8$ pc. As the region being probed is the Galactic centre, we can expect the line of sight to have multiple dust components or layers that are producing the net observed polarization at different wavelengths. Since each wavelength is sensitive to emission from dust at different temperatures, the majority of the emission at individual wavelength may come from different points along the line of sight, with the 53 \micron\ tracing the warmer outer region, 850 \micron\ originating in the inner, denser, and cooler region, and 216 \micron\ coming from the intermediate temperature regions. 

The rest of the paper is structured as follows. In Section \ref{Observations} we describe the observations used in our study. The local environmental conditions are described in Section \ref{Dust environment} followed by data analysis and results in Section \ref{Data Analysis}. The grain alignment physics in the context of RAT-A and RAT-D is discussed in Section \ref{Grain Alignment}. Finally, the discussion and summary of our results are presented in Section \ref{Discussion} and \ref{Summary}, respectively.

\section{Observations} \label{Observations}
\subsection{SOFIA/HAWC+}
We have used the publicly available observations of polarized dust emission from the Galactic centre at 53 \micron\ and 216 \micron\ from the High-resolution Airborne Wide-band Camera Plus \citep[HAWC+;][]{Harper2018} which is a far-infrared imager and polarimeter for NASA's Stratospheric Observatory for Infrared Astronomy \citep[SOFIA;][]{Temi2018}. These were part of the Guaranteed Time Observations (GTO) in Band A centered at 53 \micron\ (hereafter S53, PI: Dowell, D., ID: 70\_0511) and in Band E centered at 216 \micron\ (hereafter S216, PI: Chuss, D., ID: 05\_0018). The field of view of the S53 observation is $\sim5'$ with a beam size (FWHM) of $4.85\arcsec$ and the S216 observation has a field of view of $\sim8'$ and beam size of $18.2\arcsec$. 

We used the Level 4 data products from the SOFIA data archive which contains the flux-calibrated data along with the polarization vectors. No further reduction was performed on the data. The observed polarization degree is defined following \citet{Gordon2018}:
\begin{equation} 
   p_{\rm obs} (\%) = 100 \times \frac{\sqrt{Q^2+U^2}}{I},
\end{equation}
and the corresponding error ($\sigma_p$) is calculated as

\begin{align}
\begin{split}
        \sigma_p(\%) = \frac{100}{I}  \Bigg(\frac{1}{(Q^2+U^2)}\left[ (Q\sigma_Q)^2 + (U\sigma_U)^2 + 2QU\sigma_{QU} \right] +\\ 
        \left[ \left(\frac{Q}{I} \right)^2 + \left( \frac{U}{I} \right)^2 \right]\sigma^2_I - 2\frac{Q}{I}\sigma_{QI} - 2\frac{U}{I}\sigma_{UI} \Bigg)^{1/2}.
\end{split}
\end{align}

\begin{figure}
    \centering
    \includegraphics[scale=0.33]{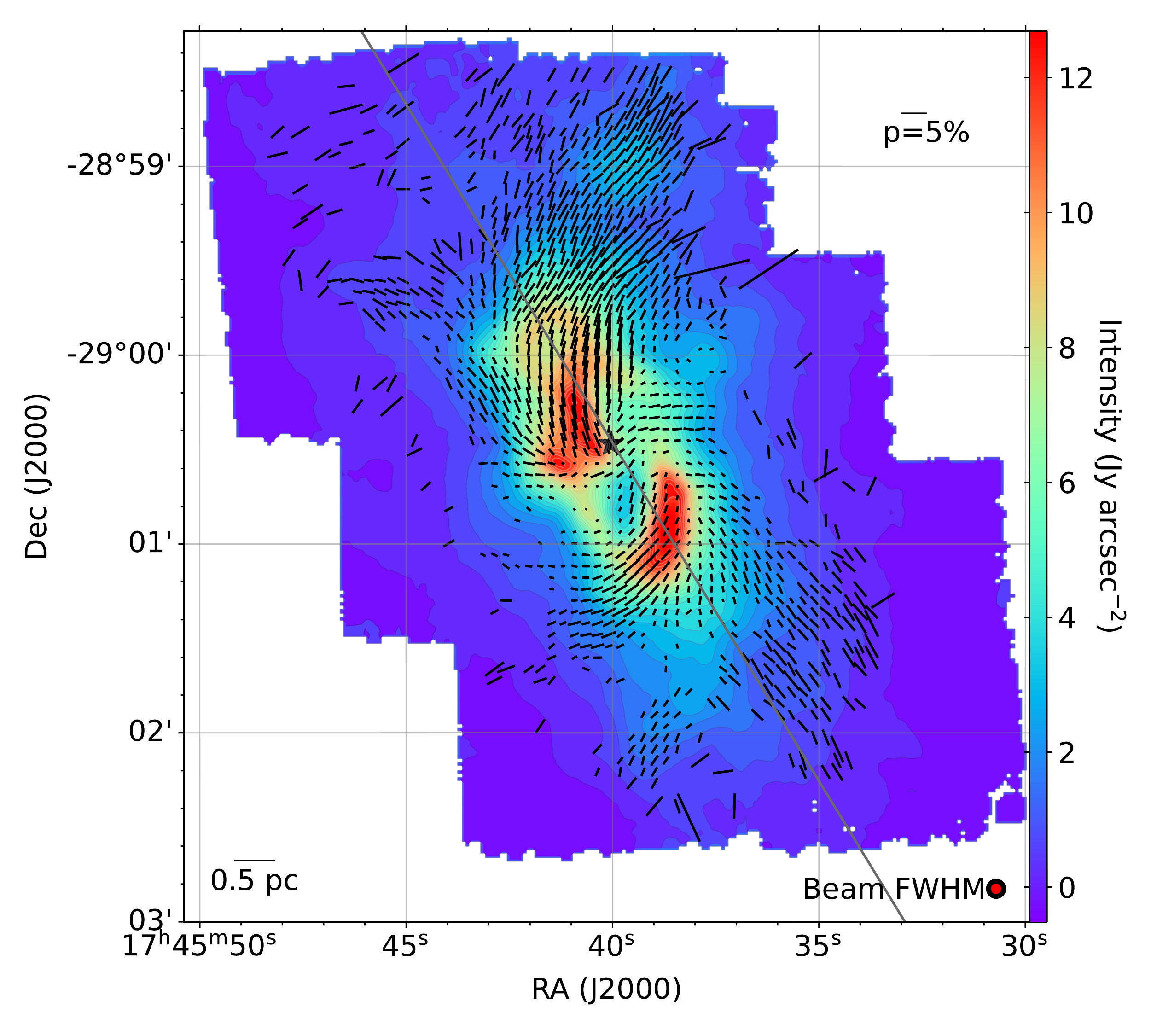}
    \caption{Polarization maps of S53 data with a field of view of $\sim5'$. The black star in the image shows the location of the Sgr A$^*$. The gray line is parallel to the Galactic plane with $b=-0.045\degree$. The background colormap represents the total intensity of the observation (Stokes $I$). The beam size of the band is shown as a red circle, and a representative scale of $p=5\%$ is also shown. The polarization vectors here trace the magnetic field (rotated by 90\degree).}
    \label{fig:Sofia_5_pol_map}
\end{figure}

\begin{figure}
    \centering
    \includegraphics[scale=0.35]{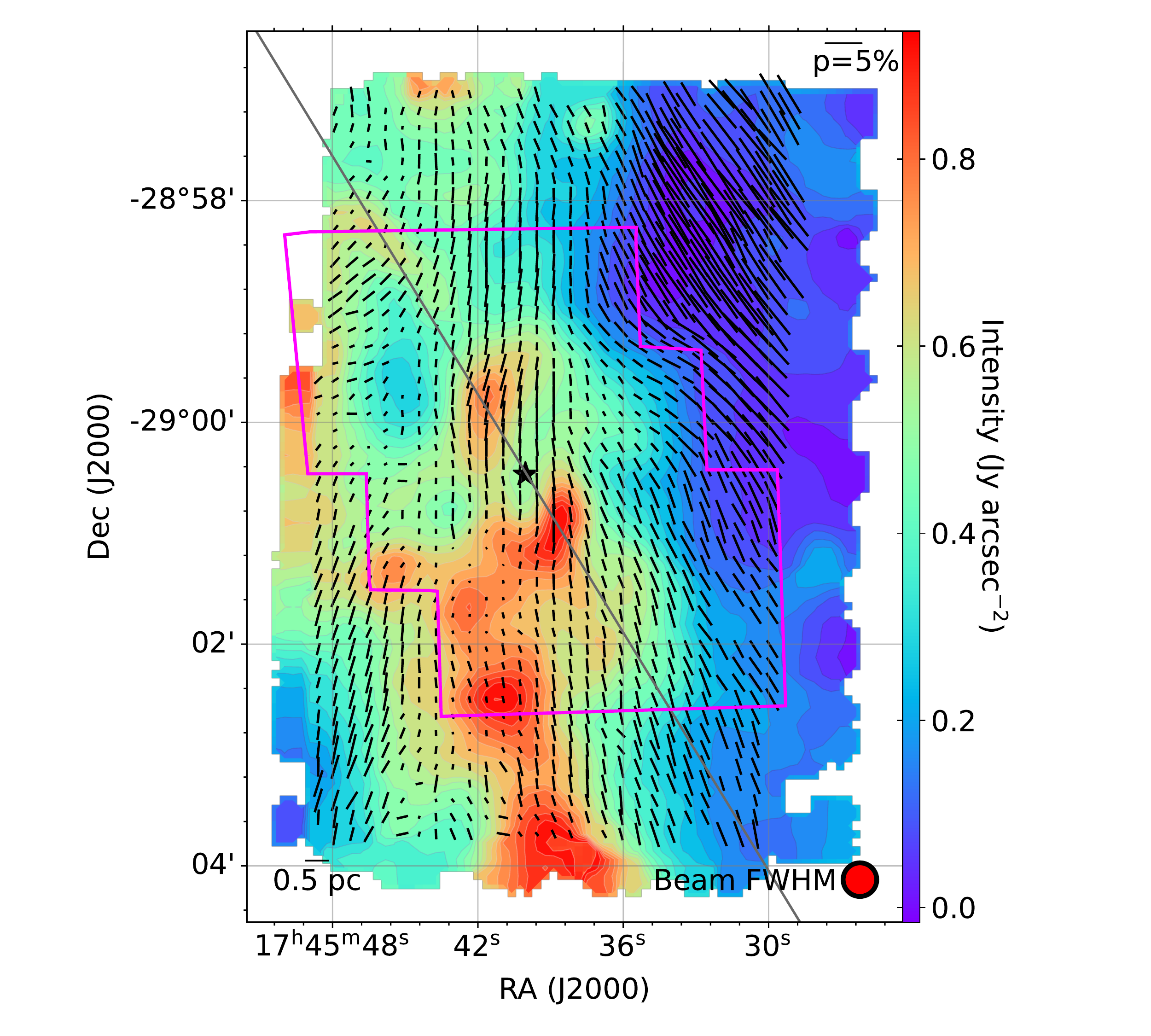}
    \caption{Same as Figure \ref{fig:Sofia_5_pol_map} but for the S216 observation with a field of view of $\sim8'$. The region covered by S53 observation shown in Figure \ref{fig:Sofia_5_pol_map} is overlaid as the magenta contour.}
    \label{fig:Sofia_10_pol_map}
\end{figure}

\begin{figure}
    \centering
    \includegraphics[trim={1cm 0 0 0},scale=0.35]{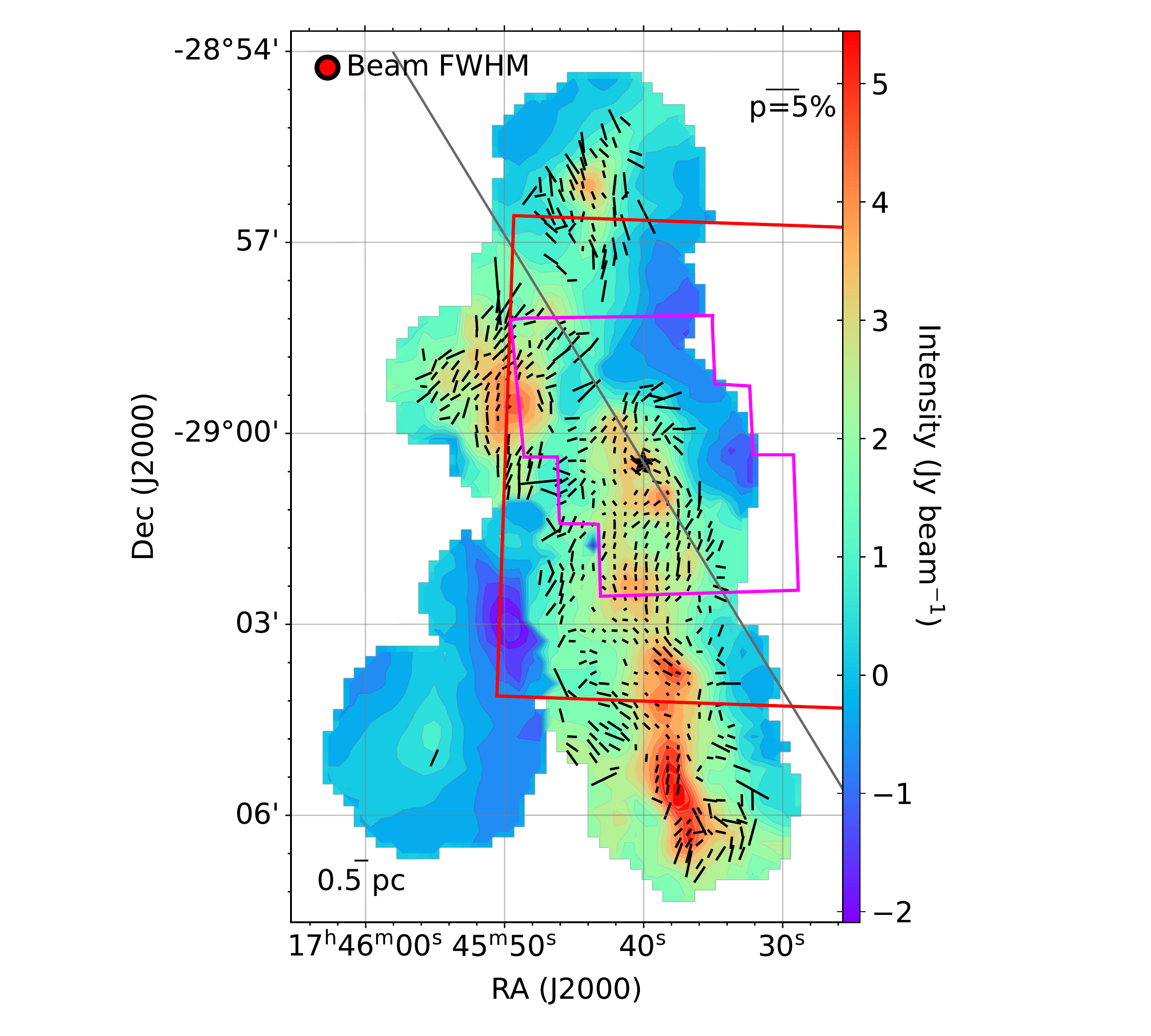}
    \caption{Same as Figure \ref{fig:Sofia_5_pol_map} but for the J850 observation with a field of view of $\sim15'$. The regions of the S53 and S216 observations shown in Figure \ref{fig:Sofia_5_pol_map} \& \ref{fig:Sofia_10_pol_map} are overlaid as magenta and red contours respectively.}
    \label{fig:JCMT_pol_map}
\end{figure}

The debiased polarization degree is given by
\begin{equation}
    p = \sqrt{p_{\rm obs}^2 - \sigma^2_p}.
\end{equation}

The polarization angle ($\theta$) and its corresponding error ($\sigma_\theta$) are defined by

\begin{equation}
    \theta = {\frac{1}{2}} {\rm tan}^{-1}\Big(\frac{U}{Q}\Big),
\end{equation}
and
\begin{equation}
    \sigma_\theta = \frac{1}{2(Q^2+U^2)} \sqrt{(U\sigma_Q)^2 + (Q\sigma_U)^2 - 2QU\sigma_{QU}}.
\end{equation}

The resulting polarization maps of S53 and S216 are shown in Figures \ref{fig:Sofia_5_pol_map} and \ref{fig:Sofia_10_pol_map} respectively. We performed quality cuts with the conditions; $p/\sigma_p<3$, $I/\sigma_I<200$ and $0<p<50\% $ on both the SOFIA/HAWC+ observations presented here.

\subsection{SCUPOL/JCMT}
SCUPOL was the polarimeter for the Submillimeter Common User Bolometer Array (SCUBA) instrument on the James Clerk Maxwell Telescope (JCMT) which operated from $1997-2005$, taking simultaneous observations at 450 \micron\ and 850 \micron\ \citep{Holland1999,Greaves2000,Jenness2000}. We have used the reprocessed SCUPOL data presented my \citet{Matthews2009} at 850 \micron\ (hereafter J850) for our study.
\citet{Matthews2009} re-reduced all the observations made by SCUPOL at 850 \micron\ spanning a period of eight years. They combined a total of 69 observations taken at the Galactic centre to create a single mosaic sampled at 10$\arcsec$ grid with an effective beam size of $20\arcsec$. 
The observed polarization intensity and its corresponding error are derived from the Stokes parameters following \citet{Doi2020} as follows:

\begin{equation}
    PI_{\rm obs} = \sqrt{Q^2 + U^2},
\end{equation}

\begin{equation}
    \sigma_{PI} = \frac{\sqrt{(Q^2\sigma_Q^2) + (U^2\sigma_U^2)}}{PI_{\rm obs}}.
\end{equation}

We need to debias the measured polarized intensity due to the squared errors in Stokes Q and U. The debiased $PI$ and the corresponding polarization degree ($p$) and its associated error ($\sigma_p$) are given by

\begin{equation}
    PI = \sqrt{PI_{\rm obs}^2 - \sigma_{PI}^2},
\end{equation}
\begin{equation}
    p (\%) = 100 \times \frac{PI}{I},
\end{equation}
and
\begin{equation}
    \sigma_p (\%) = p \times \sqrt{\Big(\frac{\sigma_{PI}}{PI}\Big)^2 + \Big(\frac{\sigma_I}{I}\Big)^2}.
\end{equation}

The polarization angle ($\theta$) and its associated error ($\sigma_\theta$) are defined as

\begin{equation}
    \theta = \frac{1}{2} {\rm tan}^{-1}\Big(\frac{U}{Q}\Big),
\end{equation}

\begin{equation}
    \sigma_\theta = \frac{1}{2} \frac{\sqrt{(Q\sigma_U)^2 + (U\sigma_Q)^2}}{PI^2}.
\end{equation}

Quality cuts were performed with $p/\sigma_p>2$ and $\sigma_p<1$ to exclude spurious polarization vectors. The resulting polarization map is shown in Figure \ref{fig:JCMT_pol_map}.

\section{Dust and Gas Environment Properties} \label{Dust environment}
\subsection{Dust Temperature, Gas Column Density, and Emissivity Spectral Index}
We use \emph{Herschel} observations in five wavebands at 70, 160, 250, 350, and 500 \micron\ to derive the dust temperature ($T_{\rm d}$), gas column density ($N_{\rm H}$), and dust emissivity spectral index ($\beta$) at the Galactic centre \citep{Battersby2011,Pokhrel2016,Lim2016}. The emission from dust in these wavelengths can be modeled as a modified blackbody spectrum (MBB) with a frequency-dependent emissivity as,
\begin{equation}
    I_\nu\ = (1 - e^{-\kappa_\nu \Sigma}) \times B_\nu(T_{\rm d}),
\end{equation}
where $B_\nu$ is the Planck function, $\kappa_\nu$ is the dust opacity at frequency $\nu$ and $\Sigma$ is the mass surface density. $\kappa_\nu$ is defined by the power law $\kappa_\nu = \kappa_0 (\nu/\nu_0)^\beta$ where $\beta$ is the emissivity spectral index and $\kappa_0$ is the opacity at reference frequency $\nu_0$ \citep{Hildebrand1983}. Assuming a canonical gas-to-dust ratio of 100 \citep{Predehl1995}, $\kappa_0$ = 0.1 cm$^2$ g$^{-1}$ at $\nu_0$ = 1000 GHz from \citet{Ossenkopf1994}. $\Sigma=\mu m_{\rm H}N_{\rm H}$ is used to derive the gas column density where $\mu\sim2.8$ is the mean molecular weight per unit hydrogen mass for a cloud with 71\% ${\rm H_2}$ gas, 27\% He, and 2\% metals \citep{Kauffmann2008,Sadavoy2013}, $m_{\rm H}$ is the mass of hydrogen atom, and $N_{\rm H}$ is the gas column density.  

\begin{figure}
    \centering
    \includegraphics[scale=0.5]{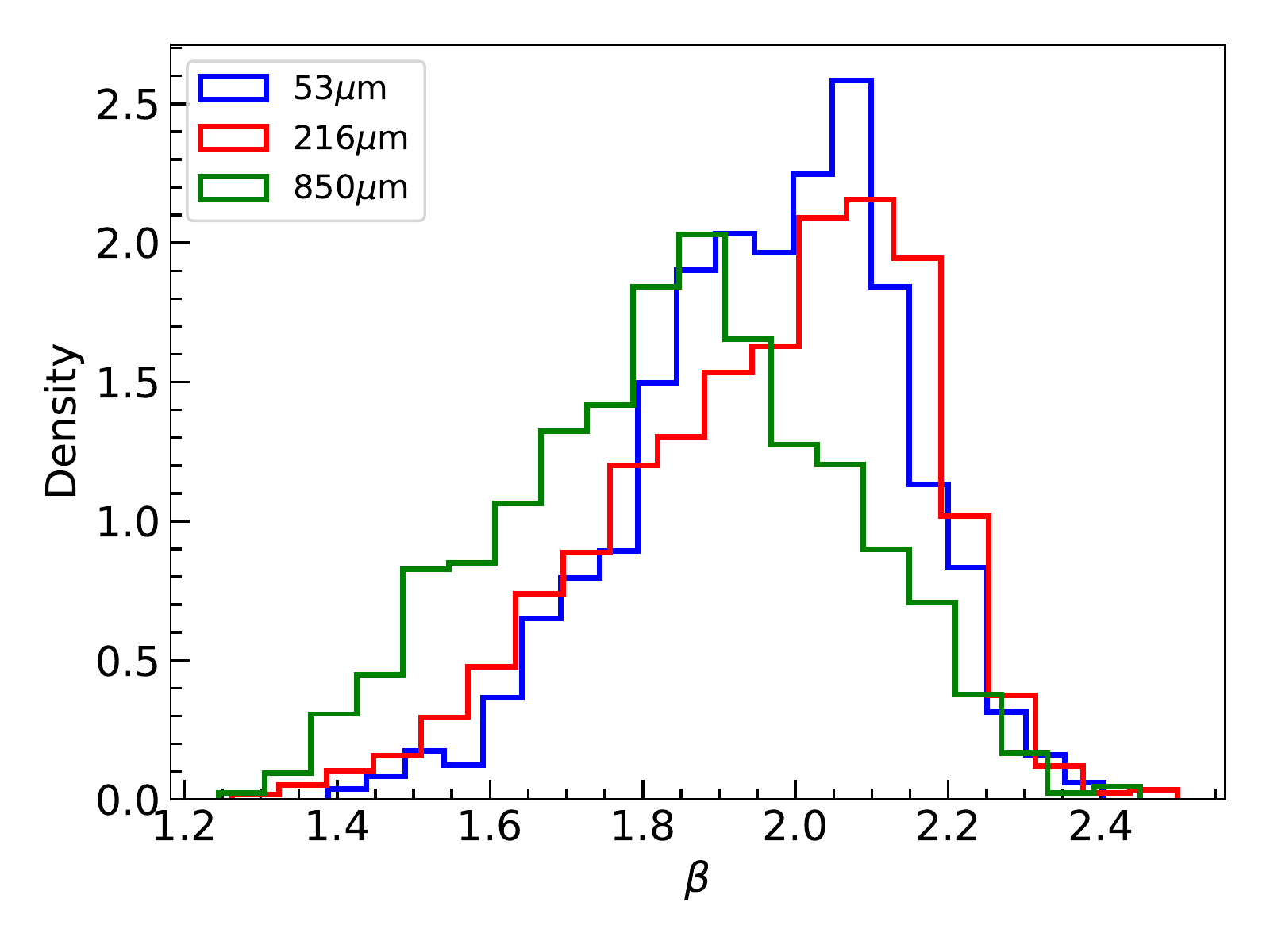}
    \caption{Histogram of the derived $\beta$ values for the three datasets considered. It can be seen from the histogram that the values are distributed around $\beta \sim 2.0$ in all the observations.}
    \label{fig:beta_hist}
\end{figure} 

\begin{figure}
    \centering
    \includegraphics[trim={2cm 0 0 0},scale=0.5]{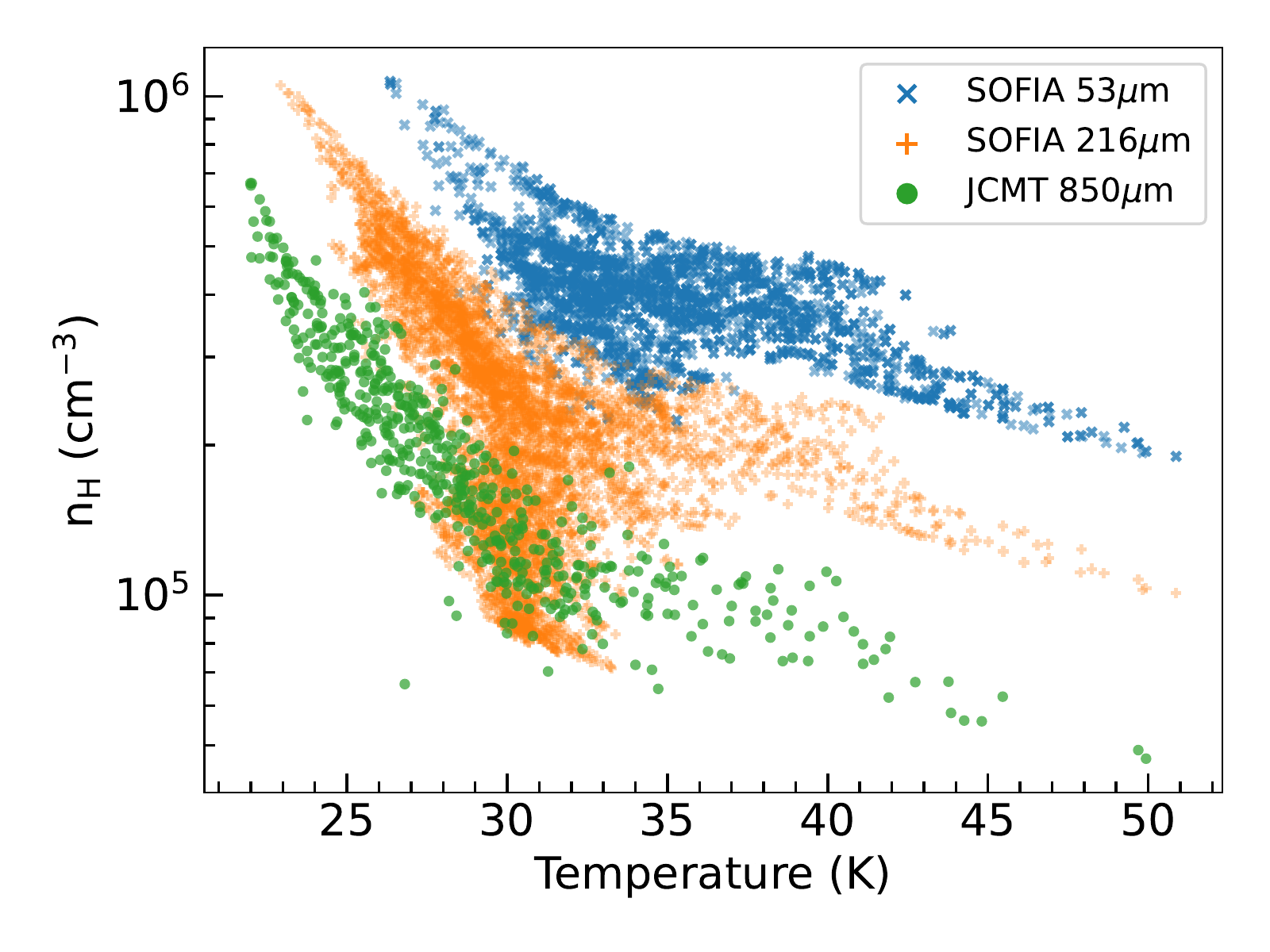}
    \caption{Density distribution of the observations derived from \emph{Herschel} data using the formulation given by \citet{Li2014}.}
    \label{fig:densities}
\end{figure}

\begin{figure*}
    \centering
    \includegraphics[trim={1.5cm 0 1cm 0},scale=0.25]{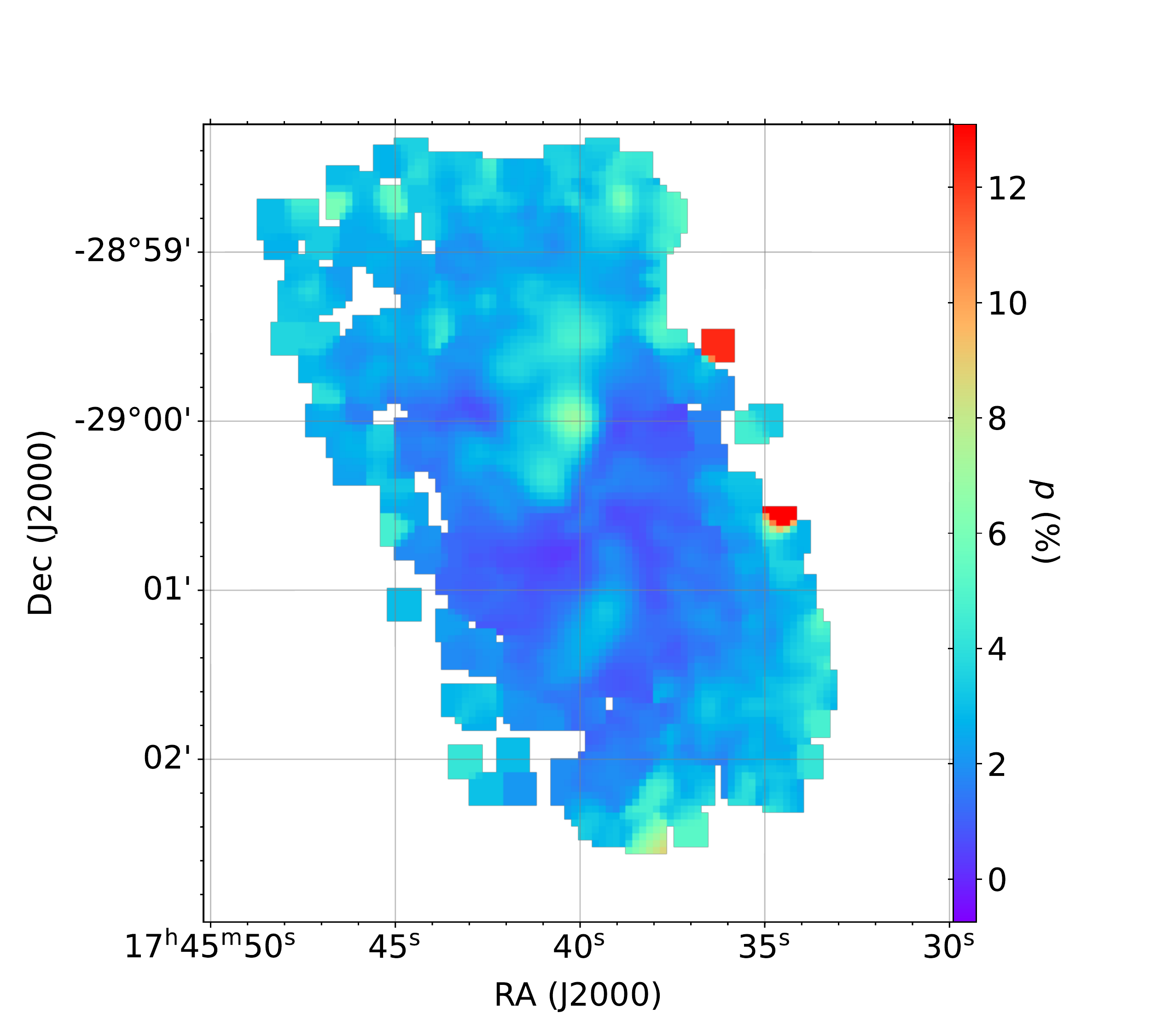}
    \includegraphics[trim={1cm 0 2cm 0},scale=0.25]{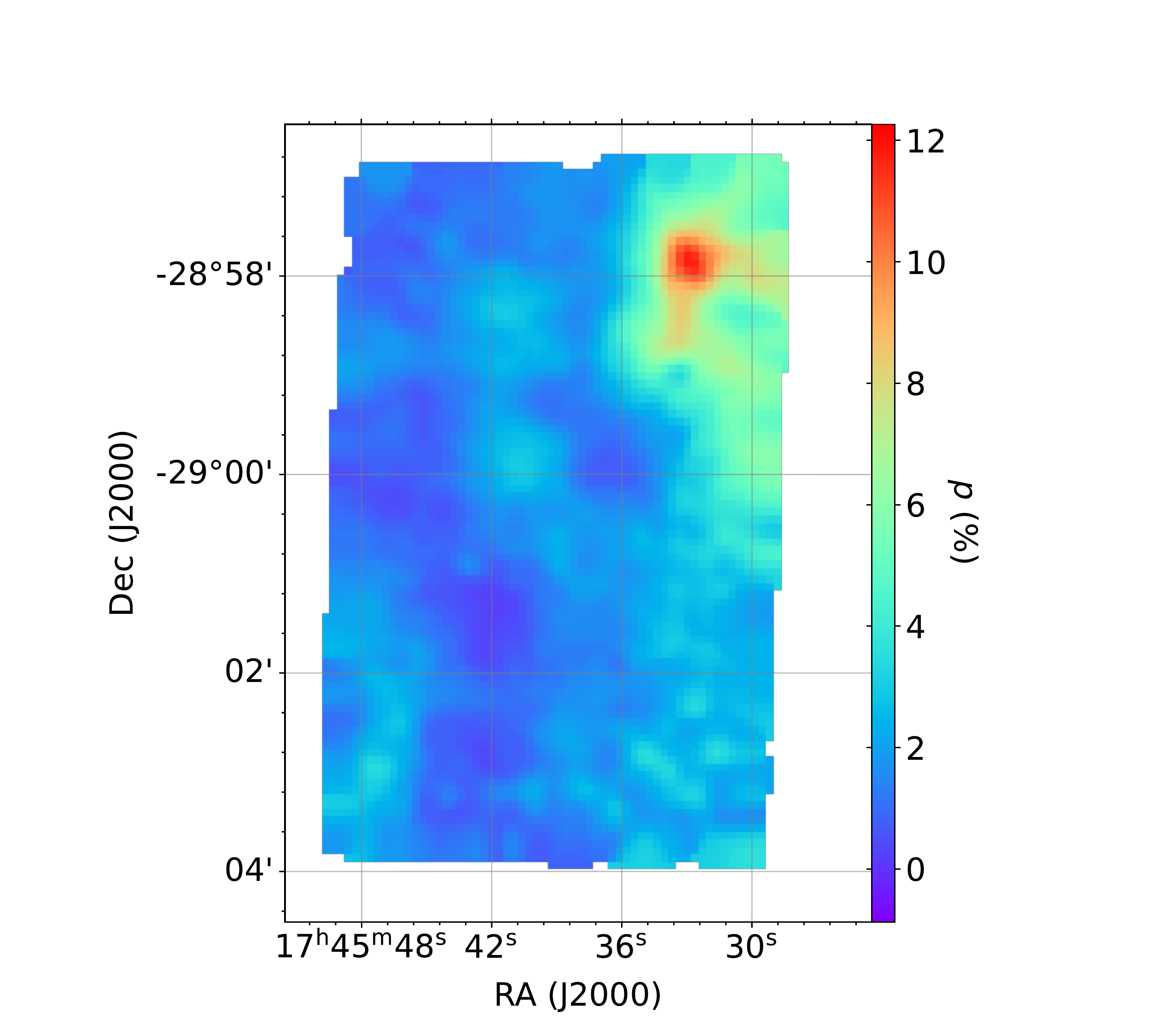}
    \includegraphics[trim={2cm 0 3cm 0},scale=0.25]{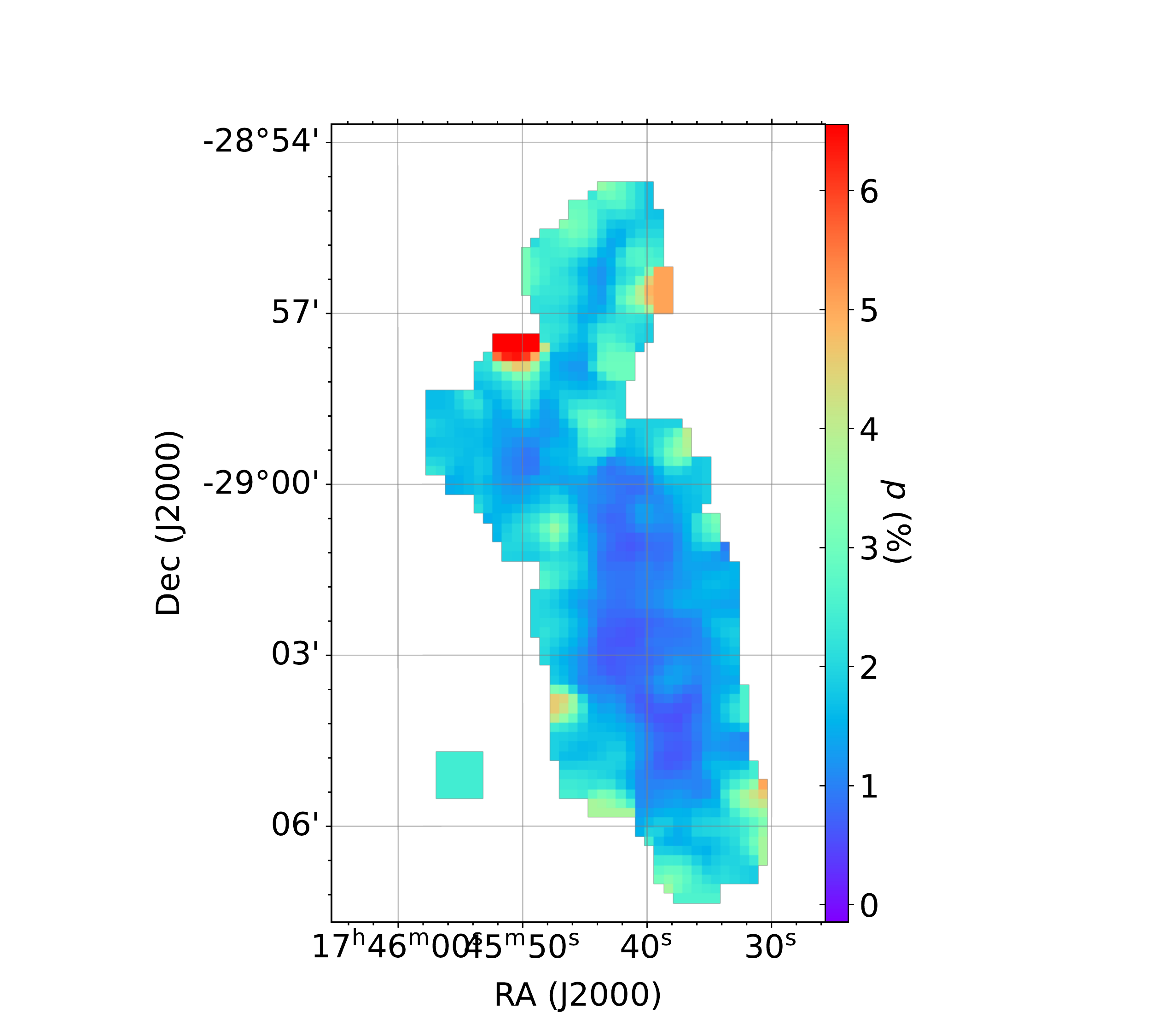}
    \caption{Colormaps showing regions of highest polarization for S53 (left), S216 (middle) and J850 (right) observations.}
    \label{fig:p_maps}
\end{figure*}

All the \emph{Herschel} observations were convolved to a common resolution of the 160 \micron\ data with pixel scale of 2.88$\arcsec$. The fit was performed pixel-by-pixel in logarithmic scale using $T_{\rm d}$, $N_{\rm H}$, and $\beta$ as free parameters. A logarithmic scale was used to reduce the impact of noise in the data on the derived parameters \citep{Shetty2009}. The common practice is to set a fixed value for $\beta$ based on the region of interest when performing a MBB fit, usually of $\beta = 2$ which is the accepted value for silicate grains in the ISM \citep{Draine1984,Draine2007beta,Draine2011}. We allowed $\beta$ to be a free parameter between $1 < \beta < 2.5$. The resulting $T_{\rm d}$, $N_{\rm H}$, and $\beta$ maps for each of the observations is shown in Figures \ref{fig:Sofia5_mbb_maps} -- \ref{fig:JCMT_mbb_maps}. The distribution of derived $\beta$ for the three regions considered is shown in Figure \ref{fig:beta_hist}. The mean value of $\beta$ was found to be between $1.85-2.0$, close to the value expected in the diffuse ISM and found by earlier studies \citep{Li2009,Schnee2010,Paradis2012,Pokhrel2016}. The range of $\beta$ also agrees well with those derived by \citet{PlanckXI2014} for the Galactic plane ($1.8<\beta<2.0$). There was not much difference between the derived values of $T_{\rm d}$ and $N_{\rm H}$ when $\beta$ is a free parameter compared to the values when $\beta = 2.0$. These plots are also shown in Figures \ref{fig:Sofia5_mbb_maps} -- \ref{fig:JCMT_mbb_maps}. We observe the general trend of anti-correlation in the $\beta-T_{\rm d}$ relation \citep{Dupac2003,PlanckXIV2014} which is most likely due to the various components and environments of dust traced along the line of sight, as the effect of noise and cosmic infrared background anisotropies (CIBA) are negligible in the bright areas of the Galactic plane \citep{PlanckXI2014}. $\beta$ was found to decrease with $T_{\rm d}$ in the regions of high column density and low temperature. This is expected due to grain growth in dense cold environments, which can drive the emissivity spectral index to values $\beta\simeq1$ \citep{Beckwith1991}. The region of maximum $T_{\rm d}$ in all the observations is always around Sgr A$^*$.

\subsection{Volume Density}
The volume density of the gas along the line of sight is an essential parameter in the RAT paradigm due to the competition between the grain's rotational damping by gas collisions and grain's spin-up by radiation \citep{Hoang2014,Hoang2021}. We have used the method presented by \citet{Lee2012} to calculate the volume density in this region. The mass of the region was calculated as $M = \beta_* m_{\rm H_2} N_{\rm H} (D\Delta)^2$, where $\beta_*=1.39$ is the conversion factor from hydrogen mass to total mass taking the helium abundance into account, $m_{\rm H_2}$ is the mass of hydrogen molecule, $N_{\rm H}$ is the total gas column density along the line of sight, $D=8$ kpc is the distance to the Galactic centre, and $\Delta$ is the pixel angular size of the data \citep{Qian2012,Li2014}. From the total mass, the gas volume density is calculated as $n_{\rm H} = 3M/(4\pi R^3 m_{\rm H_2})$, where $R$ is taken as the radius of the region encompassing an individual pixel. The estimated densities are shown in Figure \ref{fig:densities} and range from $10^4\lesssim n_{\rm H}\lesssim10^6$ cm$^{-3}$ with S53 probing higher densities at smaller scale and J850 probing the more diffuse lower densities. The derived values of $T_{\rm d}$, $N_{\rm H}$, and $n_{\rm H}$ agree well with the previous studies \citep{Etxaluze2011,Mills2013,Tsuboi2018,Henshaw2022}. The variation in the volume density is due to the different resolution of each observation as $n_{\rm H}\propto\Delta^{-1}$. Though the same $Hershel$ maps were used to derive the parameters in each case, they were binned to match the native resolution of each observations. Hence the S53 observation can probe denser regions with resolution comparable to the original resolution of \emph{Herschel} data, where as the resolution of S216 and J850 observations are about five times lower. 

\section{Data Analysis} \label{Data Analysis}
We now perform different analyses on the observed polarization at each wavelength to study the dust physics in the Galactic centre environment.

\begin{figure}
    \centering
    \includegraphics[scale=0.5]{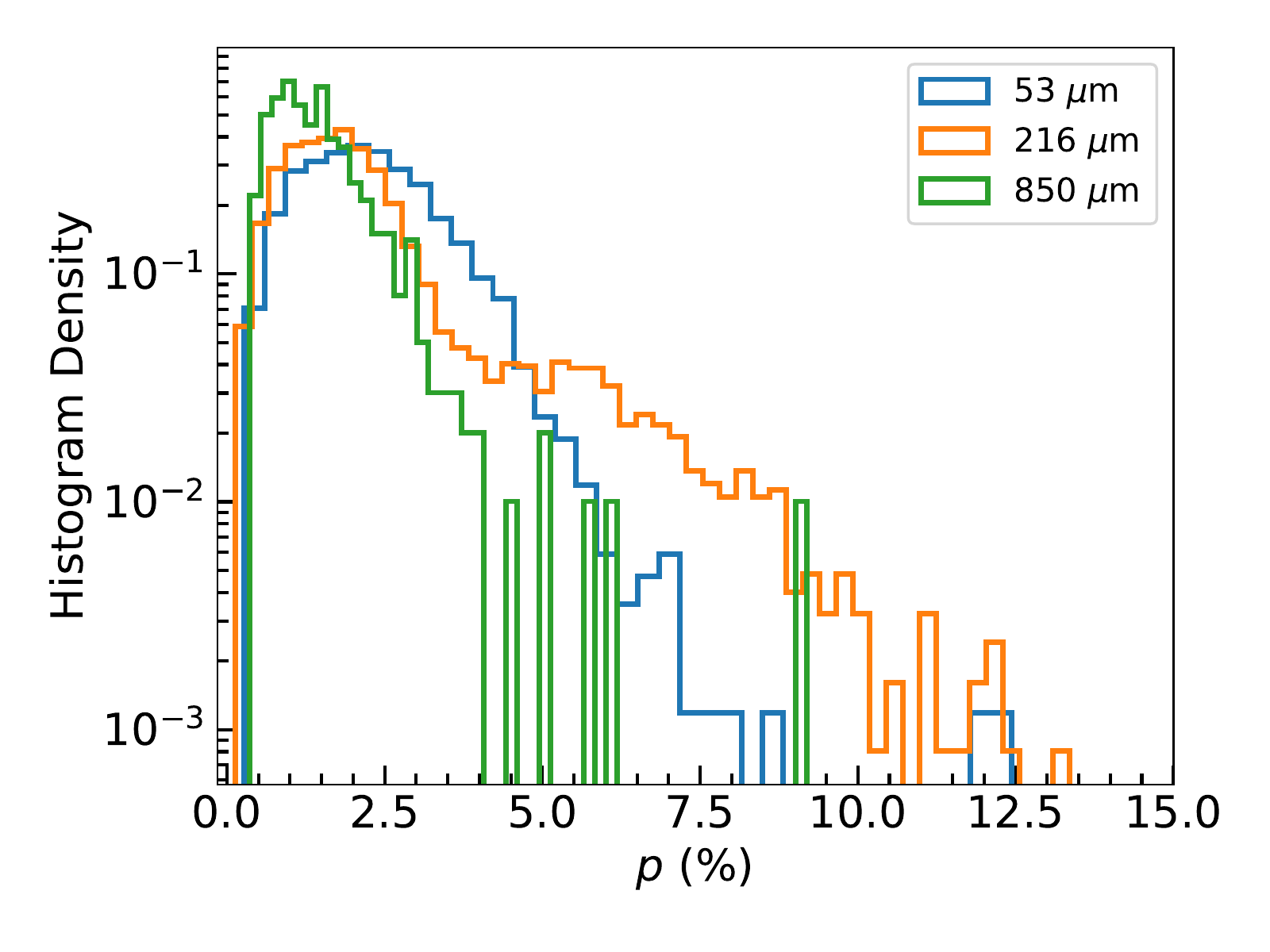}
    \caption{Histogram of the degree of polarization observed from S53, S216, and J850 observations.}
    \label{fig:p_hist}
\end{figure}

\begin{figure*}
    \centering
    \includegraphics[trim={1cm 0 0 0}, scale=0.37]{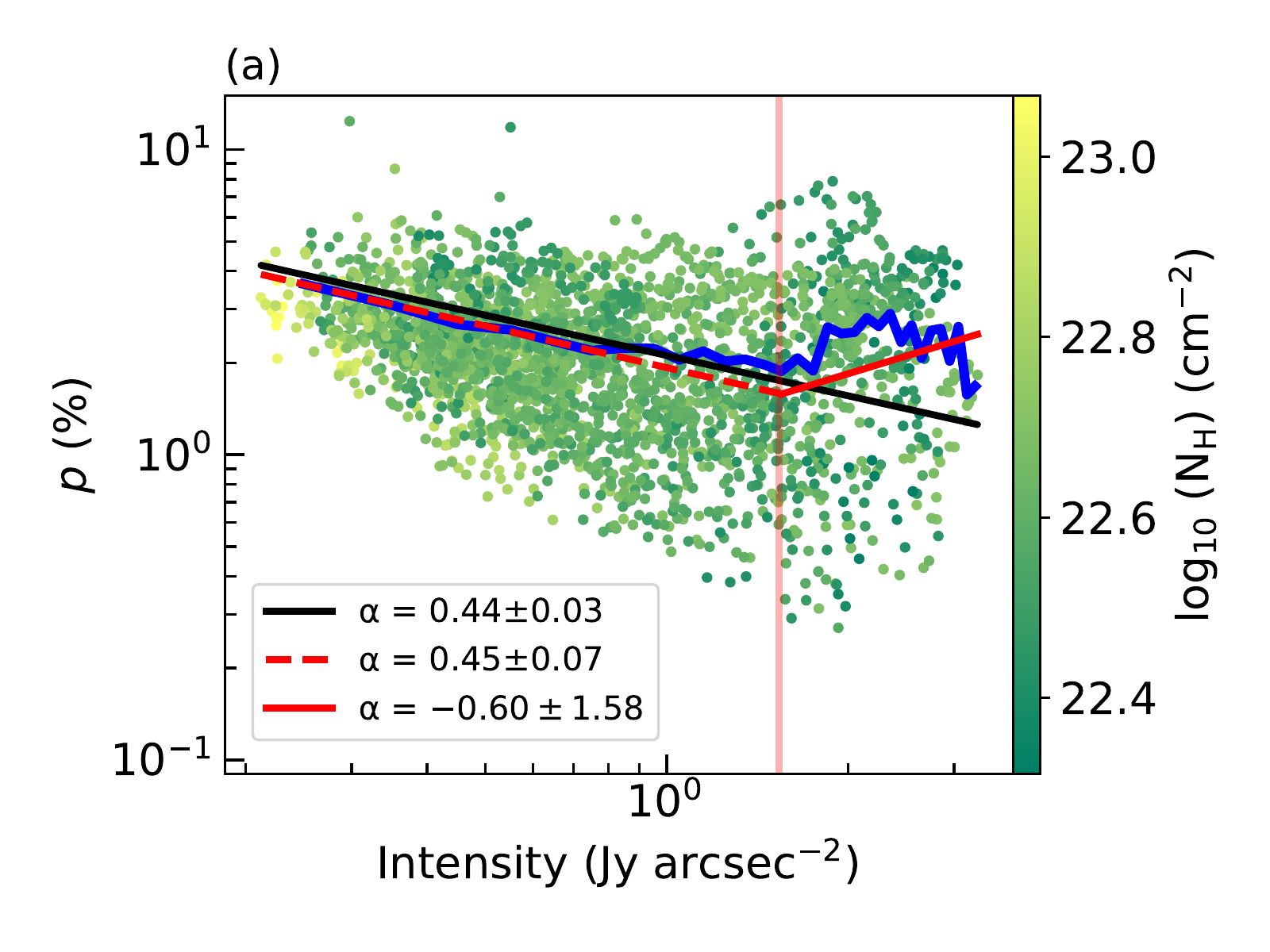}
    \includegraphics[trim={0 0 0 0}, scale=0.37]{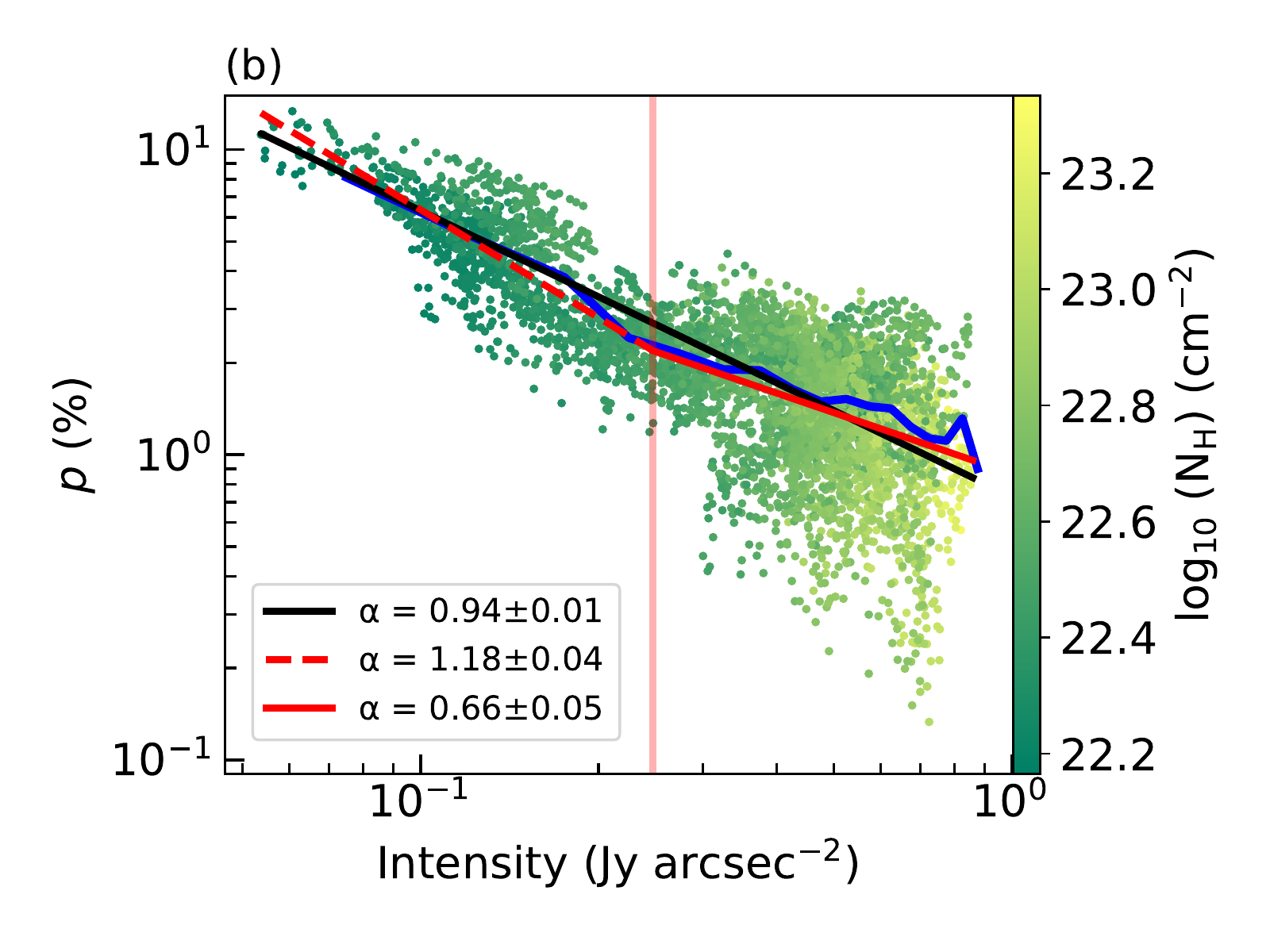}
    \includegraphics[trim={0 0 0 0}, scale=0.37]{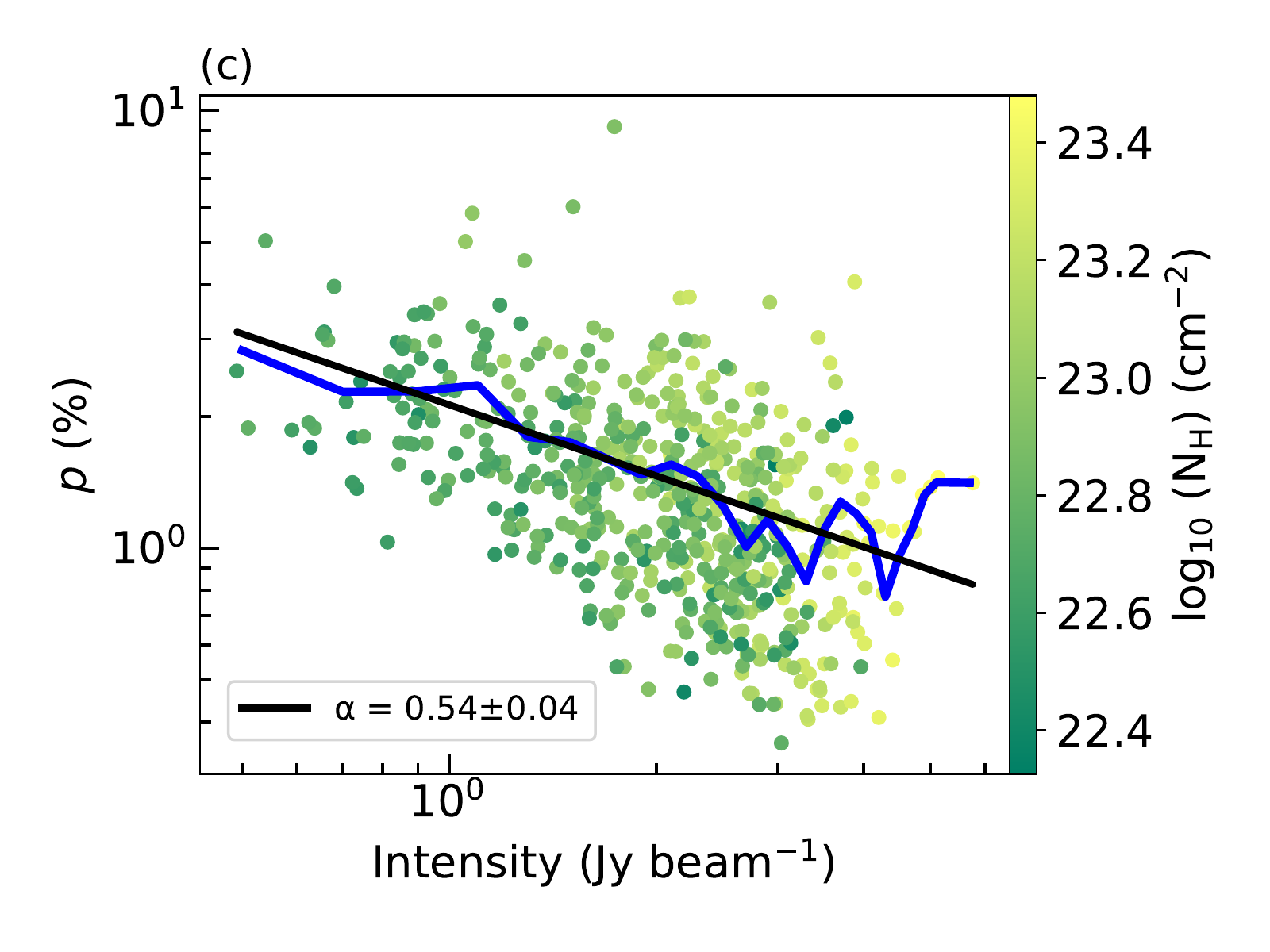}
    \caption{The observed variation of the degree of polarization with the total intensity for the S53 (a), S216 (b), and J850 (c) data. The blue line represents the weighted mean binned along the intensity. The solid black line shows the best fit to the data if all the data points are assumed to have a single slope. The red lines show the best fits when we use multiple power laws to fit the data, with the changes in $\alpha$ indicated by the dashed and solid lines. The vertical line shows the location of the breakpoints of the best fits. The color bar represents the gas column density distribution of the observations.}
    \label{fig:all_pvsI}
\end{figure*}

\subsection{Polarization Morphology}
Figures \ref{fig:Sofia_5_pol_map}-\ref{fig:JCMT_pol_map} depict the morphology of the magnetic field ($B-$field, polarization vectors are rotated by 90\degree) as observed in the respective wavelengths. The main component of each of the observation is the CND around Sgr A$^*$ and has the maximum temperature. The CND is evident in the S53 observation which has the highest resolution with the physical scale of individual pixel being $\sim0.05$ pc. The magnetic field seems to trace the CND in this case and has a spiral patter about the Galactic plane (indicated by the gray line). A similar trend is seen in the J850 observation where its $B-$field morphology is comparable to that observed in S53. However, the magnetic field in the S216 observation does not show any preferential alignment with the CND. The field is rather uniform in this case and is aligned at an angle ($\sim45$\degree) to the Galactic plane. The diffuse, low-dense regions have the magnetic field alined along the plane of the Galaxy as can be seen in the top right of Figure \ref{fig:Sofia_10_pol_map}. There are some similarities in the field traces by S216 and J850 observations in the overlapping regions, especially at high column densities. A details study of the $B-$field and its properties will be presented in our follow-up paper (Akshaya et al., in preparation). 

\subsection{Polarization Degree}
The degree of polarization ($p$) is a powerful probe of the composition, size distribution, and alignment mechanism of the dust grains as well as the local density and structure of the magnetic field \citep{Draine2009,Guillet2018,Hoang2021}. In optically thin regions, the degree of polarization is analogous to the polarization efficiency of the column of dust. The regions of maximum polarization in each of the wavebands are shown in Figure \ref{fig:p_maps} and originate from different locations, which could be due to the varied coverage of the maps. Ignoring the spurious maximum polarization vectors, at 53 \micron\ the region of maximum polarization is observed along one of the minispirals called the Northern Arm. For 216 \micron\ the value of $p$ is highest just outside and north of the CND. There is no distinct maximum region observed at 850 \micron. A histogram of the distribution of $p$ for the three wavelengths is shown in Figure \ref{fig:p_hist}. The J850 observation has the least level of polarization compared to the hotter S53 and S216 observations. From the distribution, it can be seen that the average degree of polarization in all the wavebands is around $p\sim1-3\%$ with the S216 data showing the maximum polarization degree of $p\sim13\%$. The physics of grain alignment can be studied using the relationship between the polarization degree with the observed total intensity ($I$), dust temperature ($T_{\rm d}$), the gas column density ($N_{\rm H}$), and the polarization angle dispersion function (\mS). It is important to understand how the level of polarization changes with each of these parameters to get a full picture of the dust physics in the region. The S53 data was re-binned by a factor of two to make it comparable to the resolution of \emph{Herschel} data used for the $T_{\rm d}$, $N_{\rm H}$, and $\beta$ estimates. 

\subsection{Polarization Degree vs. Intensity}
The relationship between $p$ and $I$ is a common tool used to understand the level of grain alignment from the diffuse to dense regions of the ISM, where $I$ acts as a tracer of the visual extinction and/or temperature along the line of sight, depending on the wavelength of the observation \citep{Whittet2008,Jones2016,Pattle2019}. It also traces the column density at longer wavelengths and the temperature at shorter wavelengths. Though not a perfect tracer of the grain alignment, it can be used to understand the overall dependence of $p$ on the gas density, grain size distribution, and the radiation field \citep{Tram2022}. The observed trend of the $p-I$ relationship is typically of the form of a power-law $p\propto I^{-\alpha}$, when $I$ traces the column density along the line of sight. The relationship steepens with the gradual loss of grain alignment towards denser regions. A value of $\alpha=0$ indicates uniform grain alignment at all column densities in a uniform magnetic field and $\alpha=1$ when only the grains at the outer layers of the region are aligned with the dust in the inner regions not showing any preferential alignment. The intermediate case where the alignment decreases linearly with the increase in optical depth is characterized by $\alpha = 0.5$ (e.g., \citealt{Hoang2021}). However, if the observed intensity is more correlated to the temperature instead of the column density, we would expect a raise in $p$ with an increase in the intensity. 

These interpretations of the $p-I$ relationship assume RAT-A theory, where the alignment is driven by RATs. Figure \ref{fig:all_pvsI} shows this relation for the three observations considered. A weighted mean is shown in all the plots to visualize the general trend of the relation. We have used the open-source python packages \textsc{lmfit} \citep{lmfit} and piecewise linear fit \citep[\textsc{pwlf};][]{pwlf} for our analysis. The best fit assuming a single power law is shown as the black line for all the observations. We find that a double power law (shown in red lines) fits the SOFIA/HAWC+ data better and results in lower minimized $\chi^2$ than using a single power law (black line). The breakpoints where the slope changes is indicated by the light red line at $1.55\pm0.2$ \jyarcsec\ and $0.25\pm0.03$ \jyarcsec\ for S53 and S216 respectively. There was no change in the slope observed for the J850 data. 

The color bar in Figure \ref{fig:all_pvsI} shows the distribution of the gas column density ($N_{\rm H}$), which is expected to be linearly correlated with $I$ at long wavelengths in optically thin regions. From the plots, it can be seen that this is indeed the case for the S216 and J850 data, where the polarization degree drops with a rise in $I$. We observed two slopes for S216. The first slope with $\alpha=1.18\pm0.04$ indicates a steep drop in $p$ at higher column densities and where the grains may be aligned only in the outer regions of the dust distribution. The later slope is shallower indicating a more gradual loss of polarization at higher densities with $\alpha=0.66\pm0.05$. For J850 data, $\alpha=0.75\pm0.03$ is similar to the second slope in S216. 

\begin{figure*}
    \centering
    \includegraphics[scale=0.36]{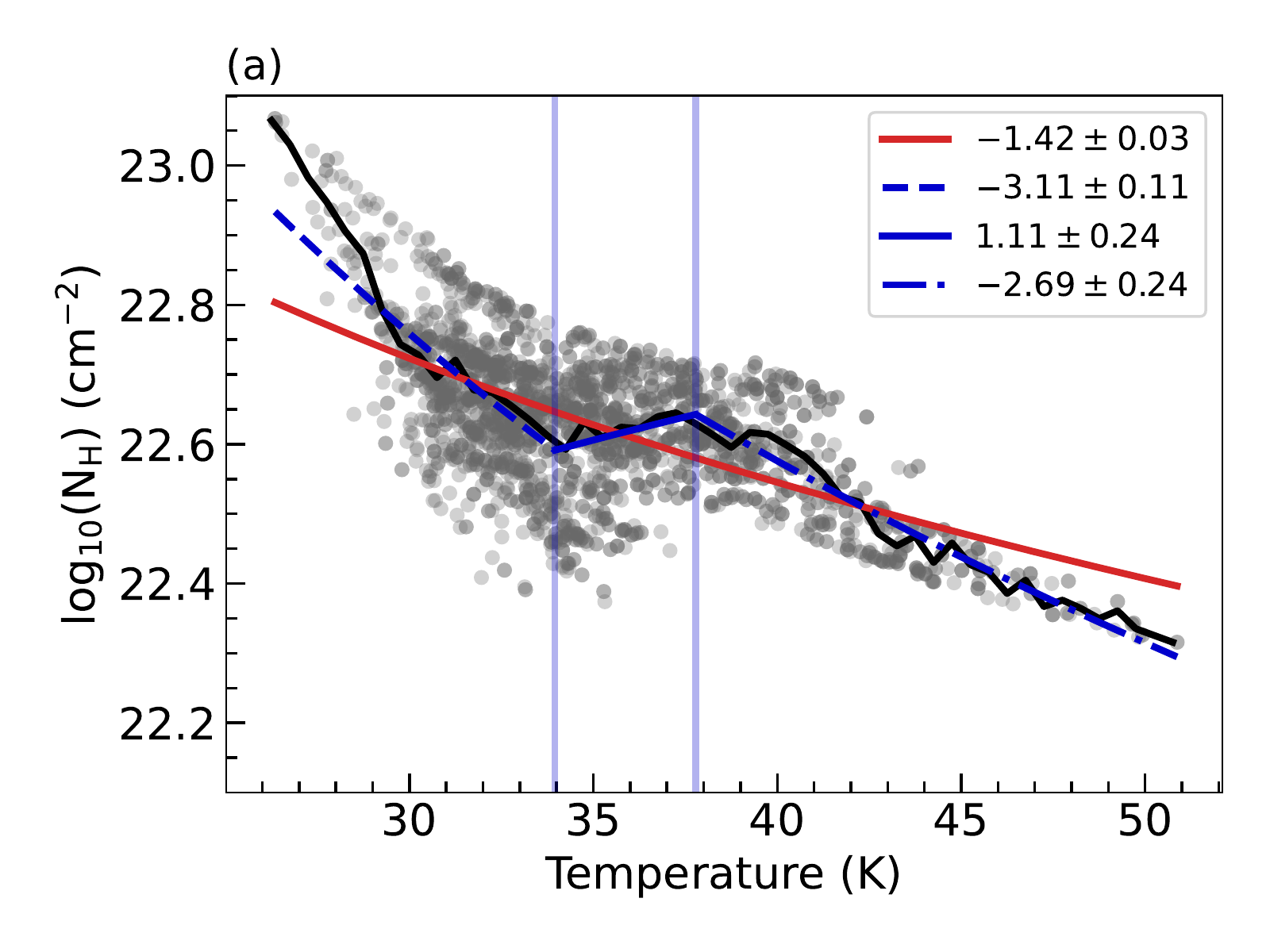}
    \includegraphics[scale=0.36]{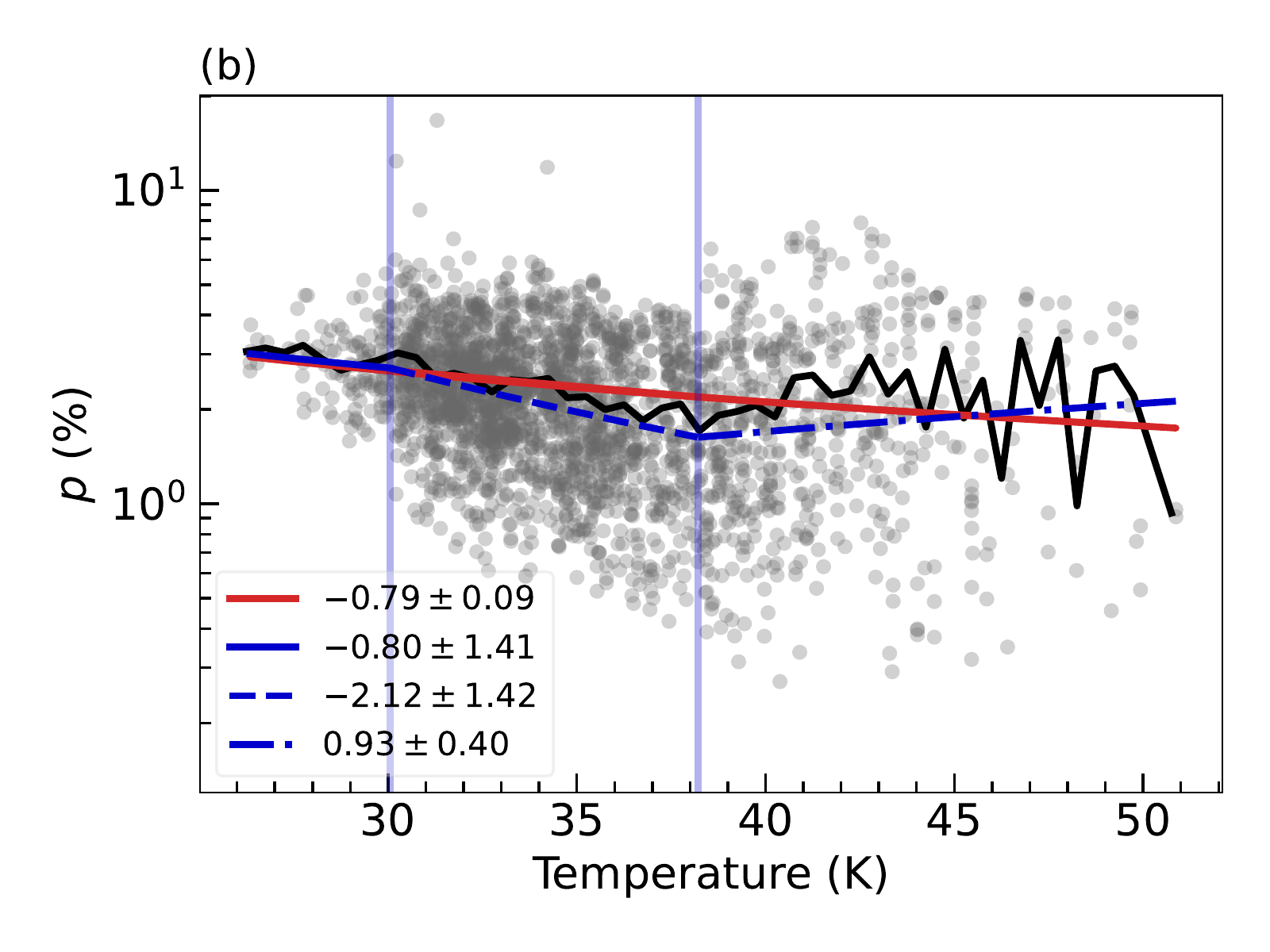}
    \includegraphics[scale=0.36]{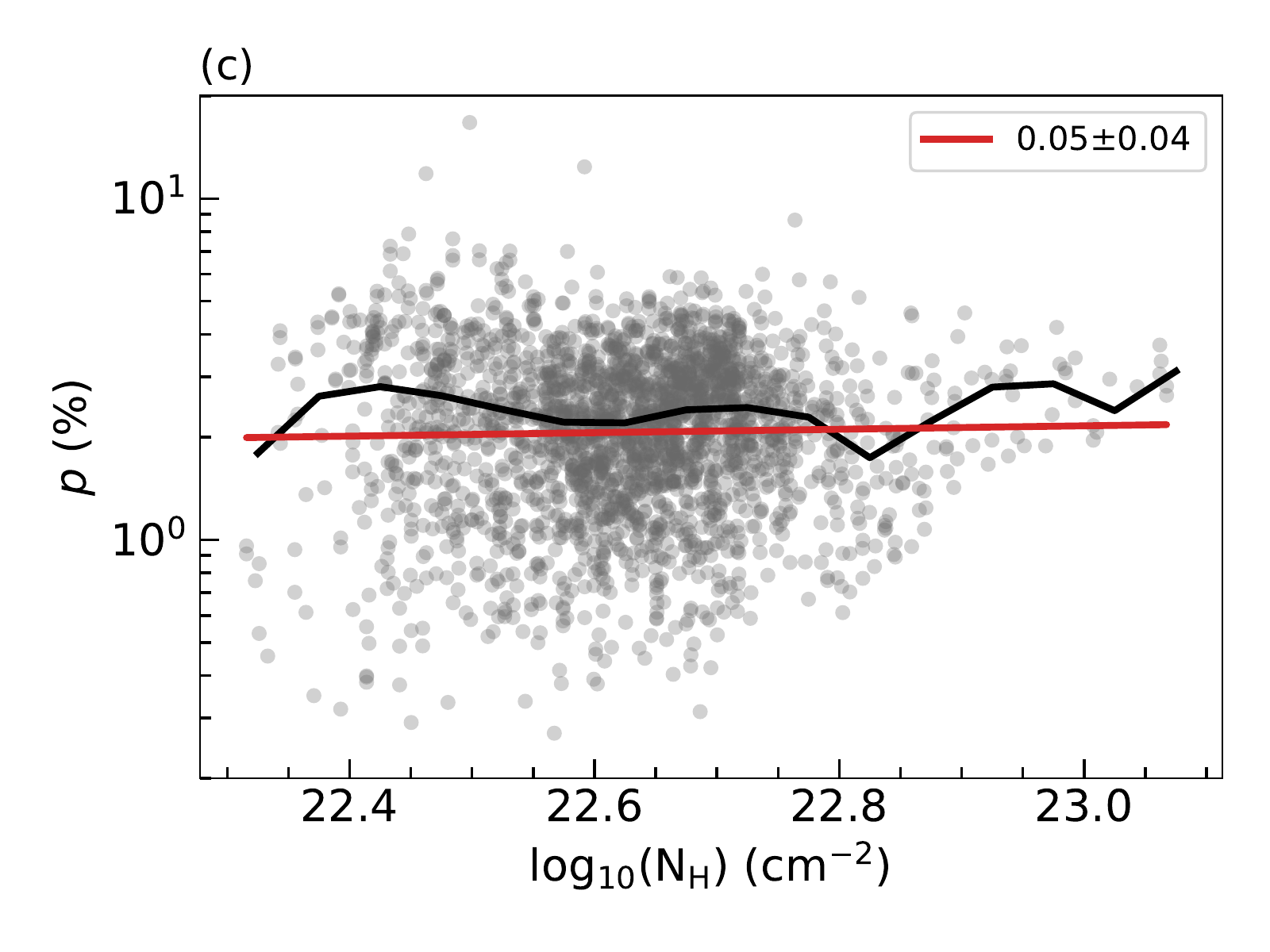}
    \caption{The relation between temperature and gas column density for the S53 data is shown in (a). Panel (b) and (c) show the variation of the degree of polarization with temperature and gas column density respectively. The black line shows the weighted means for the bins along temperature (a \& b) and column density (c). The red lines show a single linear fit to the data while the blue lines show the best fits using multiple power laws. The breakpoints where the slopes change are shown by the vertical blue lines. The corresponding slopes for each of the fits are shown in the legends.}
    \label{fig:Sofia5_T_NH_p}
\end{figure*}

\begin{figure*}
    \centering
    \includegraphics[scale=0.36]{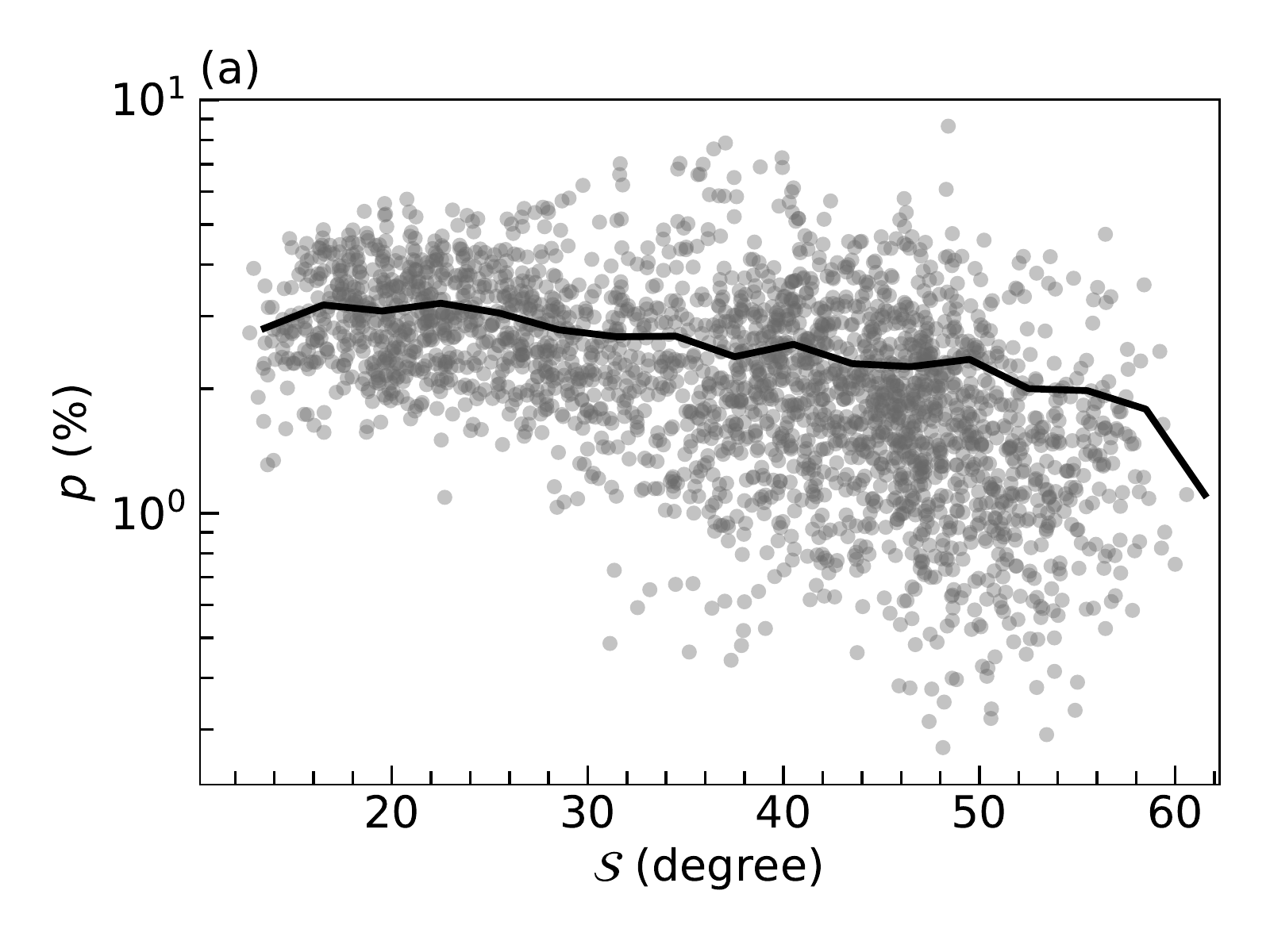}
    \includegraphics[trim={0 0 1cm 0}, scale=0.33]{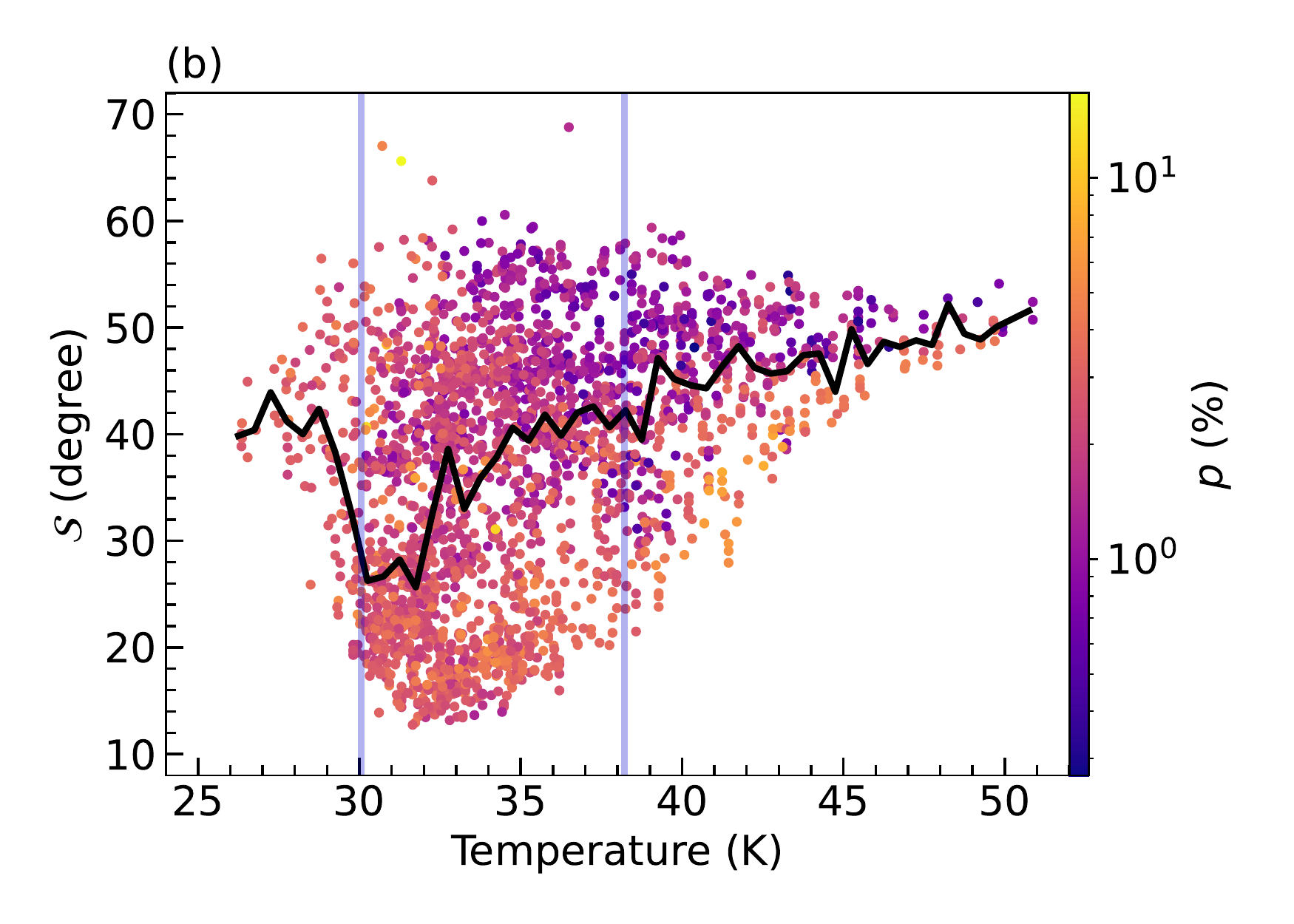}
    \includegraphics[scale=0.33]{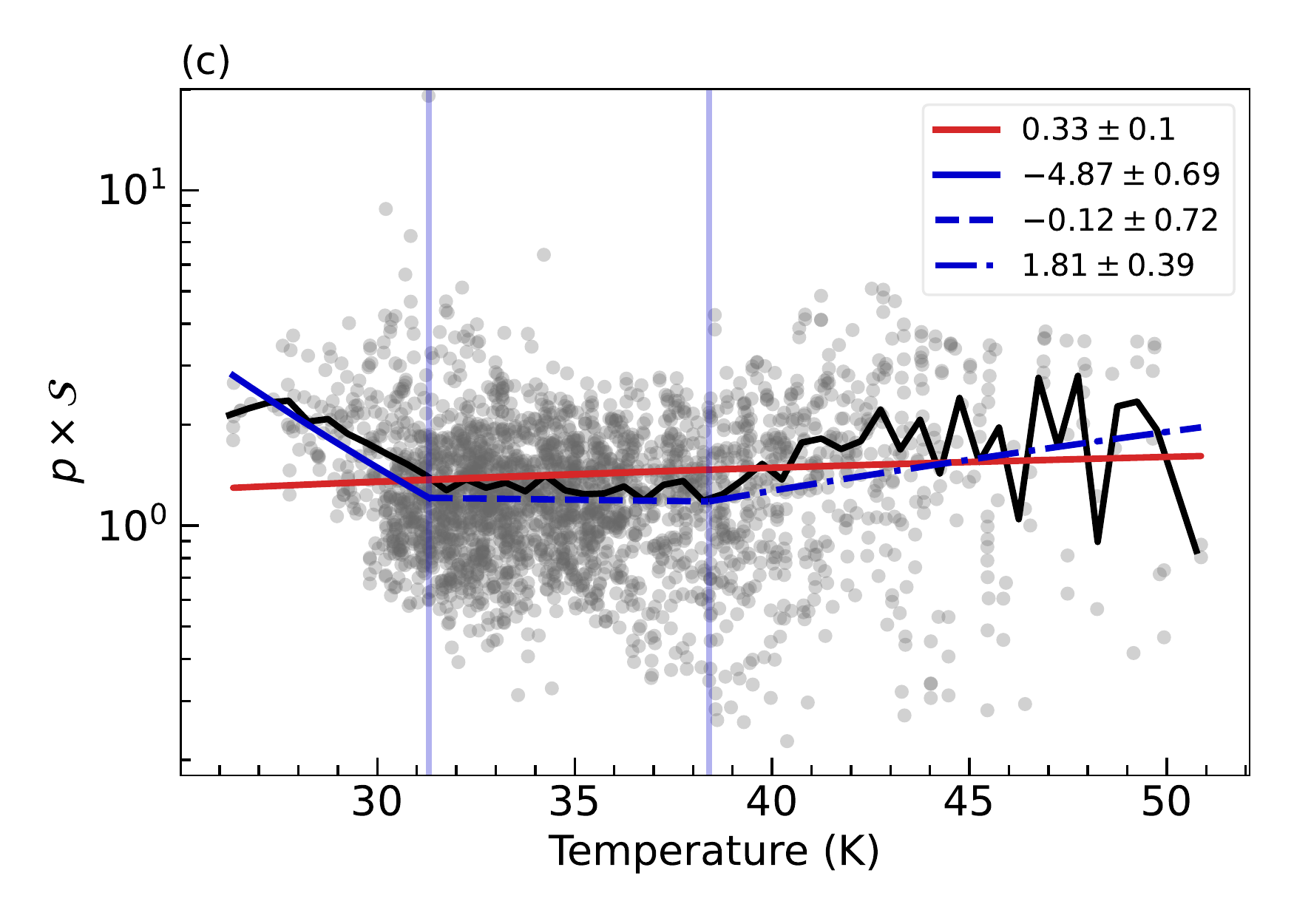}
    \caption{Analysis of the effect of magnetic field tangling on the observed polarization degree for the S53 data. Figure (a) shows the relation between the degree of polarization and polarization angle dispersion function. The inverse relation observed is the expected trend of the data as a higher order of tangling results in low alignment efficiency of the grains. The variation of the polarization angle dispersion function with temperature is shown in (b). The grain alignment efficiency can be traced by the parameter $p\times\mathcal{S}$ which removes the effect of the tangled magnetic field. Its relation with temperature is shown in (c) along with the best fit slope values.}
    \label{fig:Sofia5_S}
\end{figure*} 

The intensity of S53 traces the temperature and not the column density as can be seen from the color map. Hence, its behavior is different from S216 and J850. We first see a drop in $p$ with increasing intensity where $\alpha=0.45\pm0.07$, followed by a steep increase in $p$ with intensity where $\alpha=-0.60\pm1.53$. The data points in the second part of the fit are more scattered, but there is a definite concentration of points that show an increase in $p$ with $I$. The second slope can be explained in the context of RAT-A theory where we expect a rise in polarization at higher temperatures due to an increase in RATs on the dust grains. The drop in $p$ with $I$ in the first slope could be a result of the low temperature and high column density in this region, which makes alignment difficult due to gas randomization and low RATs. To understand the alignment physics in more detail, we look at how the polarization degree varies with the local physical parameters in the following section.

\subsection{Polarization Degree versus Dust Temperature, Gas Column Density, and Polarization Angle Dispersion Function}
The dust temperature, gas density, and the orientation of the magnetic field determine the level of dust polarization in any given environment. The $p-T_{\rm d}$ relation is a good indicator for the alignment due to RATs, while the $p-N_{\rm H}$ relation probes the effect of gas randomization due to collisions. The level of the tangled magnetic field is measured using polarization angle dispersion function (\mS). The $p-\mathcal{S}$ relation gives a measure of the effect of magnetic field tangling on the observed polarization. These relations can probe the dominant mechanism of grain alignment. We have used the $T_{\rm d}$ and $N_{\rm H}$ derived from the \emph{Herschel} data. The polarization angle dispersion function was first introduced by \citet{PlanckXIX2015} to quantify the regularity of the observed magnetic field. It is defined as,
\begin{equation}
    \mathcal{S}(r,\delta) = \sqrt{\frac{1}{N}\sum_{i=1}^{N}[\psi(r+\delta_i) - \psi(r)]^2},
\end{equation}
where the summation is taken over all the N pixels surrounding the central pixel at $r$ and displaced by a displacement vector $\delta_i$. The average for each pixel is taken over an annulus around it with a radius (lag) of $|\delta|$ containing \emph{N} pixels.

We have used lag equal to two times the beam size in each of the chosen observations. This equates to a $|\delta|$ of 9.7\arcsec, 36.4\arcsec, and 40\arcsec\ for S53, S216, and J850 respectively. The observed relations for each of the observations are discussed in detail in the following sections.

\subsubsection{SOFIA/HAWC+ 53 \micron}
The variation of derived $T_{\rm d}$ and $N_{\rm H}$ in the S53 region along with their individual relationship with $p$ in shown in Figure \ref{fig:Sofia5_T_NH_p}. Fig \ref{fig:Sofia5_S} shows the relation between $p$, $T_{\rm d}$, and \mS. As in the case of $p-I$ relationship, multiple power laws fit the data better than a single power law. There is a steep drop in the column density with the raise in temperature except in the region between 34 K $\leqslant T_{\rm d}\leqslant$ 38 K where the column density is almost a constant with temperature (Figure \ref{fig:Sofia5_T_NH_p}a). The $p-T_{\rm d}$ relationship is shown in Figure \ref{fig:Sofia5_T_NH_p}b. We observe three slopes that fit best to the data with breakpoints at 30.05 K and 38.21 K. The first slope of $-0.8\pm1.41$ shows the least variation with few data points. This is most likely due to the low temperature and high gas density in this region (as seen in Figure \ref{fig:Sofia5_T_NH_p}a) where the extinction of radiation and gas randomization could be hindering the grain alignment, leading to a shallow drop in $p$ at low $T_{\rm d}$. The second slope of $-2.12\pm1.42$ is a much steeper relation with the degree of polarization dropping rapidly with raise in temperature. This is not the expected trend for RAT-A where the grains should be more aligned at higher temperatures due to an increase in RATs. The last slope shows this expected positive trend with a slope of $0.93\pm0.4$ though the data points seem to be a lot more scattered in this region with some of the points having a lower polarization degree. 

From Figure \ref{fig:Sofia5_T_NH_p}c which shows the $p-N_{\rm H}$ relation, it can be seen that there is no direct correlation between the column density and the observed degree of polarization with an almost flat slope of $0.05\pm0.04$. The anti-correlation between \mS\ and $p$ shown in Figure \ref{fig:Sofia5_S}a has been observed before \citep{PlanckXII2020} and is a statistical property due to the topology of the magnetic field. Figure \ref{fig:Sofia5_S}b shows the \mS$-T_{\rm d}$ relation where we see a rise in \mS\ with $T_{\rm d}$ indicating an increase in magnetic field tangling with temperature. This leads to a drop in the observed polarization degree. The rise in \mS\ is steep in the region between the temperature break points from Figure \ref{fig:Sofia5_T_NH_p}b, which explains the anti-correlation observed in the $p-T_{\rm d}$ relation. 

\begin{figure*}
    \centering
    \includegraphics[scale=0.36]{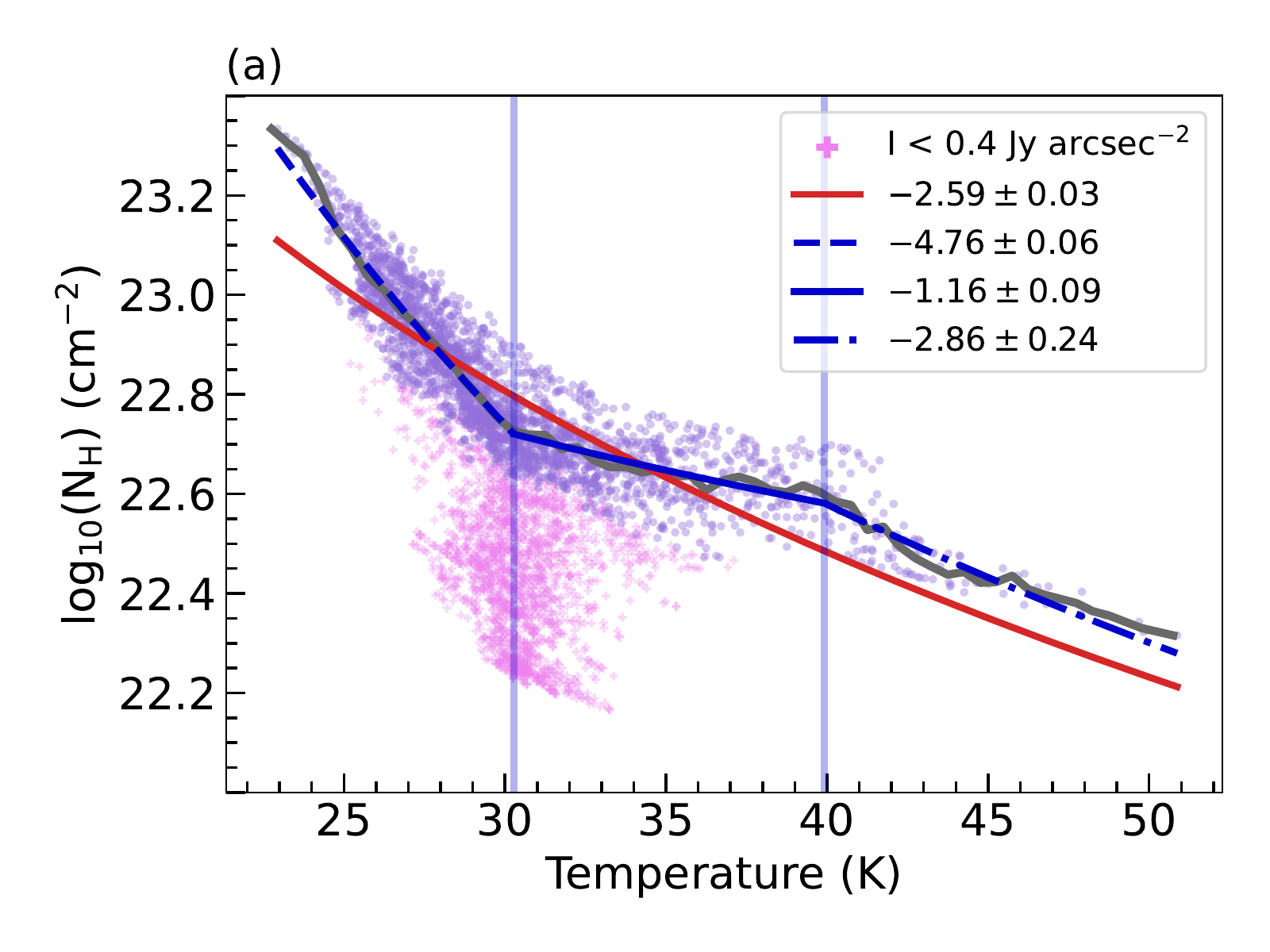}
    \includegraphics[scale=0.36]{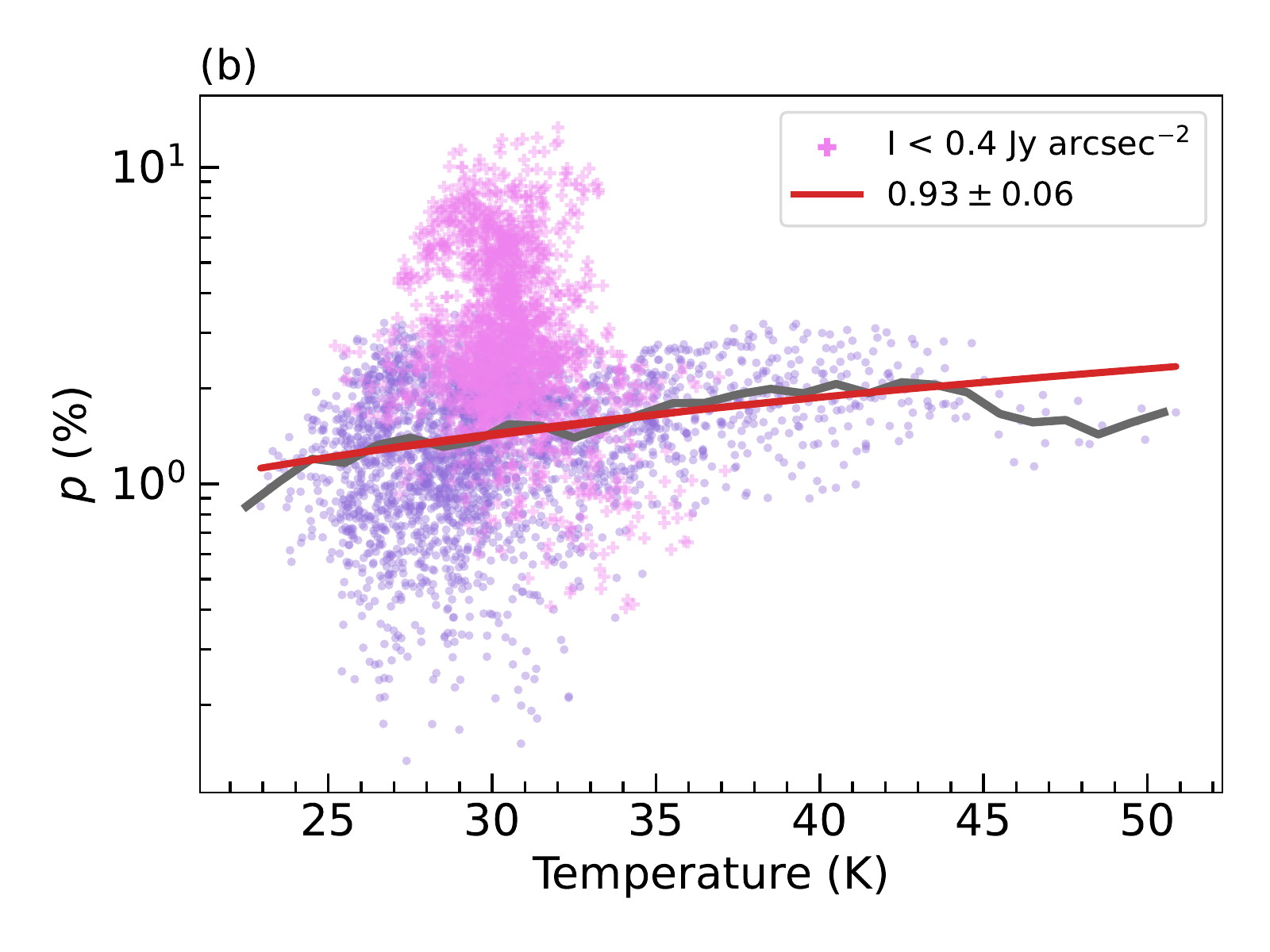}
    \includegraphics[scale=0.36]{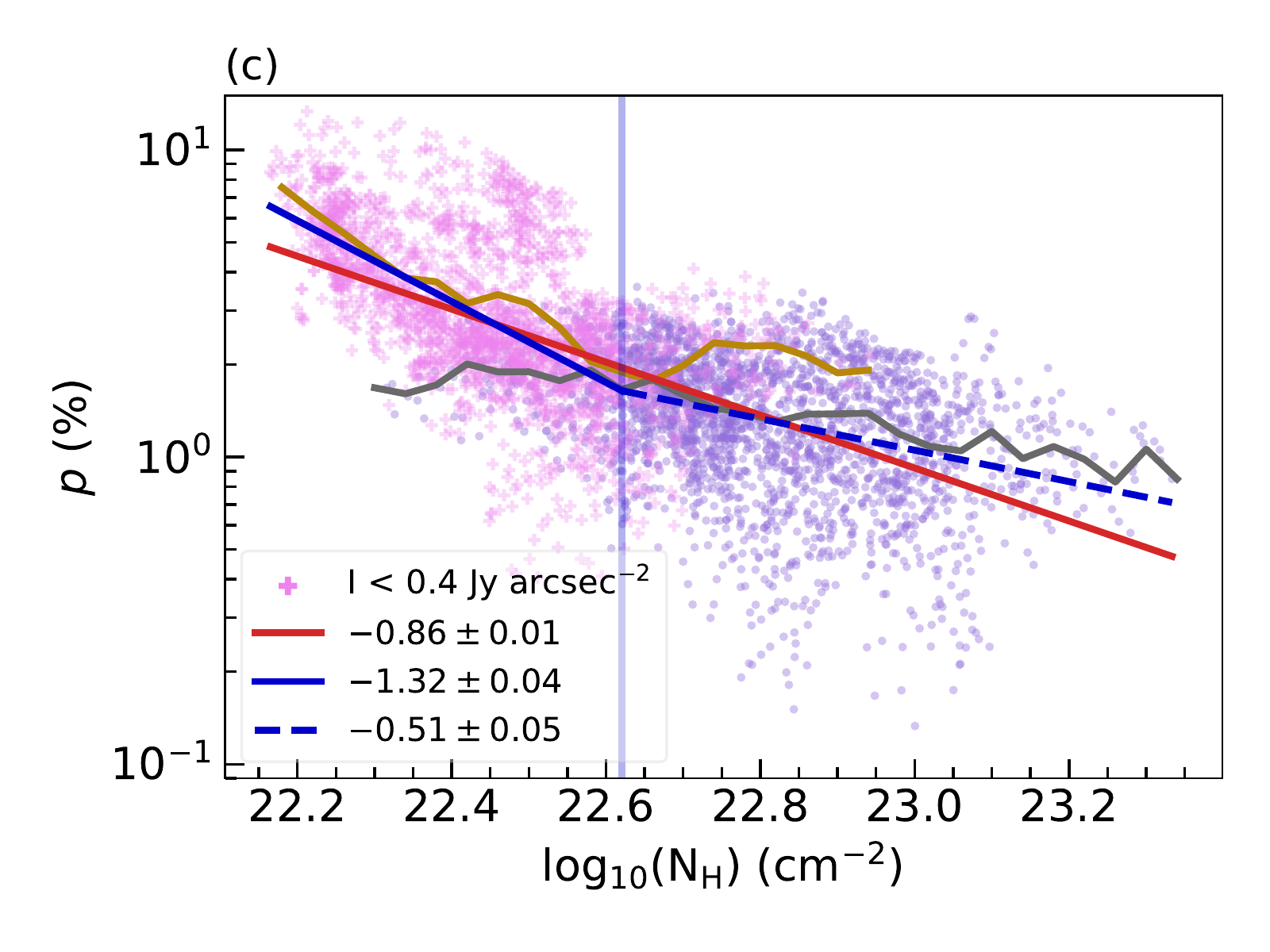}
    \caption{Same as Figure \ref{fig:Sofia5_T_NH_p} but for S216 data. The pink data points are from the regions where $I<0.4$ \jyarcsec\, which is the region in the top right corner in Figure \ref{fig:Sofia_10_pol_map} with the highest polarization degree. The weighted means are shown with gray lines for $I\geqslant0.4$ \jyarcsec\ and orange lines for $I<0.4$ \jyarcsec.}
    \label{fig:Sofia10_T_NH_p}
\end{figure*}

\begin{figure*}
    \centering
    \includegraphics[scale=0.36]{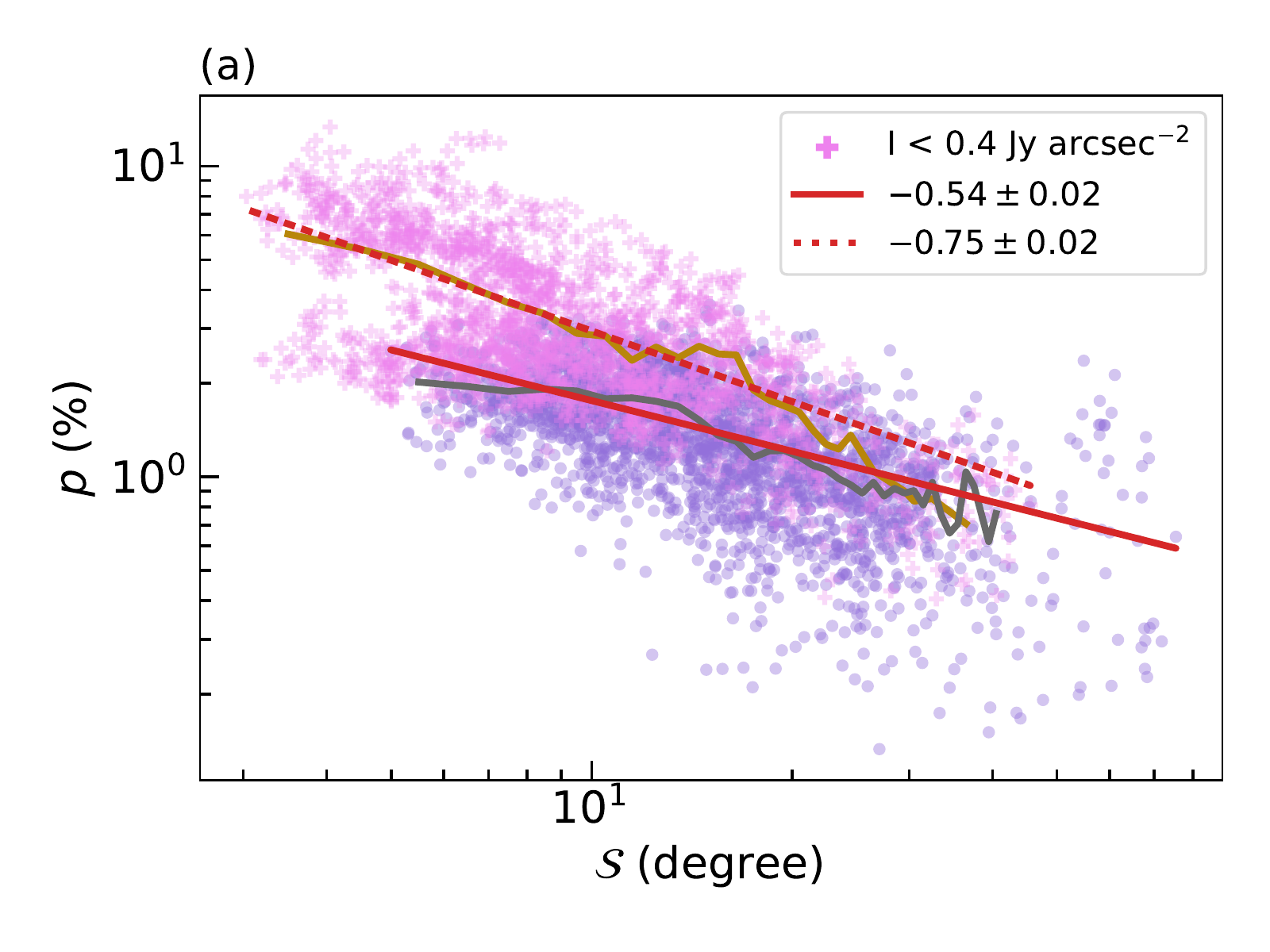}
    \includegraphics[scale=0.36]{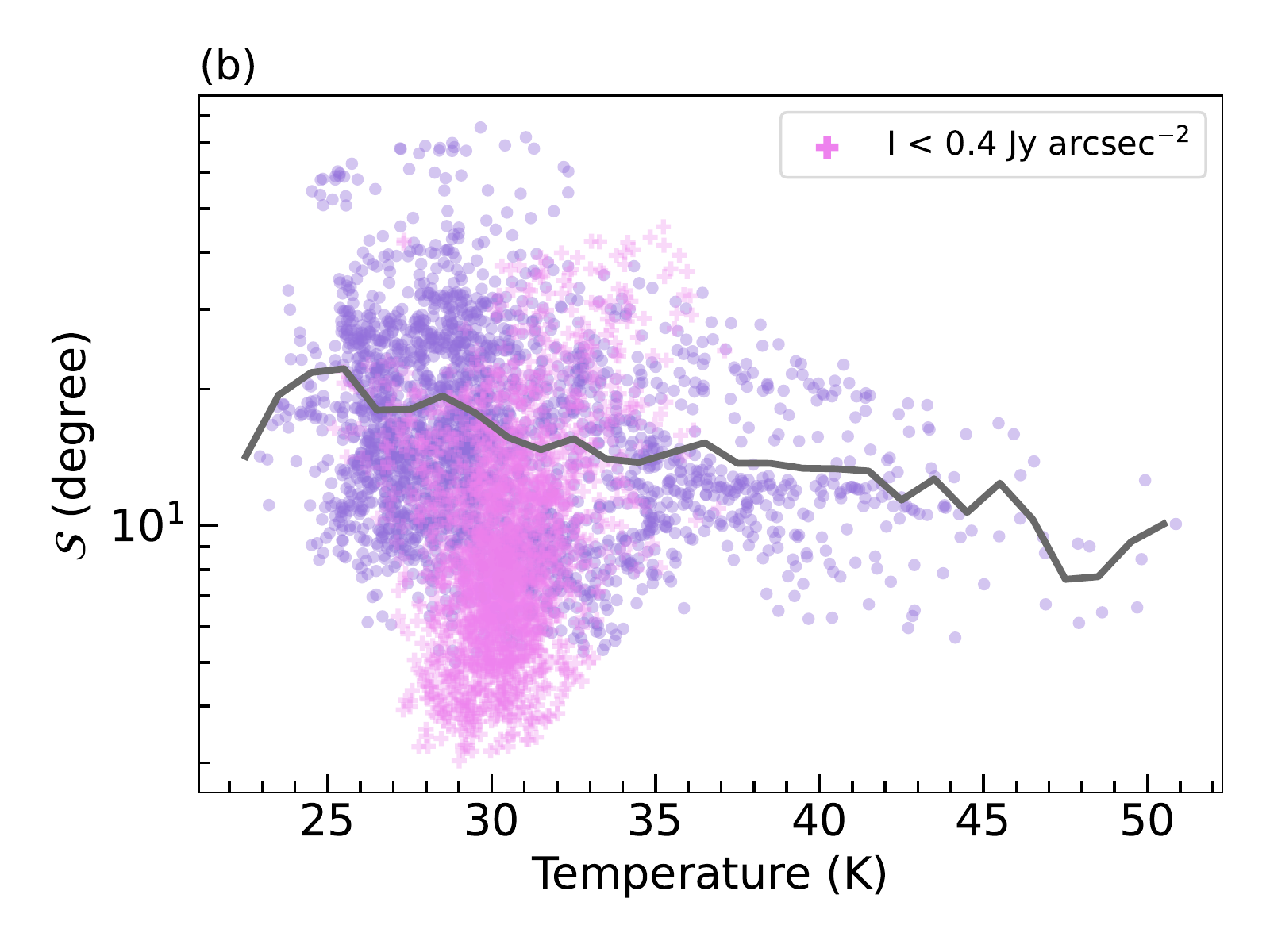}
    \includegraphics[scale=0.36]{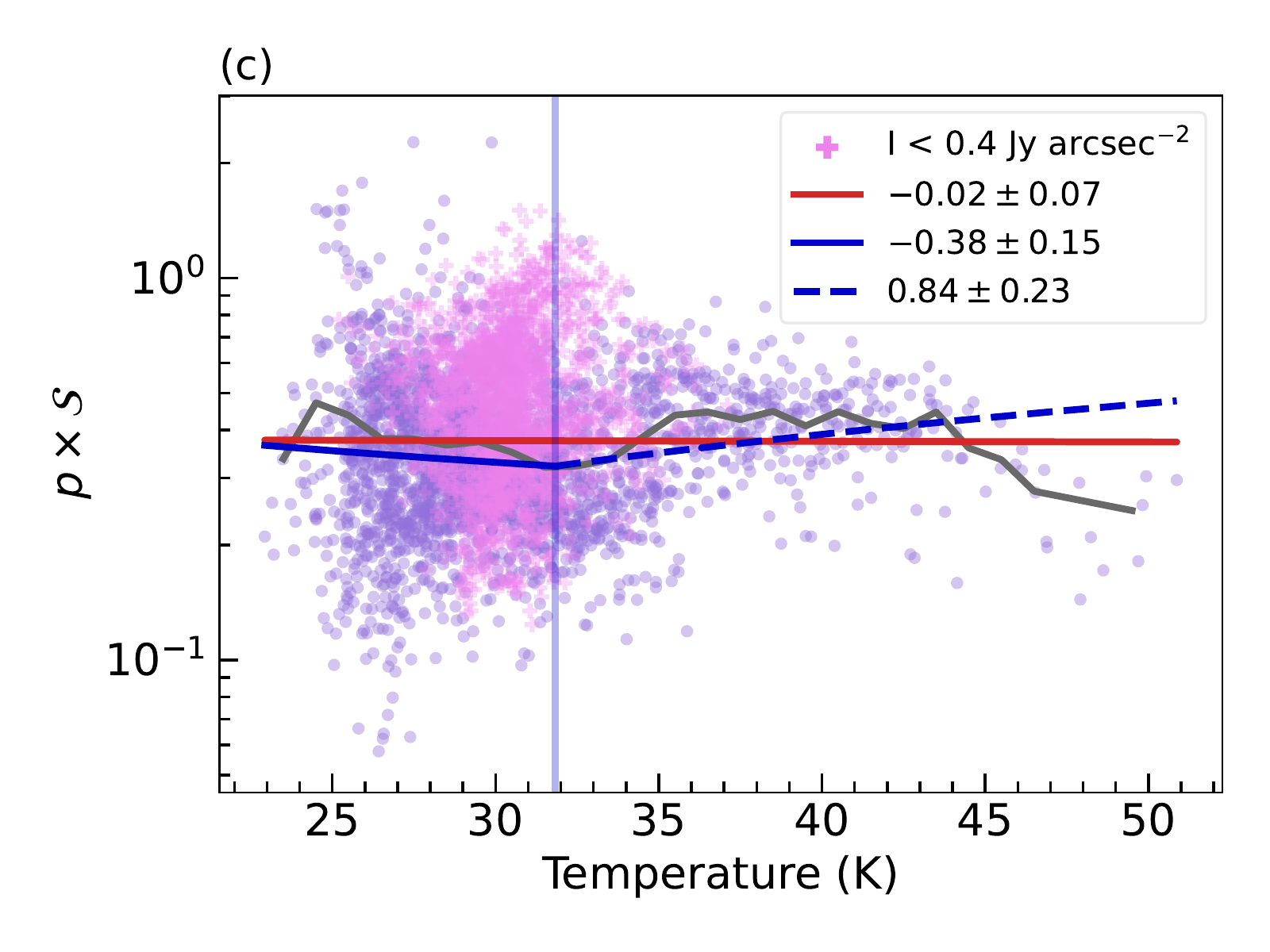}
    \caption{Same as Figure \ref{fig:Sofia5_S} but for the S216 data. The pink data points are from the regions where $I<0.4$ \jyarcsec\, which is the region in the top right corner in Figure \ref{fig:Sofia_10_pol_map} with the highest polarization degree. The weighted means are shown with gray lines for $I\geqslant0.4$ \jyarcsec\ and orange lines for $I<0.4$ \jyarcsec.}
    \label{fig:Sofia10_S}
\end{figure*} 

The impact of the properties of dust on the observed polarization is traced by the parameter $p\times$\mS\ which statistically negates the effects of a tangled magnetic field. This can be treated as a proxy to measure grain alignment, elongation, and composition. $p\times$\mS\ also depends on the depth of the dust probed along the line of sight as well as the ratio of the turbulent to ordered magnetic field \citep{PlanckXII2020}. This is shown in Figure \ref{fig:Sofia5_S}c as a function of temperature. The variation is less pronounced compared to the $p-T_{\rm d}$ and $\mathcal{S}-T_{\rm d}$ relation in the region between the break points 31.31 K $\leqslant T \leqslant 38.41$ K where we expect a significant contribution from \mS\ to the observed level of $p$. However, the relation at $T_{\rm d}\leqslant31.31$ K and $T_{\rm d}\geqslant38.41$ K is more pronounced, with a steeper drop and rise at low and high temperatures respectively. This can be interpreted as a loss of grain alignment in denser regions at low temperatures ($T_{\rm d}\leqslant31.31$ K) due to gas randomization and an enhancement of grain alignment at higher temperatures which leads to an increase in polarization ($T_{\rm d}\geqslant38.41$ K) according to RAT-A theory.

\subsubsection{SOFIA/HAWC+ 216 \micron}
We performed a similar analysis on the S216 data. The relation between $p$, $T_{\rm d}$ and $N_{\rm H}$ is show in Figure \ref{fig:Sofia10_T_NH_p}. Figure \ref{fig:Sofia10_S} shows the relation between $p$, $T_{\rm d}$ and \mS. The $T_{\rm d}-N_{\rm H}$ relation in this region is similar to that of S53 except for the data points clustered around $T_{\rm d}\sim 30$ K. These come from the region of least column density and high polarization degree and have an intensity $I<0.4$ \jyarcsec. They seem to follow a different physical trend compared to the rest of the data. Based on this observation, we divided the S216 region into two sub-regions with $I<0.4$ \jyarcsec\ and $I\geqslant0.4$ \jyarcsec.

The $p-T_{\rm d}$ relation is shown in Figure \ref{fig:Sofia10_T_NH_p}b. It can be seen that the $I < 0.4$ \jyarcsec\ does not follow the same trend in the relationship as the rest of the data and clusters around $T_{\rm d}\sim30$ K. The remaining data show a linear rise in the degree of polarization with raise in temperature which is in accordance with the RAT-A theory. The relation of $p$ with $N_{\rm H}$ is shown in Figure \ref{fig:Sofia10_T_NH_p}c. It is clear from the plot that there are two slopes in the relationship with the $I < 0.4$ \jyarcsec\ data points occupying the first slope. The $p-N_{\rm H}$ relation is sleeper for the $I<0.4$ \jyarcsec\ region (slope = $-1.32\pm0.04$). This drop in polarization can be attributed to the loss of grain alignment due to gas randomization and attenuation of radiation at higher column densities.  The second slope of $-0.51\pm0.05$ shows a gradual loss of polarization with column density applicable to the rest of the data with $I\geqslant0.4$ \jyarcsec.

\begin{figure*}
    \centering
    \includegraphics[scale=0.36]{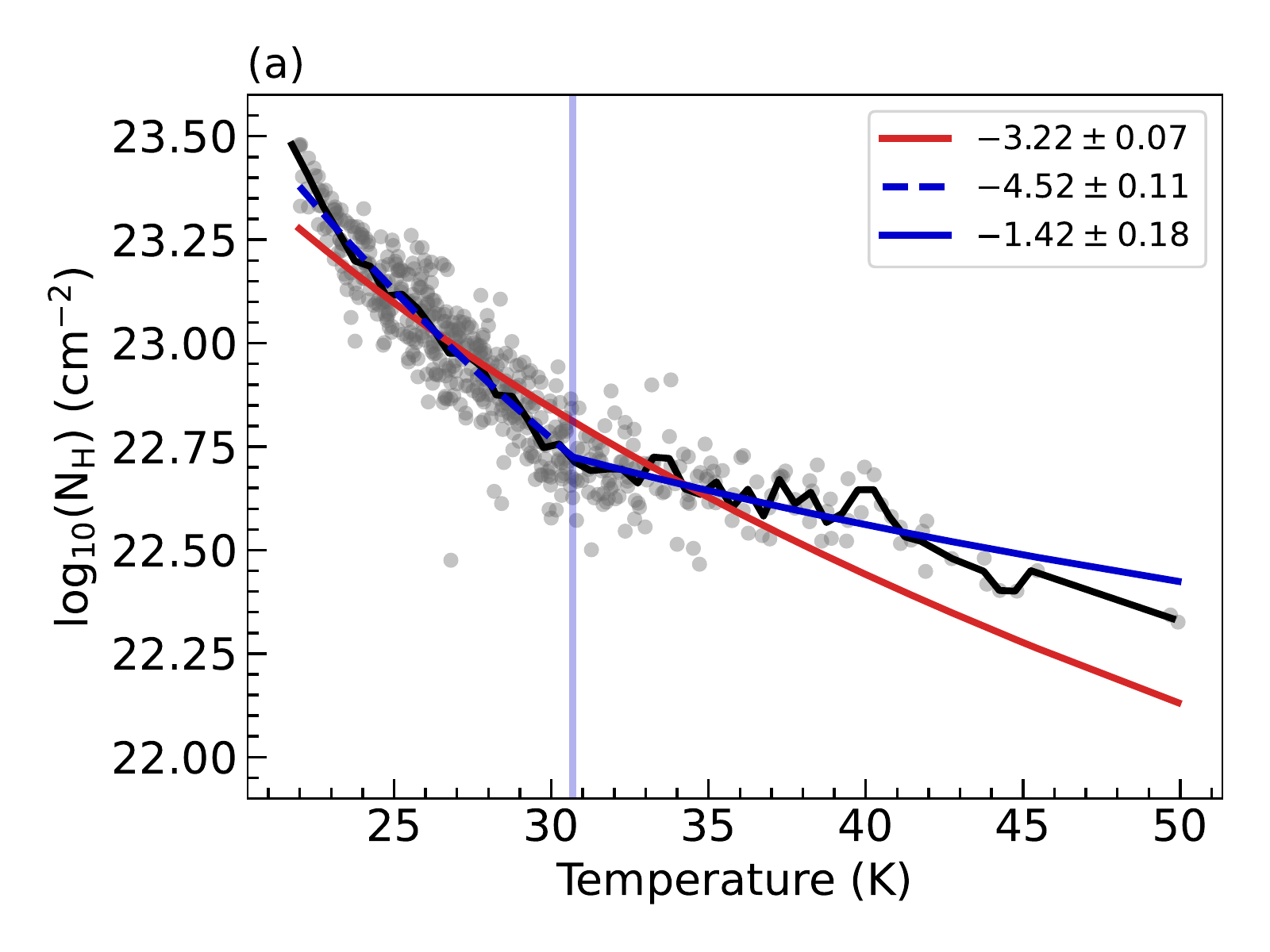}
    \includegraphics[scale=0.36]{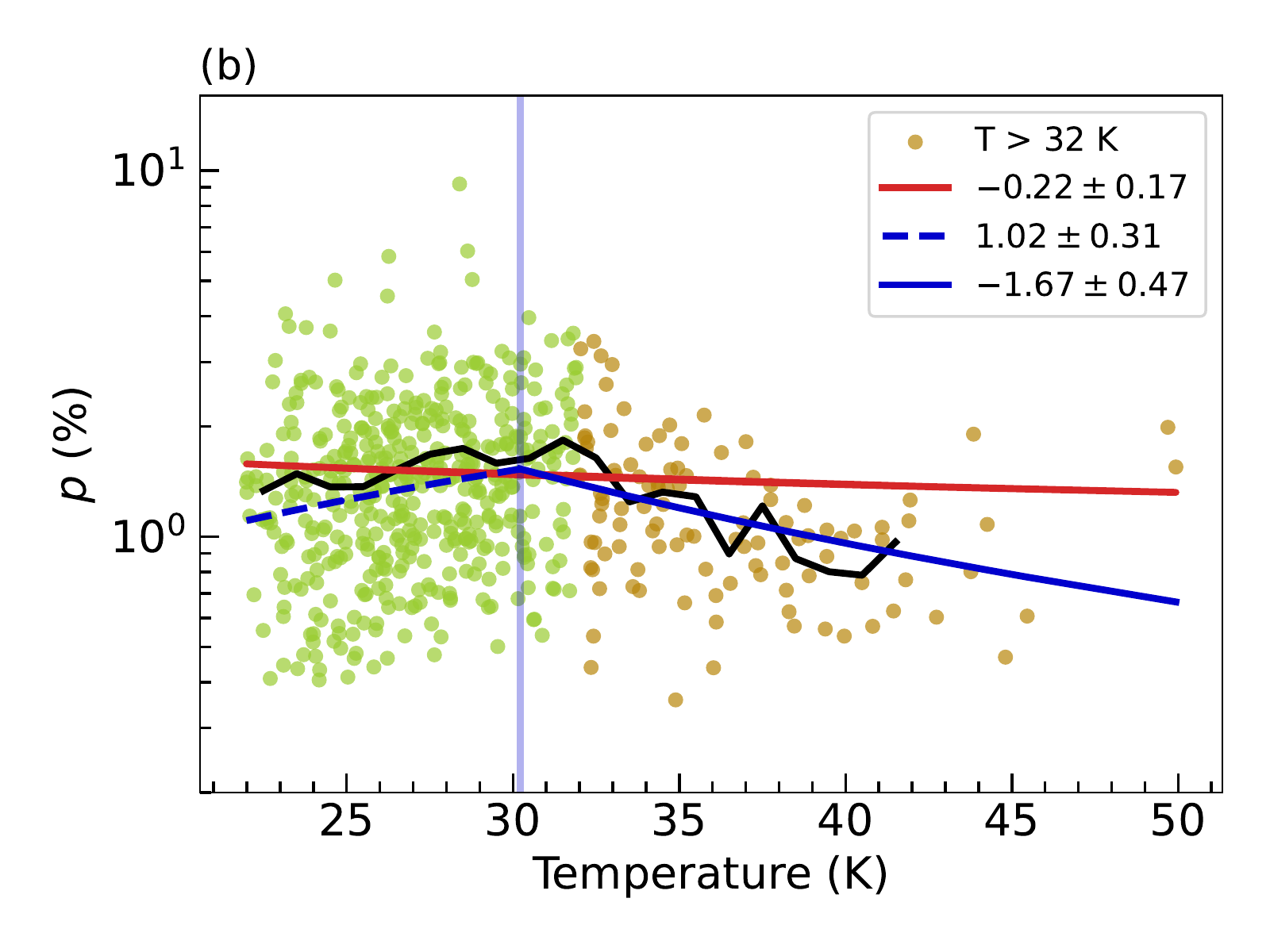}
    \includegraphics[scale=0.36]{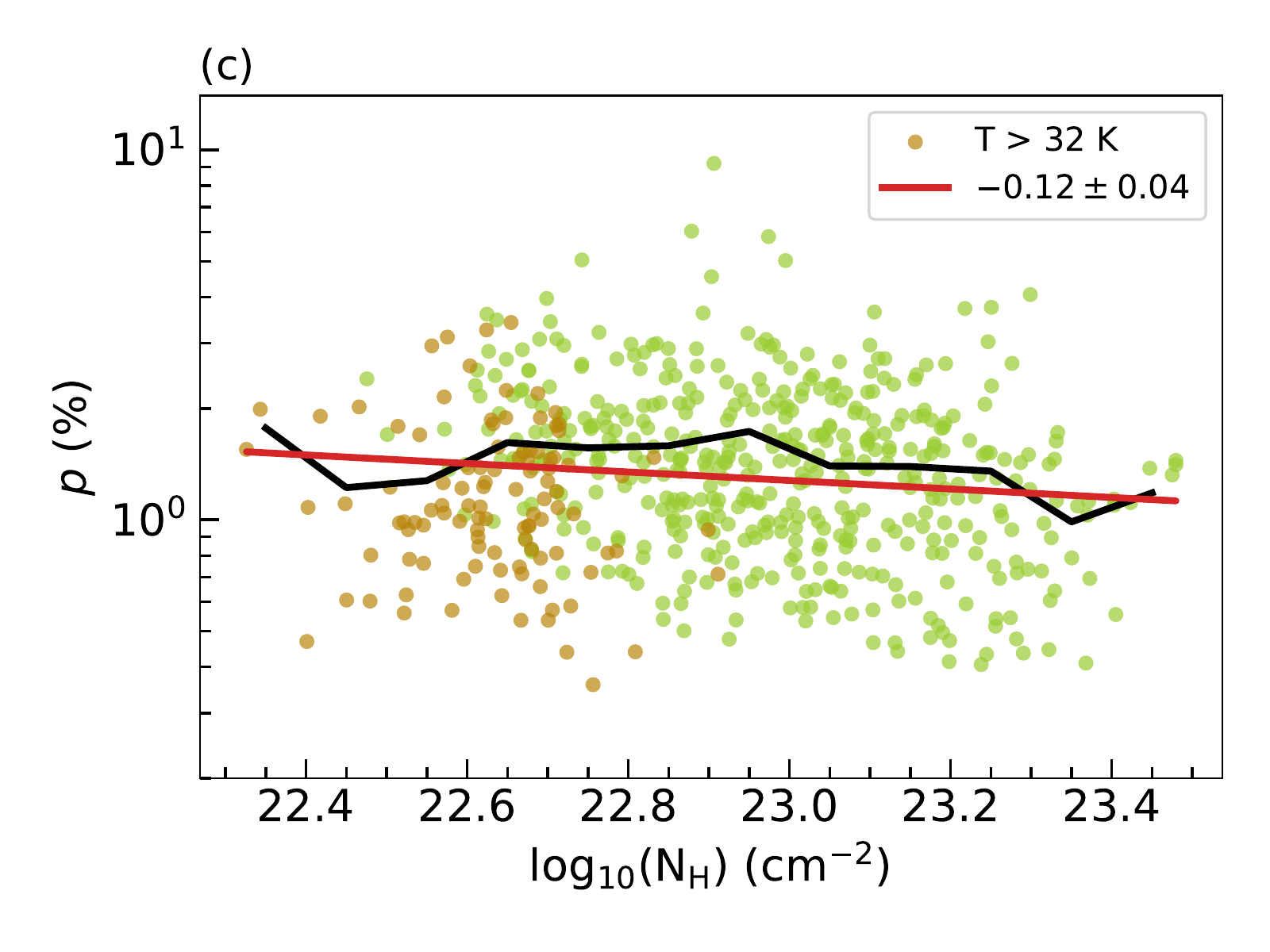}
    \caption{Same as Figure \ref{fig:Sofia5_T_NH_p} but for the J850 data. The orange data points are from a region where $T_{\rm d}>32$ K which is around the Galactic centre and the CND.}
    \label{fig:JCMT_T_NH_p}
\end{figure*}

\begin{figure*}
    \centering
    \includegraphics[scale=0.36]{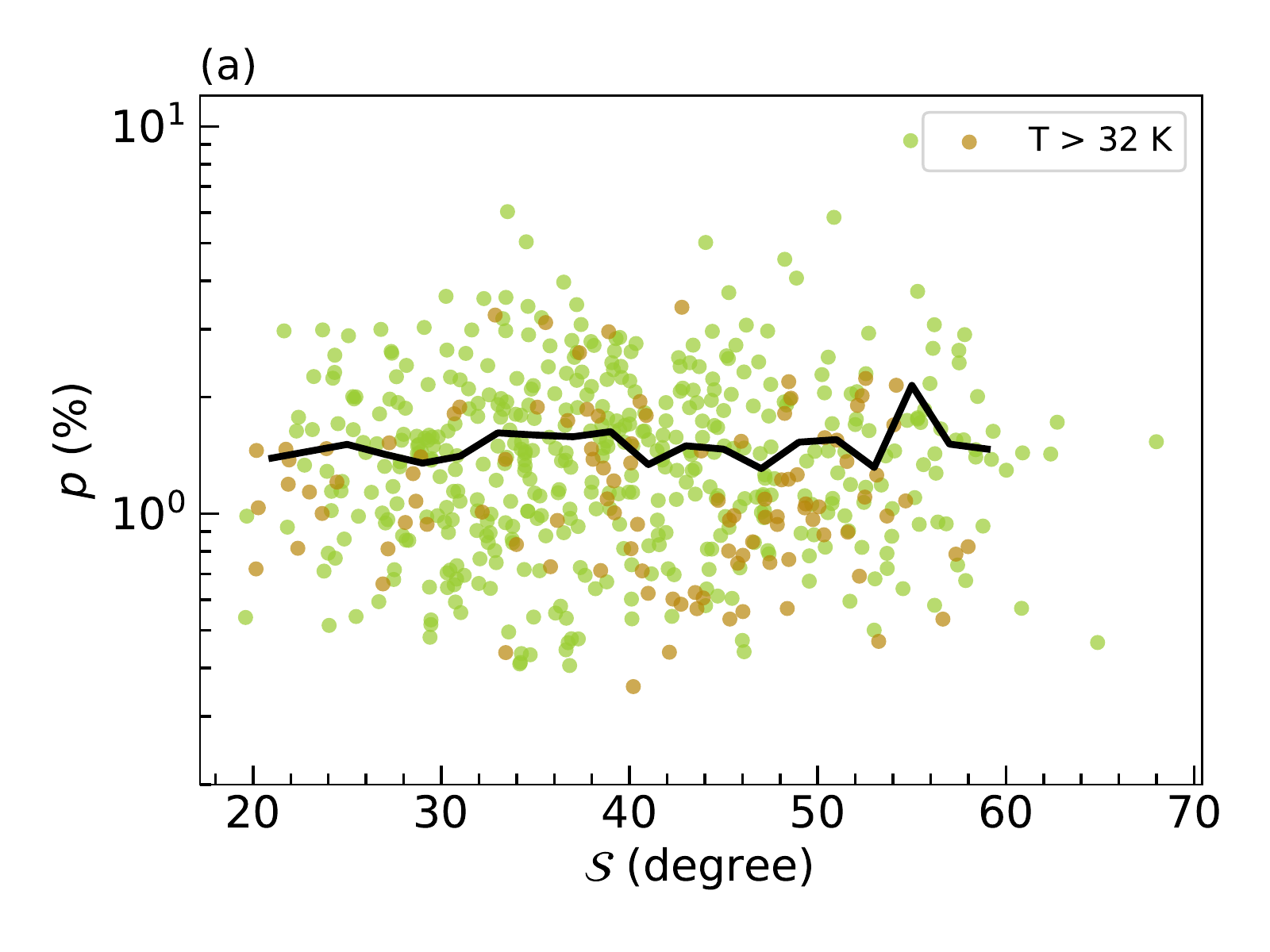}
    \includegraphics[scale=0.36]{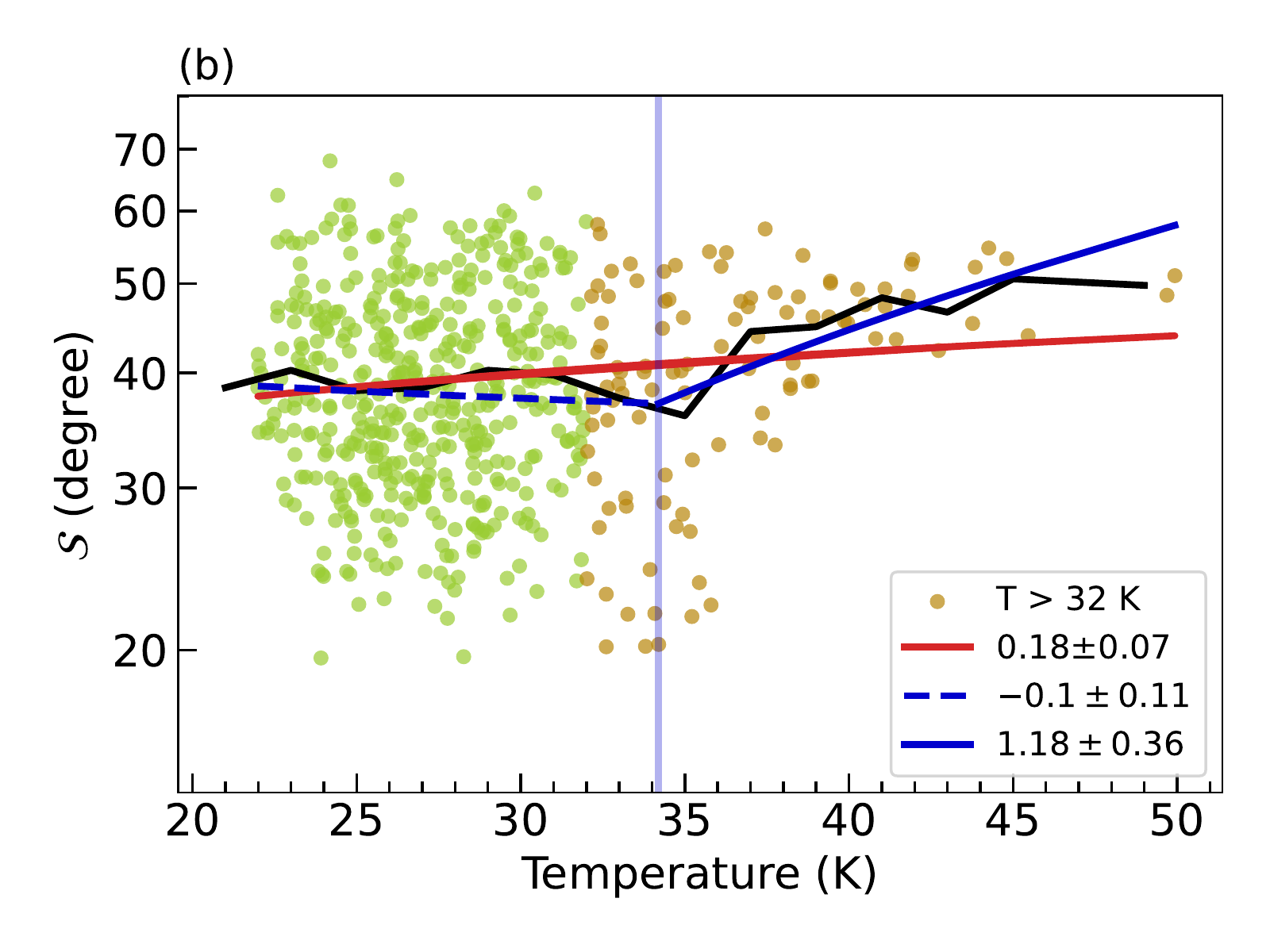}
    \includegraphics[scale=0.36]{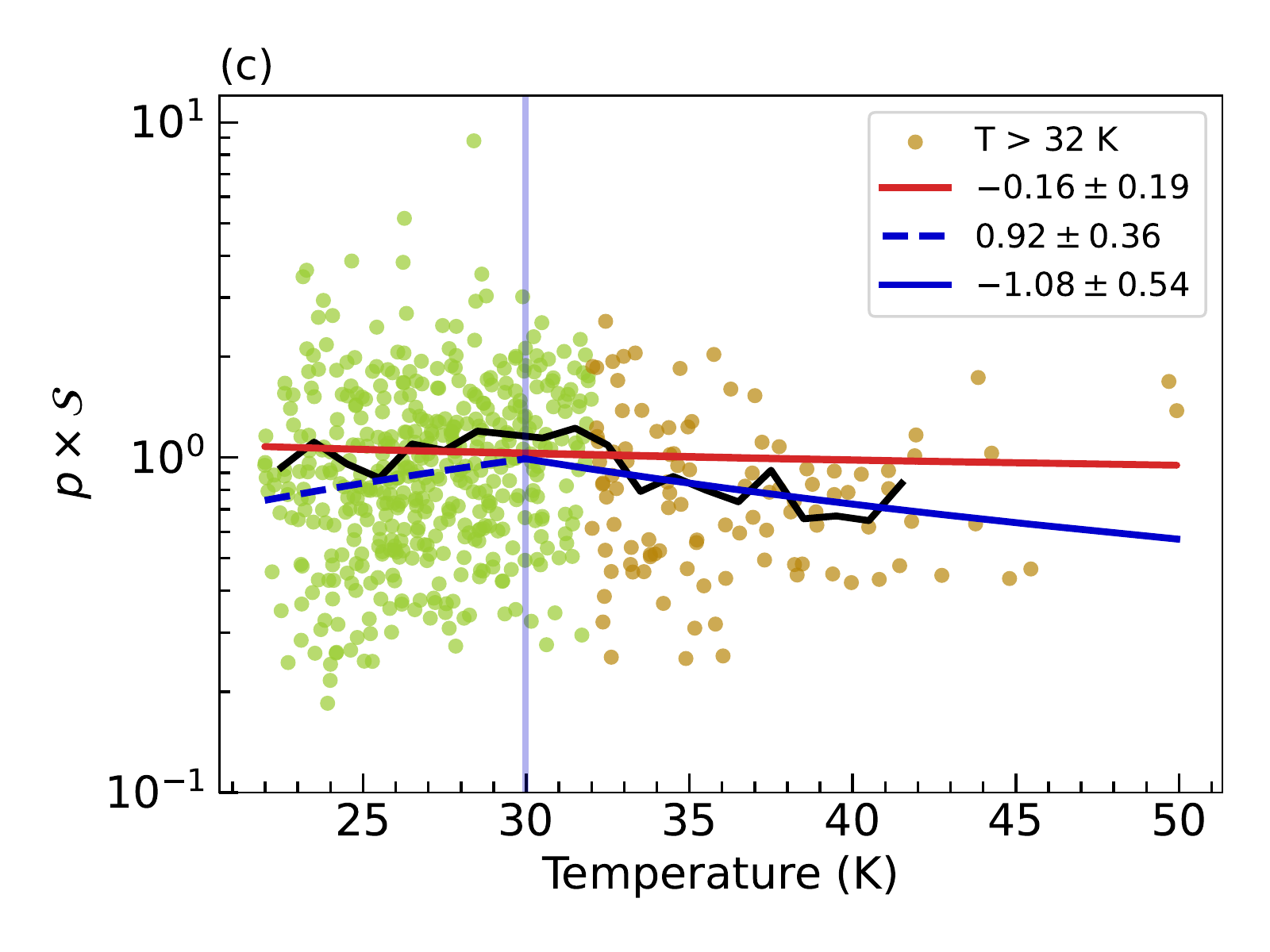}
    \caption{Same as Figure \ref{fig:Sofia5_S} but for J850 data. The orange data points are from a region where $T_{\rm d}>32$ K which is around the Galactic centre and the CND.}
    \label{fig:JCMT_S}
\end{figure*}

The $p-\mathcal{S}$ relation has the expected inverse relation with the $I<0.4$ \jyarcsec\ region showing a steeper drop in $p$ with \mS\ (slope = $-0.75\pm0.02$) compared to $I\geqslant0.4$ \jyarcsec\ region (slope = $-0.54\pm0.02$). There is a drop in \mS\ at higher temperatures as seen in Figure \ref{fig:Sofia5_S}b except for data with $I<0.4$ \jyarcsec\ which does not show any relation with \mS. But the $I<0.4$ \jyarcsec\ region does have an overall lower value of \mS\ implying a less tangled magnetic field along this line of sight, which can promote greater grain alignment and thus a higher degree of polarization. This can be seen in Figure \ref{fig:Sofia10_S}c where once the effect of magnetic fields is removed, the $I<0.4$ \jyarcsec\ region align with the rest of the data unlike in the case of Figure \ref{fig:Sofia10_T_NH_p}b. This shows that the higher polarization degree observed in the $I<0.4$ \jyarcsec\ is a result of a combination of low column density ($N_{\rm H}\lesssim4.17\times10^{22}$ \cm), less tangled magnetic field ($\mathcal{S}\lesssim20\degree$) and high enough temperature ($T_{\rm d}\sim30$ K). There is little variation in the $p\times\mathcal{S}$ relation with $T_{\rm d}$ for data with $I>0.4$ \jyarcsec\ and $T_{\rm d}<31.8\pm0.01$ K which is a high density ($N_{\rm H}\gtrsim4.17\times10^{22}$ \cm) and low temperature region. The alignment increases beyond this temperature where we also observe a drop in $N_{\rm H}$ and \mS\  promoting more efficient grain alignment. The observed features in this region follow RAT-A theory as well like in the case of S53.

\subsubsection{JCMT/SCUPOL 850 \micron}
J850 is the longest wavelength observation we have considered and also the one that covers the most physical distance around the Galactic centre ($\sim30$ pc). The resolution of J850 is much lower than S53 and S216 data with the physical pixel scales ranging from $\sim0.046$ pc, 0.18 pc, and 0.39 pc for S53, S216, and J850 data respectively. The relation between $T_{\rm d}-N_{\rm H}$ for the J850 region is shown in Figure \ref{fig:JCMT_T_NH_p}a. There is a steep drop in column density with raise in temperature upto $T_{\rm d}\sim30.8$ K beyond which the column density is almost a constant at $N_{\rm H}\sim4.5\times10^{22}$ \cm. The $p-T_{\rm d}$ relation from Figure \ref{fig:JCMT_T_NH_p}b clearly shows two different trends with the polarization degree increasing with temperature in the beginning as expected from RAT-A theory. But beyond $T_{\rm d}>30.22$ K there is a drop in $p$ with increasing temperature, which is contradictory to what we would expect from RAT-A theory at higher temperatures. Based on this declining trend we have selected the data points which show a clear drop in $p$ with $T_{\rm d}$, that is at about $T_{\rm d}>32$ K to trace their behavior with other parameters (shown as brown points in Figure \ref{fig:JCMT_T_NH_p} \& \ref{fig:JCMT_S}). The $p-N_{\rm H}$ relation from Figure \ref{fig:JCMT_T_NH_p}c does not show two distinct regions as in the case of $p-T_{\rm d}$ relation. The region with $T_{\rm d}>32$ K is clustered at lower column densities, again where we expect it to have a higher degree of polarization. The overall trend of $p-N_{\rm H}$ is a shallow decline in $p$ with $N_{\rm H}$ with a slope of $-0.12\pm0.04$. 

From Figure \ref{fig:JCMT_S}a we can see that there is no apparent correlation between \mS\ and $p$. There is an increase in \mS\ at higher temperatures as seen in Figure \ref{fig:JCMT_S}b, especially for the $T_{\rm d}>35$ K region which can explain the drop in $p$ at $T_{\rm d}>35$ K. If \mS\ is the main contributor to the drop in $p$, there should be a flattening in the relation between $p\times\mathcal{S}$ and $T_{\rm d}$ where the effect of \mS\ on $p$ is removed. But as seen in Figure \ref{fig:JCMT_S}c, though the slope beyond $T_{\rm d}>35$ K is smaller than in the case of $p-T_{\rm d}$ slope, there is still a significant drop in $p$ with $T_{\rm d}$ which cannot be attributed to \mS\ alone. This might be evidence for the RAT-D mechanism, where the large grains are disrupted due to RATs. Since the thermal emission at 850 \micron\ predominantly comes from large grains, this is a reasonable explanation. The drop in $p$ with $T_{\rm d}$ is reported in earlier studies by \citet{PlanckXII2020,Santos2019,Tram2021Doradus1,Tram2021Ophi,HoangM172022}. We will look at the RAT-A and RAT-D mechanisms in detail in the following section.

\section{Grain Alignment and Rotational Disruption by RATs} \label{Grain Alignment}
Here we discuss the implications of observational data for constraining the grain alignment and rotational disruption by RATs. According to the RAT-A theory, the degree of polarization is determined by the size distribution of the dust grains that are aligned due to the radiation field \citep{Lee2020,Hoang2021}. This distribution depends on the local gas density and the dust temperature as the RAT-A is a result of the balance between the spin induced by the RATs and damping caused by gas collisions. The minimum size of the grains that can be aligned due to RATs follows the relation $a_{\mathrm{\rm align}} \propto n_{\rm H}^{2/7} T_{\rm d}^{-12/7}$ \citep{Hoang2021,Tram2021Doradus1}. Thus the grains have a smaller minimum alignment size in hotter and low-dense environments compared to denser and cooler environments. Assuming the maximum size of grains to be constant, the size distribution of grains that can be aligned is broader when the minimum size of the grains that can be aligned is smaller, resulting in a higher polarization degree. The amount of polarized emission from a population of grains depends on this minimum size of grain alignment as well as the maximum size of gains that can be aligned by various physical processes. The RAT-D mechanism can also set an upper cutoff on the maximum size of grains, beyond which they are disrupted into smaller grains ($a_{\mathrm{disr}}$). The critical sizes of grains that determine the level of the observed polarization at the Galactic centre are described below.

\subsection{Critical sizes of grain alignment: $a_{\mathrm{\rm align}}$ and $a^{\mathrm{Lar}}_{\mathrm{max}}$}
\begin{figure}
    \centering
    \includegraphics[trim={0.5cm 0 0 0}, scale=0.55]{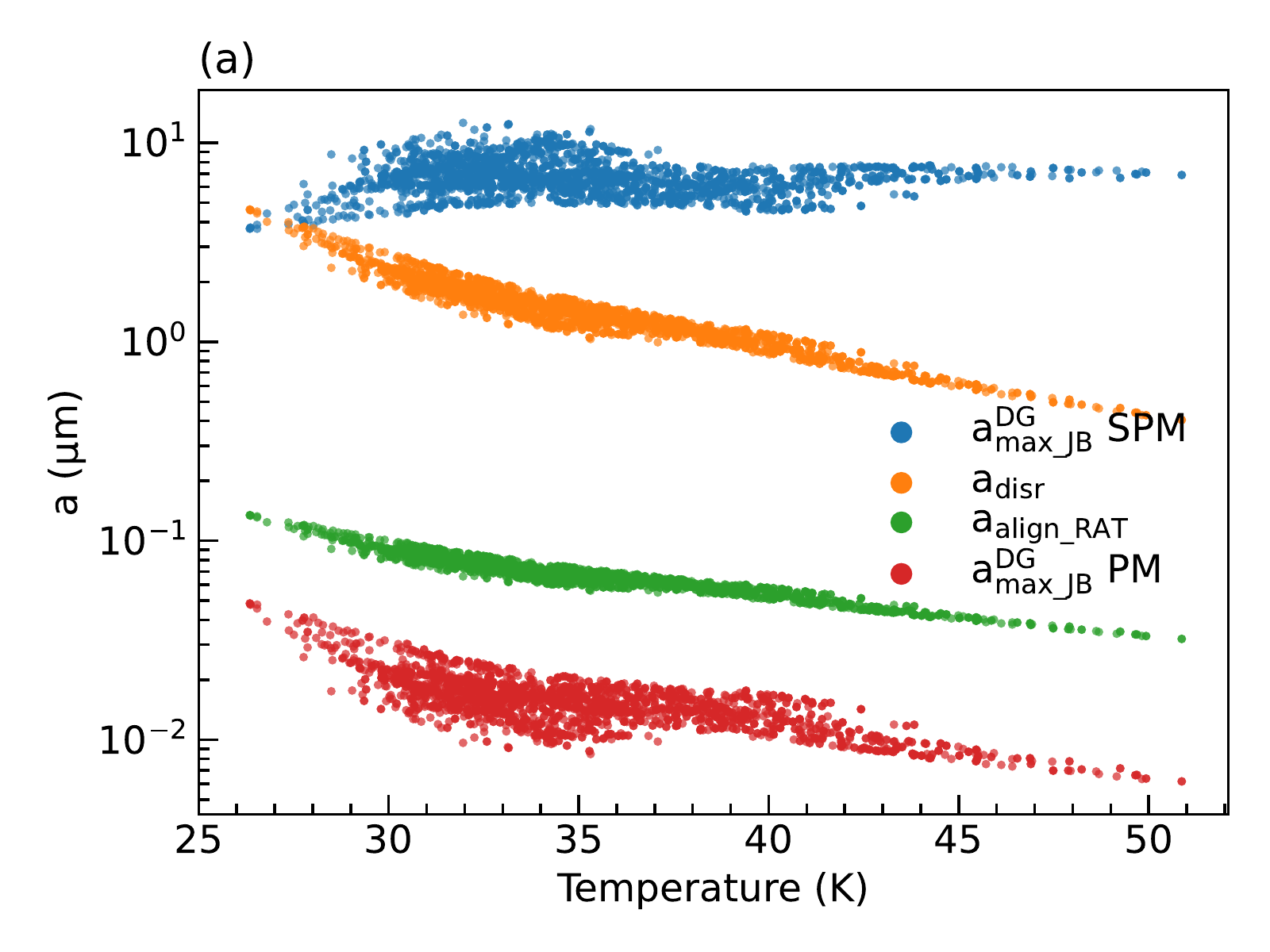}
    \includegraphics[trim={0.5cm 0 0 0}, scale=0.55]{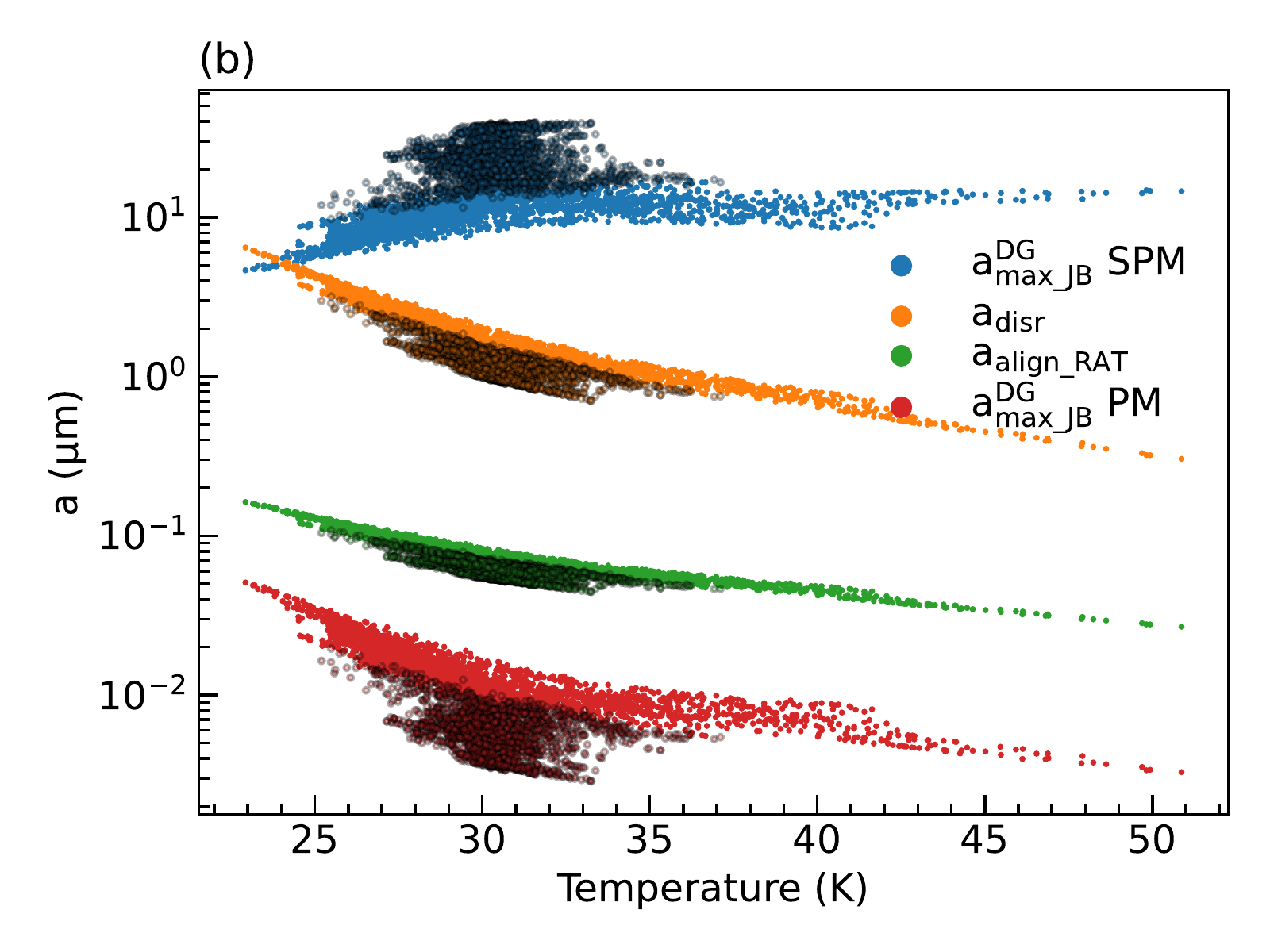}   
    \includegraphics[trim={0.5cm 0 0 0}, scale=0.55]{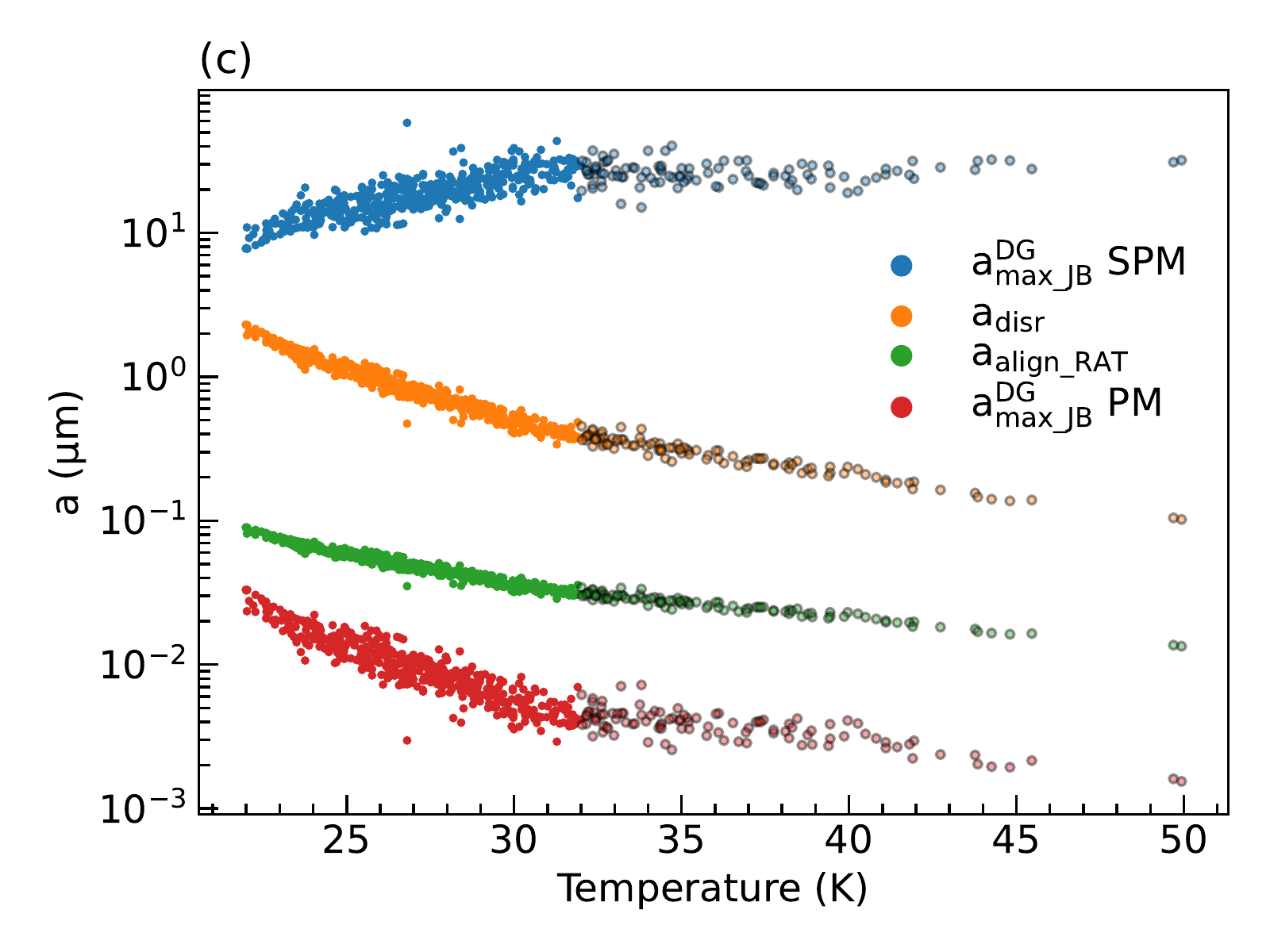}
    \caption{Grain size distribution for RAT-A, RAT-D, and MRAT mechanisms as a function of temperature calculated for S53 (a), S216 (b), and J850 (c) environments respectively. The points with black circles correspond to $I<0.4$ \jyarcsec\ from S216 and $T_{\rm d}>32$ K from J850 data (refer Figure \ref{fig:Sofia10_T_NH_p} \& \ref{fig:JCMT_T_NH_p}).}
    \label{fig:a-distribution}
\end{figure}

Grain alignment due to radiative or mechanical torques occurs efficiently only when the grains are rotating suprathermally, i.e., at angular velocities ($\omega$) greater than the thermal value at the given temperature ($\omega_T=\sqrt{kT_{\rm gas}/I_{\rm a}}$, where $k$ is the Boltzmann constant, $T_{\rm gas}$ is the temperature of the gas, and $I_{a}$ is the principal moment of inertia of a spherical grain). Using the suprathermal rotation threshold of $\omega = 3\omega_T$ \citep{Hoang2008}, \citet{Hoang2021} derived the analytical formulae to calculate the minimum size of the aligned dust grains, denoted by $a_{\mathrm{\rm align}}$. We have used the same to calculate $a_{\mathrm{\rm align}}$ for each observation, assuming the typical values for the dust grain parameters with mean wavelength $\bar{\lambda}=1$ \micron\ and anisotropy degree $\gamma=0.5$. The values obtained are shown in Figure \ref{fig:a-distribution}. The expected drop in the $a_{\mathrm{\rm align}}$ with raise in temperature is observed for all the observations. For S216 data shown in Figure \ref{fig:a-distribution}b, the region which shows the highest polarization ($I<0.4$ \jyarcsec), marked by black circles has a lower alignment size than the other data points at the same temperature as a result of low density in this region. As in the case of S53 and S216, J850 also shows a drop in $a_{\mathrm{\rm align}}$ with temperature (Figure \ref{fig:a-distribution}d). This again cannot explain the drop in the $p-T_{\rm d}$ relation at $T_{\rm d}>32$ K observed in Figure \ref{fig:JCMT_S}c.

Dust grains with magnetic moment like paramagnetic (PM, with diffusely embedded iron atoms) or super-paramagnetic grains (SPM, containing iron atoms in the form of clusters) interact with the external magnetic field, producing Larmor precession of the grain angular momentum ($J$) around the direction of the ambient magnetic field ($B$). When this precession is faster than the gas randomization (also damping), magnetic alignment occurs, i.e., the magnetic field becomes the preferred direction of grain alignment (Fig \ref{fig:time_scale}). As in the case of the diffuse ISM where the Larmor precession timescale ($\tau_{B}$) is much shorter than the gas damping ($\tau_{\rm gas}$) or the radiative alignment timescales ($\tau_{k}$), in the region of strong magnetic fields like the Galactic centre, magnetic alignment dominates because $\tau_{B}\ll\tau_{\rm gas}<\tau_{k}$. 

\renewcommand{\thefootnote}{\fnsymbol{footnote}}
The maximum size of grains that can be magnetically aligned constrained by the Larmor precession is given by $a^{\mathrm{Lar}}_{\mathrm{max}}\propto Bn_{\rm H}^{-1}T_{\rm gas}^{-1/2}T_{\rm d}^{-1}$\footnote[1]{throughout this paper we assume $T_{\rm gas}=T_{\rm d}$} \citep{Hoang2022AJ,GiangPolaris2022}. For the Galactic centre region, we assume the magnetic field $B\sim5$ mG based on previous estimates \citep{Aitken1986,Aitken1998} and our measurement using SOFIA/HAWC+ data (Akshaya et al. in preparation). In the presence of this strong magnetic field, $a^{\mathrm{Lar}}_{\mathrm{max}}$ has a large value of about $\sim$10 mm even for PM grains. From the RAT paradigm alone, $a_{\rm align}$ and $a^{\mathrm{Lar}}_{\mathrm{max}}$ define the lower and upper limits of the grain size distribution that can be aligned with the magnetic field.

\begin{figure}
    \centering
    \includegraphics[scale=0.5]{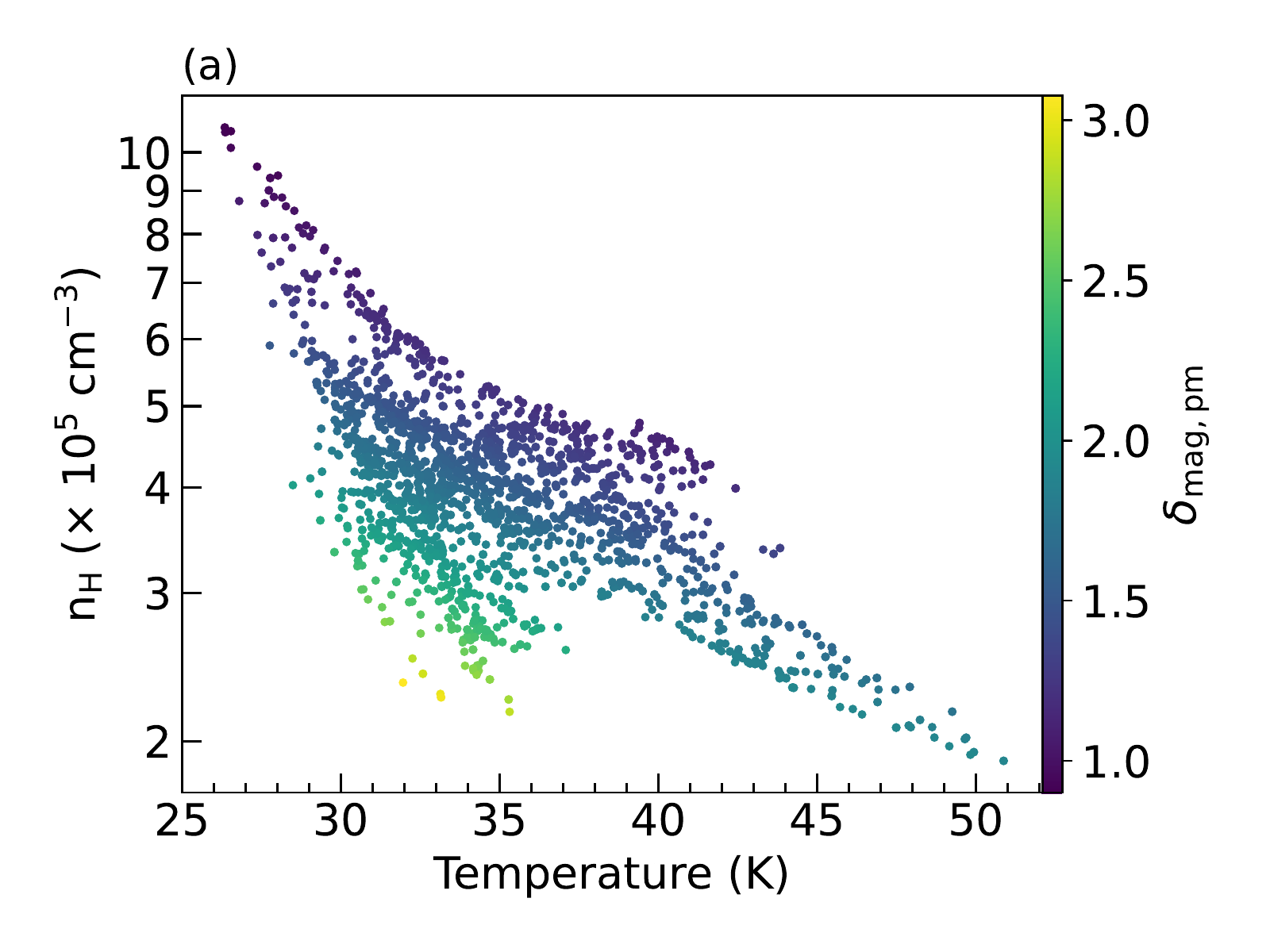}
    \includegraphics[trim={2cm 0 0 0}, scale=0.45]{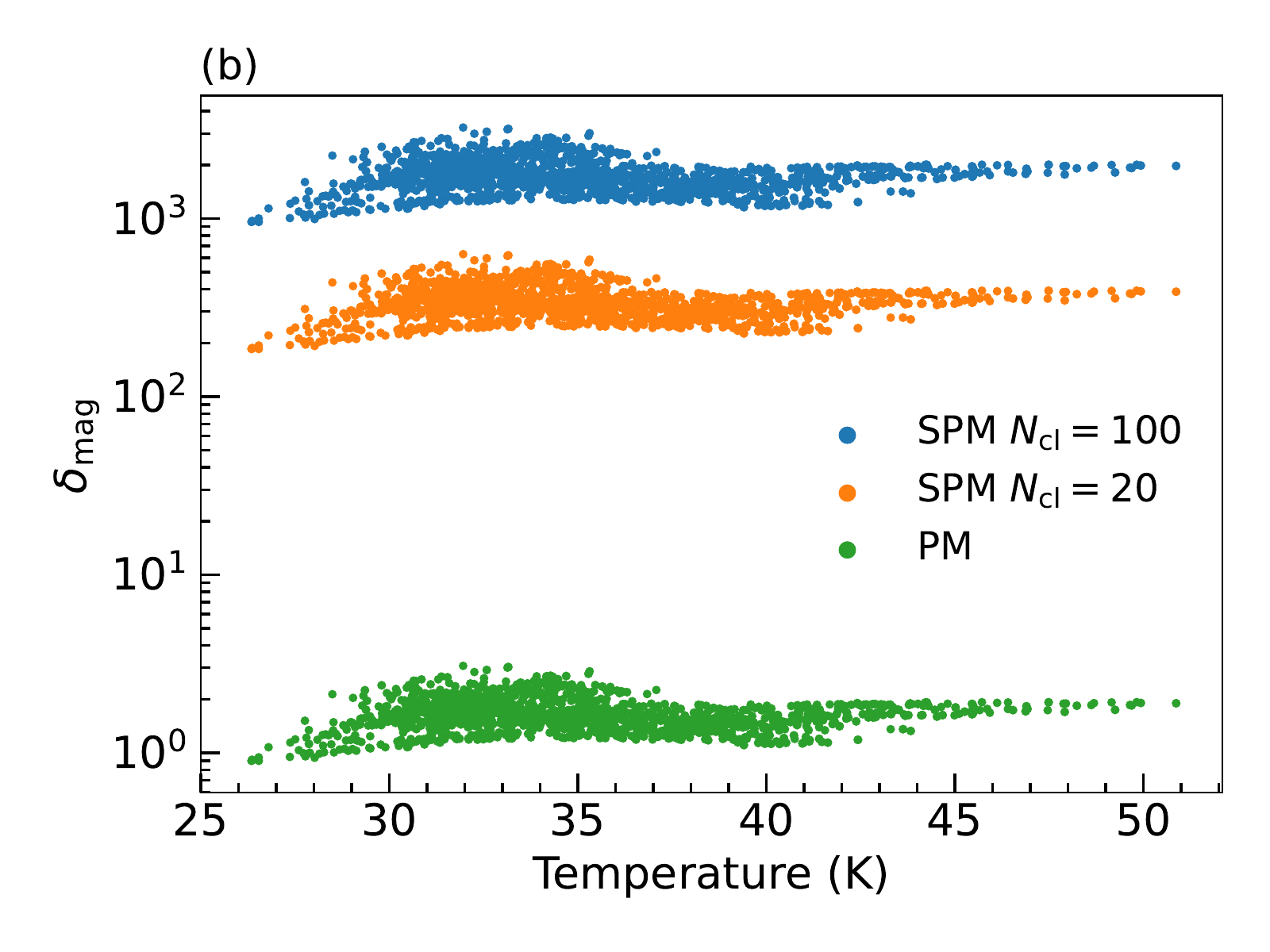}
    \caption{Magnetic relaxation calculated for the S53 data with $B=5$ mG. Plot (a) shows the estimated value of $\delta_{\mathrm{mag,pm}}$ for PM grains. The plot (b) shows the estimated $\delta_{\mathrm{mag}}$ for PM and SPM grains with $N_{\mathrm{cl}}=$ 20, and 100. $\delta_{\mathrm{mag}}>10$ for the SPM grains indicate perfect alignment with MRAT mechanism for grains at suprathermal rotation.}
    \label{fig:delta-mag}
\end{figure}

\subsection{Effect of magnetic relaxation: $\delta_{\mathrm{mag}}$ and $a^{\mathrm{DG}}_{\mathrm{max\_JB}}$}
Another important process that affects the alignment of PM and SPM grains is the magnetic relaxation suggested by \citet{Davis1951}, also called Davis-Greenstein (DG) alignment mechanism. When a magnetized grain is rotating at an angle with respect to the ambient magnetic field, it loses a part of its rotational energy in the form of heat due to rotating magnetization with respect to the grain body. This eventually leads to the alignment of the grain angular momentum along the magnetic field, which is the configuration of least rotational energy. Detailed analytical studies of the effect of iron inclusions on the grain alignment have been carried out by \citealt{Hoang2016,Hoang2022AJ,Hoang2022ApJ}. They show that the DG alignment depends strongly on the strength of the magnetic field and the level of iron inclusions in the dust grains, and that even with small levels of iron inclusions, the effect of the magnetic relaxation on RAT-A mechanism cannot be ignored. \citet{Hoang2016} introduced a dimensionless parameter $\delta_{\mathrm{mag}}=\tau_{\rm gas}/\tau_{\rm m}$ (where $\tau_{\rm m}$ is the magnetic relaxation timescale), which determines the efficiency of the grain alignment by magnetic relaxation against the disalignment by gas collisions. For a PM grain, $\delta_{\mathrm{mag}}$ can be approximated as $\delta_{\mathrm{mag,pm}}\sim aB^2n_{\rm H}^{-1}T_{\rm gas}^{-1/2}$ \citep[Eq. 6,][]{Hoang2016} and for SPM grains with iron clusters,  $\delta_{\mathrm{mag,sp}}\sim a^{-1}N_{\mathrm{cl}}\phi_{\mathrm{sp}} B^2n_{\rm H}^{-1}T_{\rm d}^{-1}T_{\rm gas}^{-1/2}$ \citep[Eq. 51,][]{Hoang2022AJ}, where $N_{\mathrm{cl}}$ is the number of iron atoms per cluster and $\phi_{\mathrm{sp}}$ is the  volume filling factor of iron clusters. Assuming typical parameters for the grains, we estimate $\delta_{\mathrm{mag,pm}}$ for S53 data with $B=5$ mG and $a=0.1$ \micron, shown in Figure \ref{fig:delta-mag}a. Except few data points from the region of least $T_{\rm d}$ and highest $n_{\rm H}$, the rest of the regions show $\delta_{\mathrm{mag,pm}}>1$, indicating that even for PM grains with a low level of iron inclusion, the magnetic field in this region is strong enough for PM relaxation to play an important role in grain alignment.

MRAT mechanism discussed earlier leads to an enhanced degree of grain alignment, beyond that which could be achieved by RAT-A alone. The PM dust grains subject to RATs tend to get aligned at low angular momentum attractor points (thermal rotation) called low-$J$ attractors and/or at high angular momentum attractor points (suprathermal rotation) called high-$J$ attractors, depending on the strength of the RATs \citep{Hoang2008}. The net polarization can be determined by the fraction of the grains at high-$J$ attractors as these are perfectly aligned. \citet{Lazarian2008} found that the combined effect of suprathermal rotation induced by RATs and enhanced magnetic relaxation leads to a greater fraction of the grains at high-$J$ attractors, as the grains slowed down due to RATs are spun-up due to magnetic relaxation. Magnetic relaxation acts to stabilize the high-$J$ attractor points of RATs. Efficient magnetic relaxation is expected for suprathermally rotating grains such that MRAT can lead to perfect alignment if $\delta_{\mathrm{mag}}>10$ \citep{Hoang2016,Hoang2022AJ}. From Figure \ref{fig:delta-mag}a it can be seen that the PM grains in this region are not perfectly aligned as $\delta_{\mathrm{mag,pm}}<10$. We estimated $\delta_{\mathrm{mag,sp}}$ for the same region with $N_{\mathrm{cl}}=20$ and 100, and $\phi_{\rm sp}=0.03$ (assuming that 10\% of Fe abundance is depleted into the dust, \citealt{Bradley1994,Martin1995,Hoang2016}). The derived values are shown in Figure \ref{fig:delta-mag}b and it can be seen that $\delta_{\mathrm{mag}}>10$ even for small iron clusters of $N_{\mathrm{cl}}=20$. Thus, due to the strong magnetic field at the Galactic centre, perfect alignment can be achieved if the dust grains are SPM in nature. 

\begin{figure*}
    \centering
    \includegraphics[scale=0.45]{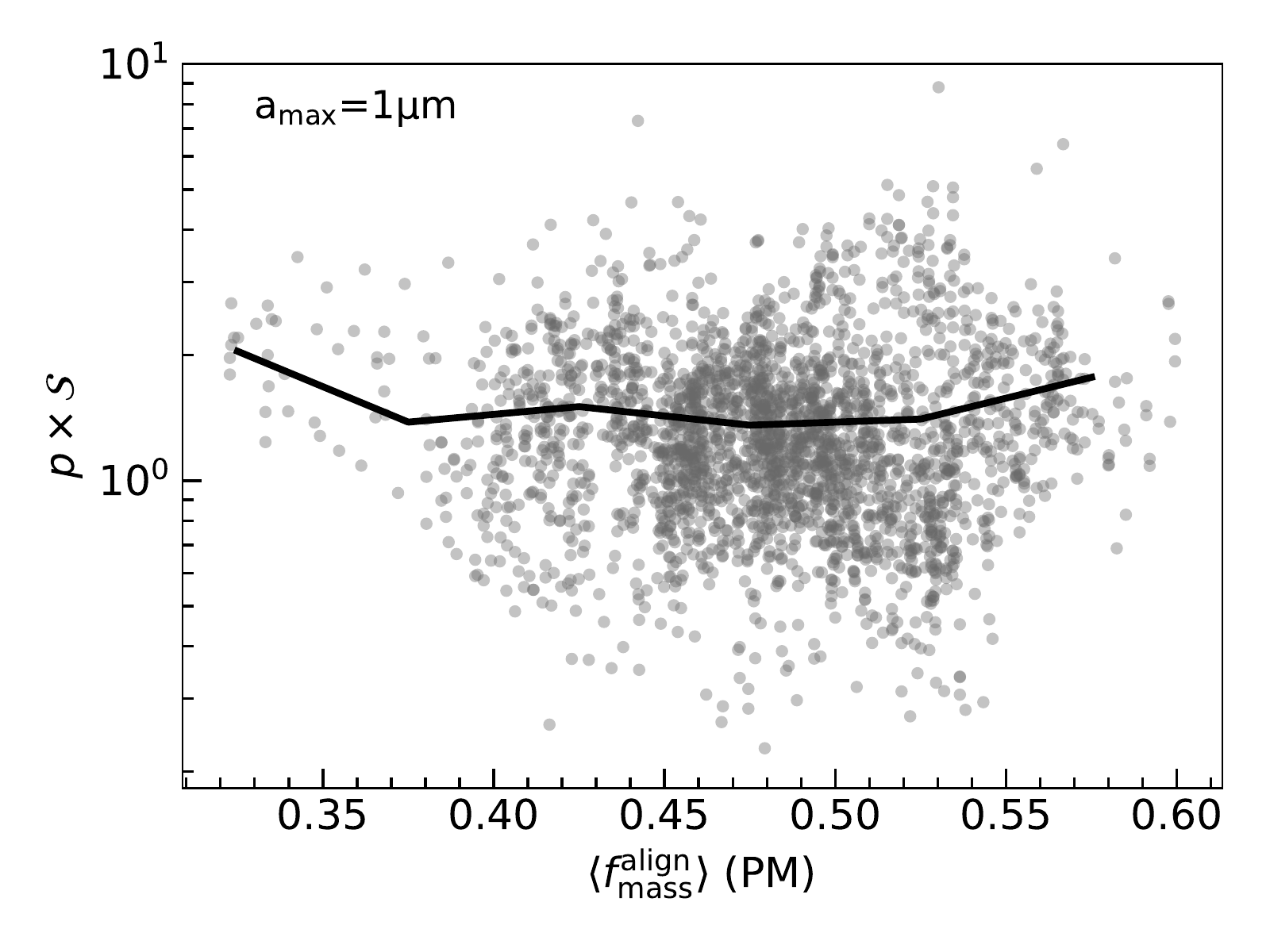}
    \includegraphics[scale=0.45]{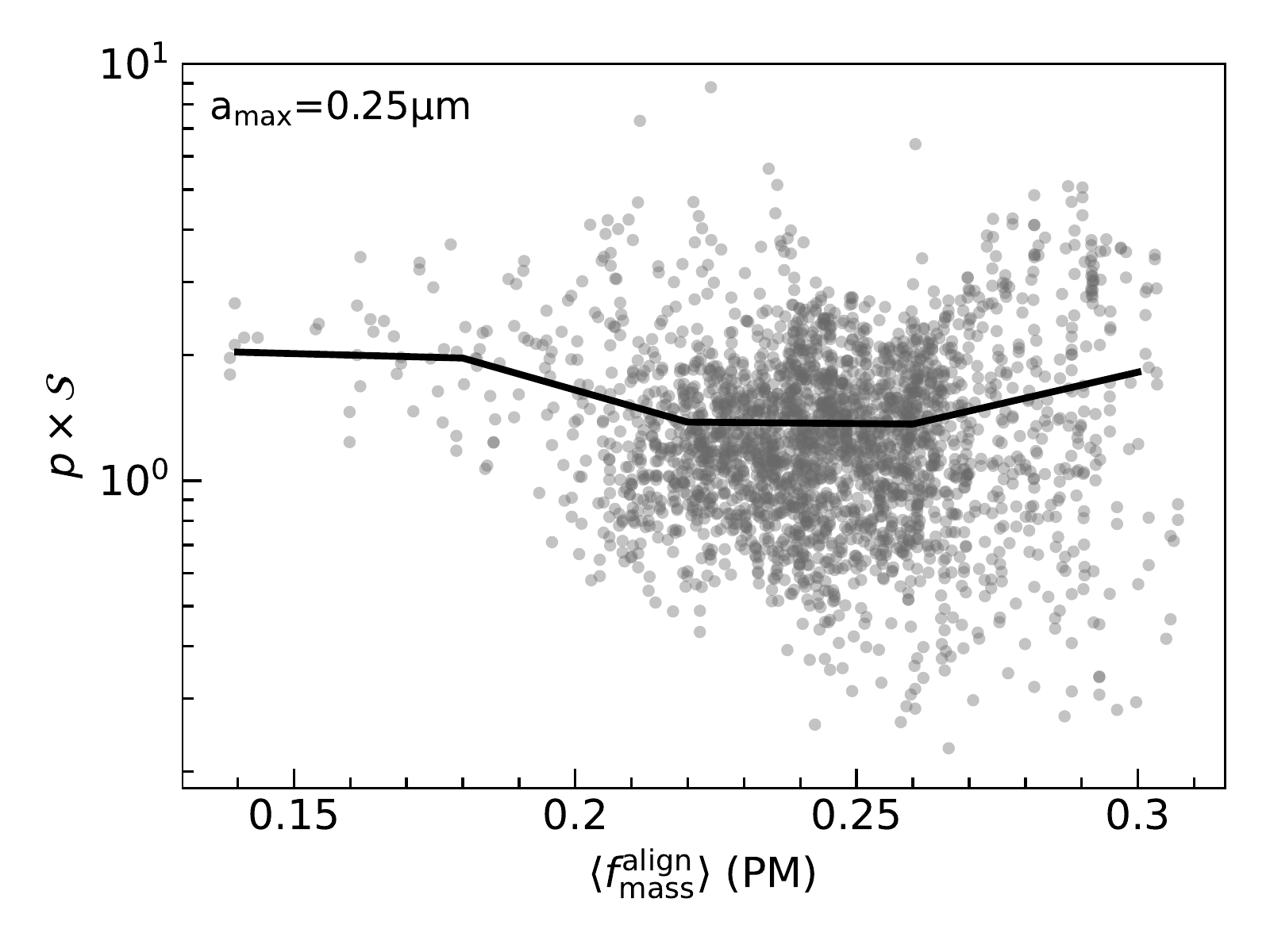}
    \includegraphics[scale=0.45]{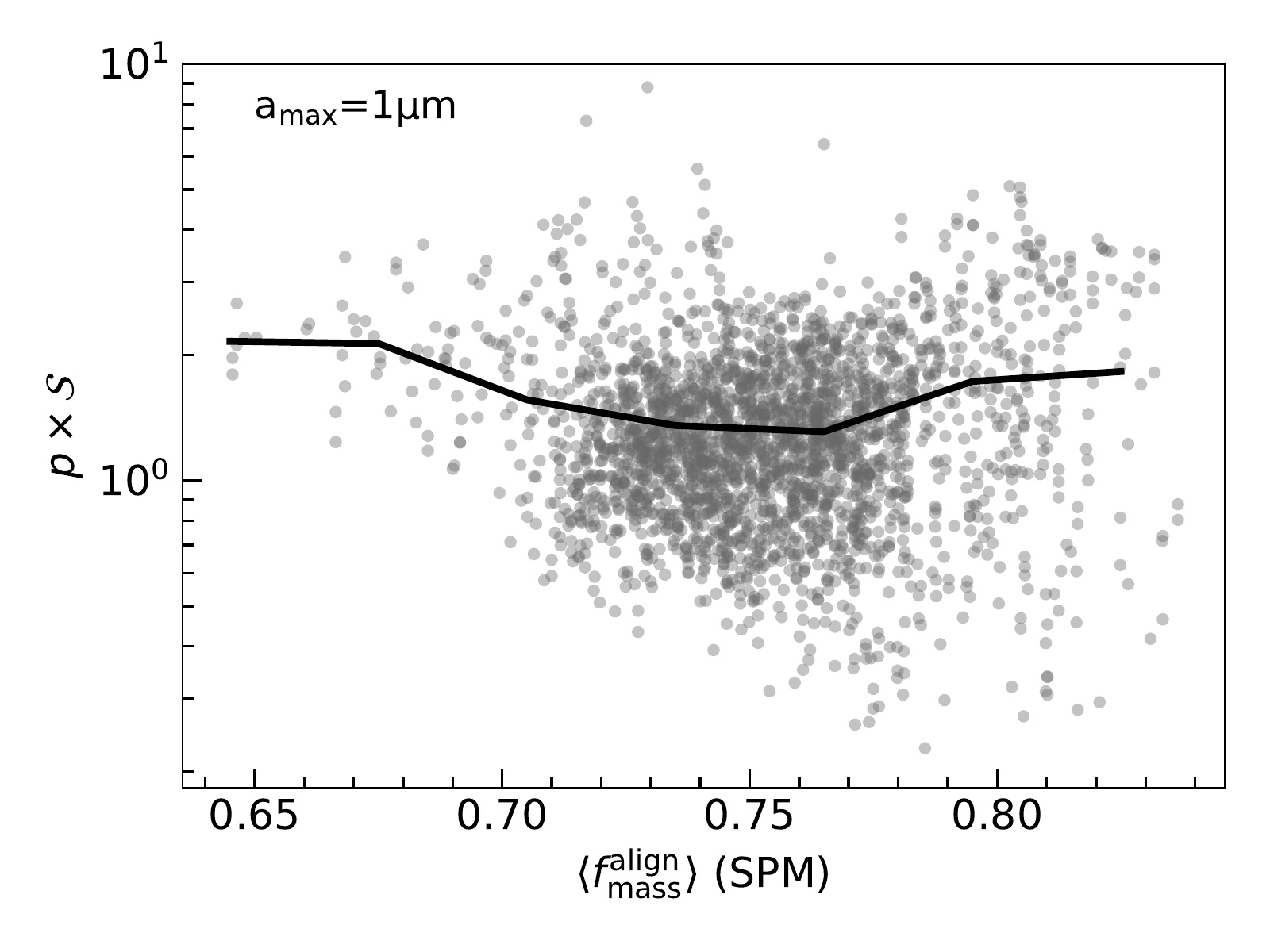}
    \includegraphics[scale=0.45]{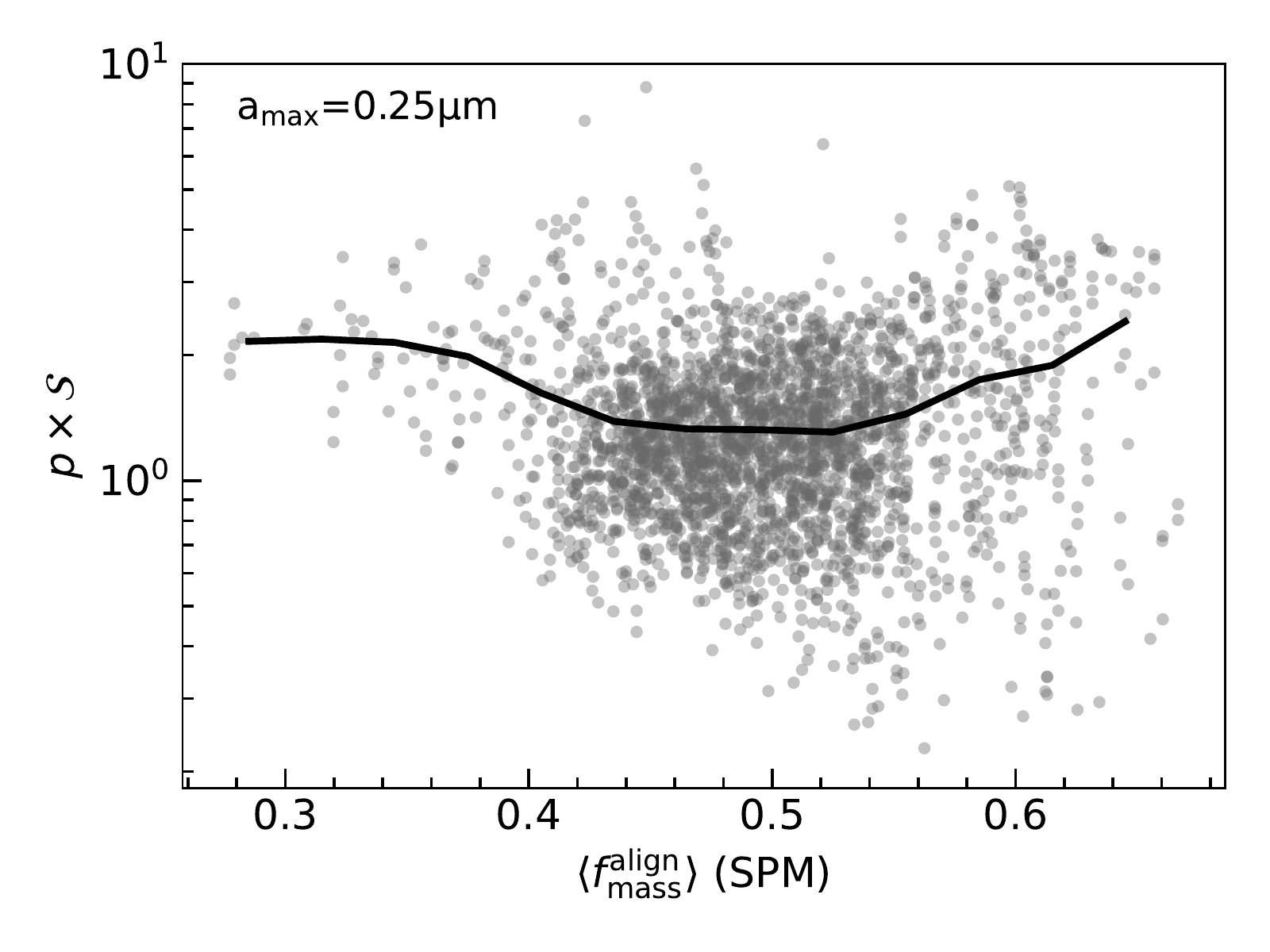}
    \caption{Mass fraction of the aligned PM (top) and SPM (bottom) grains as a function of $p\times$\mS\ which traces the grain alignment after removal of the effect of the magnetic field tangling. The left figures are for $a_{\rm max}=1$ \micron\ while the right is for $a_{\rm max}=0.25$ \micron. The weighted mean binned along $\langle f^{\mathrm{\rm align}}_{\mathrm{mass}} \rangle$ in shown by the solid black line in each plot.}
    \label{fig:f_align}
\end{figure*}

\citet{Hoang2022AJ} define the maximum size of grains for which magnetic relaxation is important as $a^{\mathrm{DG}}_{\mathrm{max\_JB}}$ using the condition $\delta_{\mathrm{mag}}>1$ \citep[Eq. 52,][]{Hoang2022AJ}. This is shown in Figure \ref{fig:a-distribution} for PM and SPM grains with $B=5$ mG. For PM grains $a^{\mathrm{DG}}_{\mathrm{max\_JB}}<a_{\rm align}$ and is of the order of $a^{\mathrm{DG}}_{\mathrm{max\_JB}}\sim10^{-2}$ \micron\ even though $\delta_{\rm mag, PM}>1$. This again reflects the fact that even with the strong magnetic field of the region, perfect alignment of grains is not attained if the grains are assumed to be PM in nature. RAT-A still remains the dominant mechanism for grain alignment with a low fraction of grains at high-$J$ attractors due to little aid from magnetic relaxation. However, if the grains are assumed to be SPM, $a^{\mathrm{DG}}_{\mathrm{max\_JB}}>a_{\rm align}$ and is of the order of $a^{\mathrm{DG}}_{\mathrm{max\_JB}}\sim10$ \micron. Grains in this case have perfect alignment through MRAT mechanism, with all the grains reaching high-$J$ attractors. It is also interesting to note that in the case of S216 shown in Figure \ref{fig:a-distribution}b, the range of SPM grains that can be aligned ($a_{\rm align}-a^{\mathrm{DG}}_{\mathrm{max\_JB}}$) is maximum for the black circled points which show the highest polarization degree ($p\sim13\%$).

\subsection{Mass fraction of aligned dust grains}
The degree of dust polarization depends crucially on the critical size for RAT-A, $a_{\rm align}$ and the fraction of grains aligned at the high-$J$ attractor, termed $f_{\mathrm{high-J}}$. Smaller $a_{\rm align}$ or larger $f_{\rm high-J}$ correspond to a broader range of grain sizes that can be aligned, which implies a larger mass fraction of grains that are aligned. To directly relate the grain alignment to dust polarization, we define the mass fraction of dust grains that are aligned with the magnetic field,
\begin{equation}
    \langle f^{\mathrm{\rm align}}_{\mathrm{mass}} \rangle = \frac{\int_{a_{\mathrm{\rm align}}}^{a_{\mathrm{max}}} \Big(\frac{dm}{da}\Big) da \times f_{\mathrm{high-J}}(a)}{\int_{a_{\mathrm{min}}}^{a_{\mathrm{max}}} \Big(\frac{dm}{da}\Big) da} , \label{eq:fmass}
%    \frac{\int_{a_{\mathrm{\rm align}}}^{a_{\mathrm{max}}} m(a) da \times f_{\mathrm{high-J}}(\delta_{\mathrm{mag}})}{\int_{a_{\mathrm{min}}}^{a_{\mathrm{max}}} m(a) da} \\
\end{equation}
where $dm/da = (4\pi/3)\rho sa^3 n(a)$ is the mass density of dust grains in the size range $a$ to $a+da$, and $n(a)$ is the number density of grains of size $\le a$ as given by $n(a) = dn/da=Cn_{\rm H}a^{-3.5}$ for the grain size distribution from \citet[][MRN distribution]{Mathis1977}. Here $s$ is the axial ratio of the oblate spheroidal grain shape, $\rho$ the grain mass density which we assume to be 3 g cm$^{-3}$, and $C$ is the normalization constant.

The exact value of $f_{\rm high-J}$ is uncertain and difficult to quantify due to the uncertainty in grain shapes and composition. $f_{\mathrm{high-J}}$ also depends on the orientation of the grains with respect to the radiation and magnetic field. Numerical calculations of MRAT alignment by \citet{Hoang2016} showed that $f_{\mathrm{high-J}}$ increases for SPM grains as a result of enhanced magnetic relaxation, leading to perfect alignment of $f_{\mathrm{high-J}}=1$ when $\delta_{\mathrm{mag}}> 10$ \citep{Lazarian2021}. Using their calculations, \citet{GiangPolaris2022} proposed a parametric model that defines $f_{\mathrm{high-J}}$ based on the observed level of $\delta_{\mathrm{mag}}$ as,
\begin{equation}
    f_{\mathrm{high-J}}(\delta_{\mathrm{mag}}) = 
    \left\{
    \begin{array}{ll}
        0.25 & \mbox{for $\delta_{\mathrm{mag}}<1$} \\
        0.5 & \mbox{for $1\leq\delta_{\mathrm{mag}}\leq10$} \\
        1 & \mbox{for $\delta_{\mathrm{mag}}>10$} \\
    \end{array}
    \right\},\label{eq:fhighJ}
\end{equation}
where $f_{\mathrm{high-J}}=0.25$ for the case when the grains are aligned with $B$ through RATs alone with negligible contribution from magnetic relaxation. The second case of $f_{\mathrm{high-J}}=0.5$ is applicable for regions between alignment only through RATs and perfect alignment through MRAT. 

For a given $a_{\rm align}$ and $\delta_{\rm mag}$, plugging Equation (\ref{eq:fhighJ}) in Equation (\ref{eq:fmass}) and assuming $s=0.5$, $a_{\mathrm{max}}=1$ \micron, and $a_{\mathrm{min}}=3.5$ \AA\ \citep[determined by thermal sublimation due to temperature fluctuations of very small grains;][]{Draine2007} we obtain the mass fraction of aligned grains for the S53 region. This is shown in Figure \ref{fig:f_align} for the case of both PM and SPM grains ($N_{\rm cl}=20; \phi_{\rm sp}=0.03$), and $a_{\rm max}=1$ \micron\ and 0.25 \micron. For the case with $a_{\rm max}=1$ \micron\ about 50\% of the grain mass is aligned if they are assumed to be PM in nature whereas if the grains are assumed to be SPM even with a low level of iron inclusions of $N_{\rm cl}=20$, about 75\% of the grain mass is aligned for the considered grain size distribution. Aligned mass depends greatly on $f_{\rm high-J}$ which is close to 1 for most of the sizes of SPM grains in this region due to the strong magnetic field. The aligned mass seem to have a weak correlation with $p\times$\mS\ at regions of high polarization indicating the alignment of larger grains that contribute to most of the mass. When the same estimates are performed with $a_{\rm max}=0.25$ \micron\ we see a similar correlation with $p\times$\mS\ but for a much lower mass range. This indicates the dominance of large grains in the alignment and also shows enhancement in smaller grains that are aligned as they follow a similar trend with $p\times$\mS.

\subsection{Rotational disruption by RATs: $a_{\mathrm{disr}}$ and $a_{\mathrm{disr\_max}}$}
RATs acting on dust grains not only lead to their rapid rotation and subsequent alignment but can also disrupt them into smaller fragments as proposed by \citet{Hoang2019NatA}. This disruption called the RAT-D mechanism occurs when the centrifugal stress on the grains due to its rapid rotation exceeds the tensile strength of the material, leading to the dust grains being instantaneously disrupted into smaller fragments. The disruption of the grains is highly dependent on their tensile strength, denoted by $S_{\rm max}$. This is an uncertain parameter and is determined by the grain structure and composition. Composite grains are known to have a lower tensile strength ($S_{\rm max}\sim10^6-10^7$erg cm$^{-3}$) compared to compact grains ($S_{\rm max}\sim10^9-10^{11}$erg cm$^{-3}$). The disruption of grains due to RAT-D has an impact on the observed degree of polarization as a result of the depletion of large grains, which leads to a drop in the polarization degree, especially at longer wavelengths. This can be observed as a negative correlation in the $p-T_{\rm d}$ relationship at high $T_{\rm d}$.

\citet{Hoang2021} derived analytical formula to estimate the minimum and maximum size of grains that can be disrupted as a function of the local physical conditions, denoted by $a_{\rm disr}$ and $a_{\rm disr\_max}$ respectively. The disruption size is determined by $a_{\rm disr}\sim n_{\rm H}^{1/2}T_{\rm d}^{-3}S_{\rm max}^{1/4}$ indicating that grains in hot and diffuse environments are disrupted more easily than grains in dense and cool environments. We have used this formulation by \citet{Hoang2021} to estimate $a_{\rm disr}$ for the Galactic centre observations assuming composite grains with $S_{\rm max}=10^6$ erg cm$^{-3}$. This is shown in Figure \ref{fig:a-distribution}.  The mean disruption size seems to be about $a_{\rm disr}\sim1$ \micron\ with a maximum size of disruption about $a_{\rm disr\_max}\sim20$ \micron. From the analysis of S53 and S216 observations, we do not see strong evidence of RAT-D in the $p-T_{\rm d}$ relationship. Though there are a few data points with low $p$ at high $T_{\rm d}$, they are not statistically significant. However, in the case of J850 data, there is a drop in $p$ for regions with $T_{\rm d}>32$ K even after the removal of the effect of the tangled magnetic field (Figure \ref{fig:JCMT_S}c). This could be clear evidence for the RAT-D mechanism, as the large grains which can be disrupted are the dominant contributors to the observed radiation at 850 \micron. From Figure \ref{fig:a-distribution}c it can be seen that $a_{\rm disr}$ is about 1 \micron\ when $T_{\rm d}>32$ K. 

\section{Discussion} \label{Discussion}

\subsection{Evidence for grain alignment and disruption by RATs}
Though the polarized thermal emission from the Galactic centre has been studied before \citep{Aitken1986,Aitken1998,Novak2000,Roche2018}, the grain alignment physics is not very well understood. The DG alignment mechanism was suggested as the leading cause of grain alignment in this region by \citet{Aitken1998}. However, numerical simulations by \cite{Hoang2016} demonstrated that grain alignment through paramagnetic relaxation alone is not efficient for thermally rotating grains even when they are SPM, due to strong thermal fluctuations that suppress the internal alignment of the grain axis of maximum moment of inertia with its spinning axis. Only the combined effect of magnetic relaxation with other process which drive grains to suprathermal rotations can cause efficient grain alignment, comparable to the observed level of polarization.

%{\bf Polarized thermal dust emission was previously observed toward the Galactic centre (GC). However, the question of how dust grains at GC get aligned is not well understood. Earlier studies by \cite{Aitken1998} suggested that grains at GC get aligned by paramagnetic relaxation via Davis-Greenstein mechanism \citep{Davis1951}. However, numerical simulations by \cite{Hoang2016} demonstrated that paramagnetic relaxation alignment is not efficient for thermally rotating grains due to strong thermal fluctuations that suppress internal alignment of the grain axis with its spinning axis. The leading mechanism for grain alignment is based on RATs, which can drive grains to suprathermal rotation and align grains with the magnetic field.}

In this paper, we studied the alignment and disruption of dust grains at the Galactic centre using multi-wavelength observations of dust polarization from SOFIA/HAWC+ and JCMT/SCUPOL. The observations covered different physical scales and were analyzed individually to understand the change in the alignment properties of dust grains from the scale of within 0.2 pc to about 1 pc. Based on the analyses of $p-I$, $p-T_{\rm d}$, and $p\times S-T_{\rm d}$ relations, we find that the alignment mechanism is most consistent with the predictions from RAT-A theory. In particular, the 850 \micron\ observation shows an increase in polarization degree with dust temperature for $T_{\rm d}<35$ K followed by a decrease at higher temperatures. This $p-T_{\rm d}$ trend is most consistent with the RAT-D effect, where the $B-$field tangling is not the main cause. Though we see traces of RAT-D trends in the 53 and 216 \micron\ observation, there are very few data points showing this feature. We have also used the recent analytical models of grain alignment in the RAT paradigm derived by \citet{Hoang2021,Hoang2022AJ} to understand the effect of iron inclusion on the observed polarization degree. We find that grains can have perfect alignment through MRAT when they are considered to be SPM in nature. 

%It has long been proposed that the observed changes in the polarization vectors along the CND could be a result of a change in the magnetic field orientation from parallel to perpendicular to the plane of the sky. The 53 \micron\ observation traces the CND in great detail while it is a very small part of the observation at 216 \micron, and 850 \micron. The observations were not registered to the same resolution due to the loss of data points as each trace a different scale of the region. 

%{\bf Polarization saturation, perfect alignment ...???}

\subsection{Constraining the dust grain model with multi-wavelength polarization} \label{Grain composition}
Interstellar dust comprises dominantly of silicate and carbonaceous material. Yet, how these materials reside in the dust grains is poorly understood. Polarization degree observed at multiple wavelengths can be used to constrain the model of the dust grains which are distinguished by their composition and internal structure. Assuming silicate and carbon to be the major constituents of the grains, we consider two popular dust models for our analysis. The first one is the simplest model in which the silicate and carbon grains are in separate populations (i.e., separate dust model) where only the silicate grains containing Fe atoms are aligned \citep{Mathis1977,Hoang2016} and carbonaceous grains are assumed to be randomly oriented. The latter is justified due to the non-detection of C-H polarization feature \cite{Chiar2006} and the diagmagnetic nature of carbon dust that makes them weakly align with the magnetic field \citep{Hoang2016,Hoang2023}. The second model is the composite dust model in which the silicate and carbon grains are loosely bound together in a single dust population such that both grains can be aligned \citep{Mathis1989,Draine2007,Jones2013,Hoang2016}. We only consider the composite dust model for a mixed population of carbonaceous and silicate grains as they have the least tensile strength ($S_{\rm max}\sim10^6-10^7$erg cm$^{-3}$) and are disrupted more easily compared to the other models with compact or core-mantle structure \citep{Greenberg1996,Li2002,Jones2013,Hoang2019,Tram2021Ophi}, and the tensile strength further reduces with the inclusion of porosity in the grains. Additionally in the context of RAT-D, the disruption of core-mantle grain can be treated as similar to that of a compact grain, where the mantle and the core can have different tensile strengths \citep{Hoang2019}.

The first physical modeling of dust polarization using the RAT theory for the separate dust model was performed by \citet{Lee2020},  which simultaneously accounts for the effects of RAT-A and RAT-D for different radiation fields (dust temperatures). Later, \citet{Tram2021Ophi} modeled the polarization spectrum for both the separate and composite dust models. The resulting polarization spectrum is radically different for the two dust models, especially beyond the peak of the spectrum. The composite dust model exhibits a flatter polarization spectrum after the peak (at around $\lambda\simeq100$ \micron), compared to the separate dust model where the polarization drops beyond the peak. As a result, the ratio of polarization degree at two wavelengths is different for the two models and we can use this to constrain the possible model of the dust grains. This can help us conclude if the dust grains are in separate or mixed populations, while the internal structure of the grains is determined from the tensile strength the best fits the observed spectrum. Here we only test the population of the grains and leave the modeling of the tensile strength and consequently, the internal structure of the grains to a future study. 

\begin{figure*}
    \centering
    \includegraphics[scale=0.5]{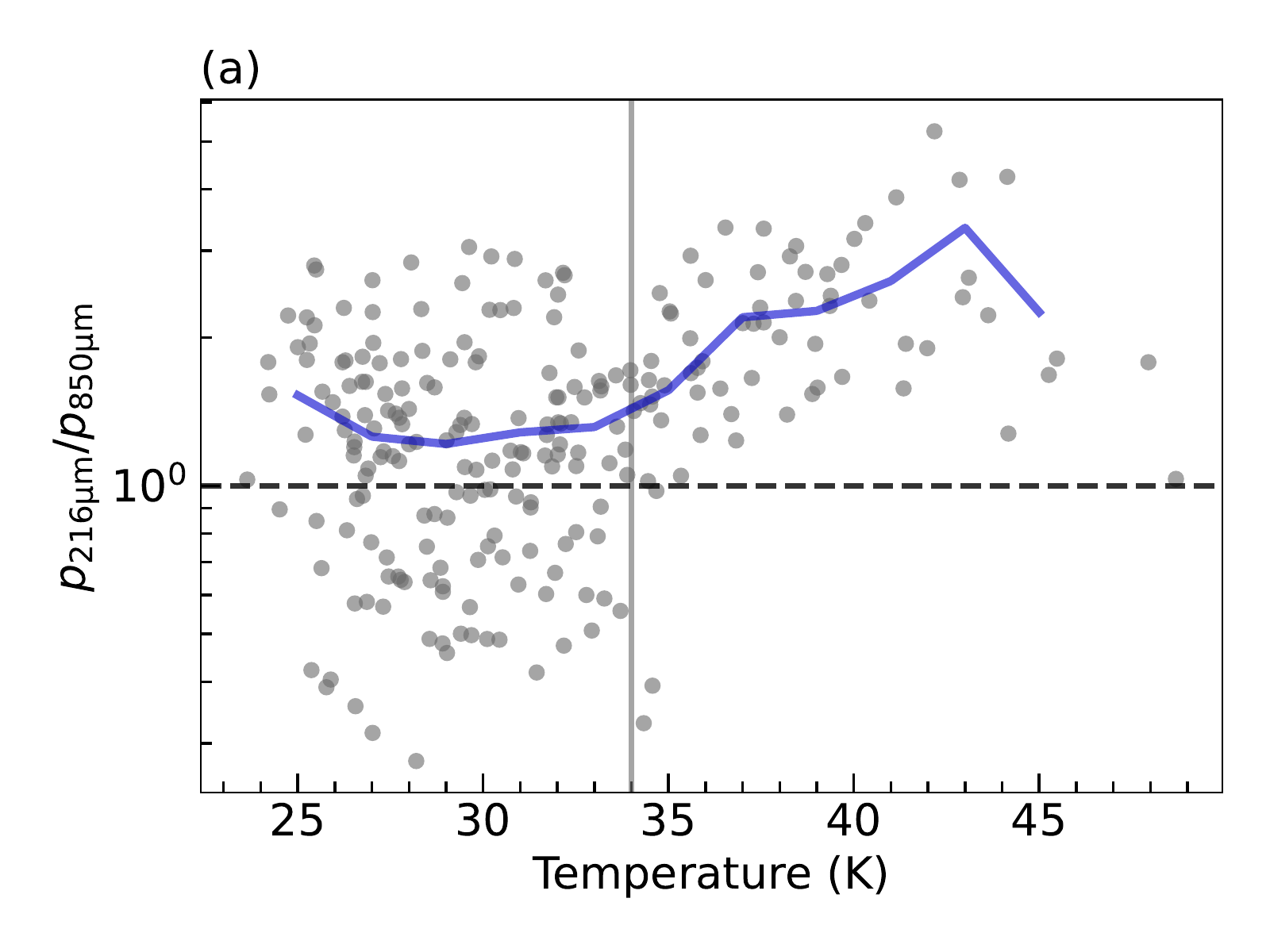}
    \includegraphics[scale=0.5]{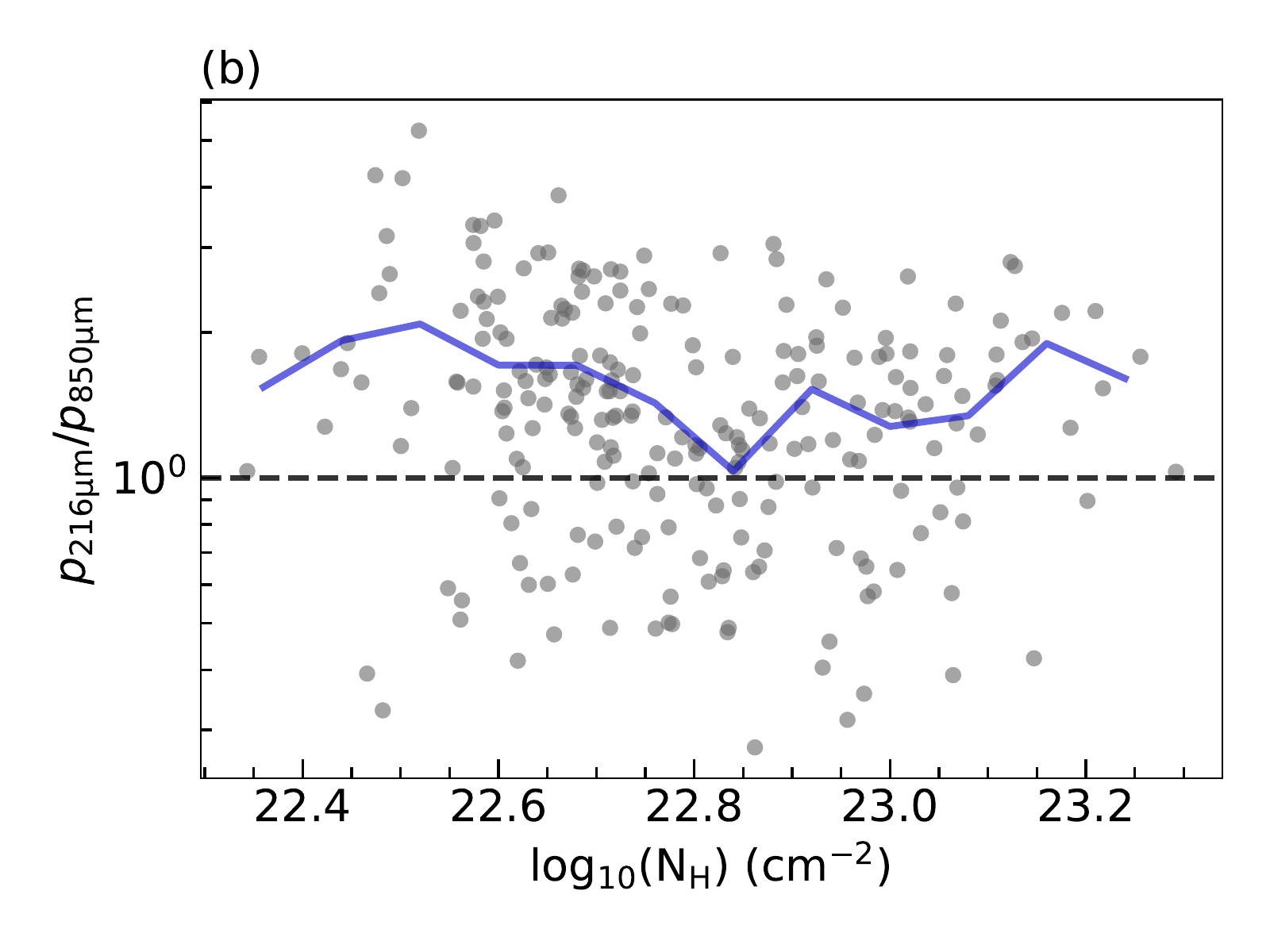}
    \includegraphics[scale=0.5]{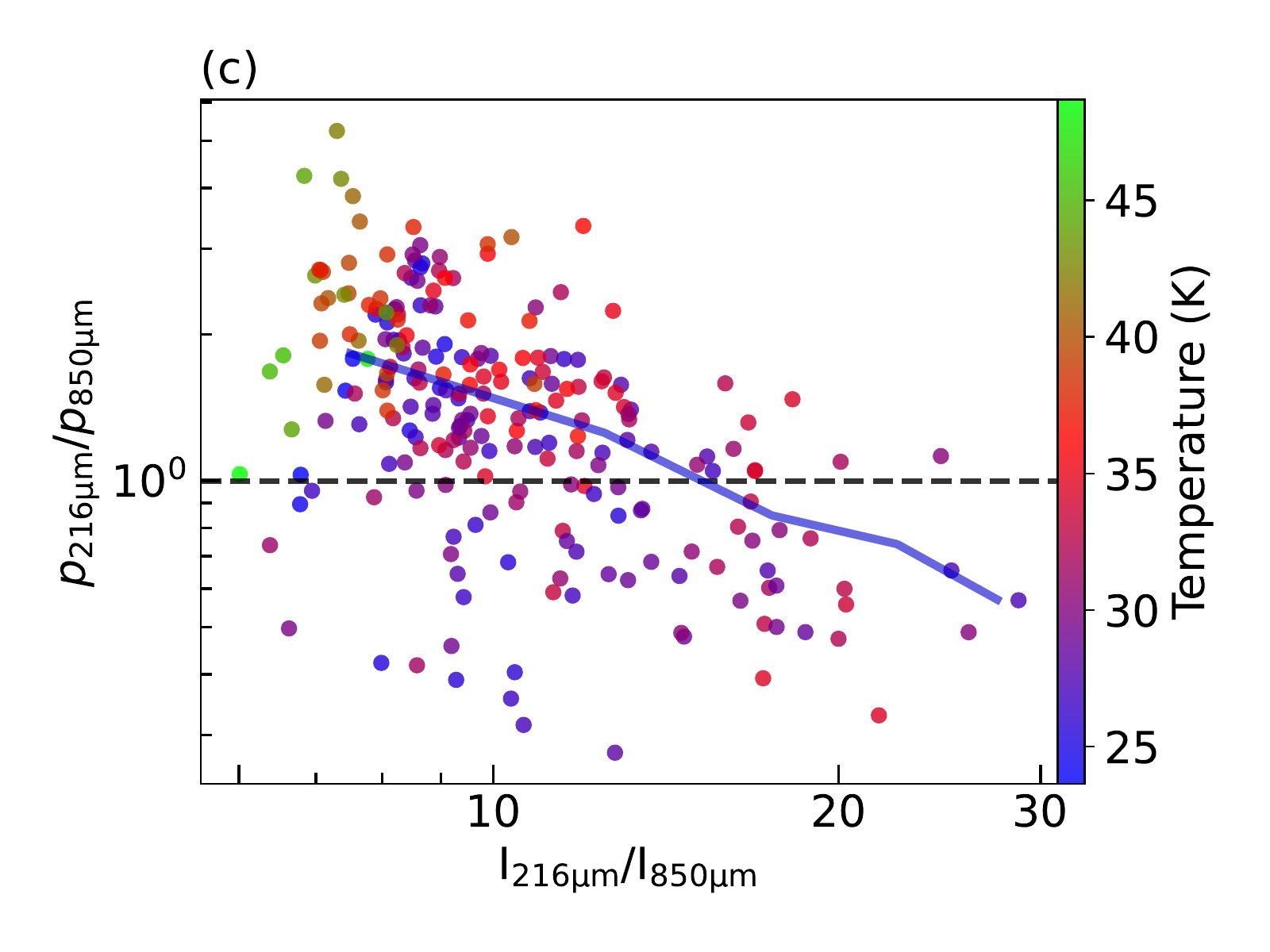}
    \includegraphics[scale=0.5]{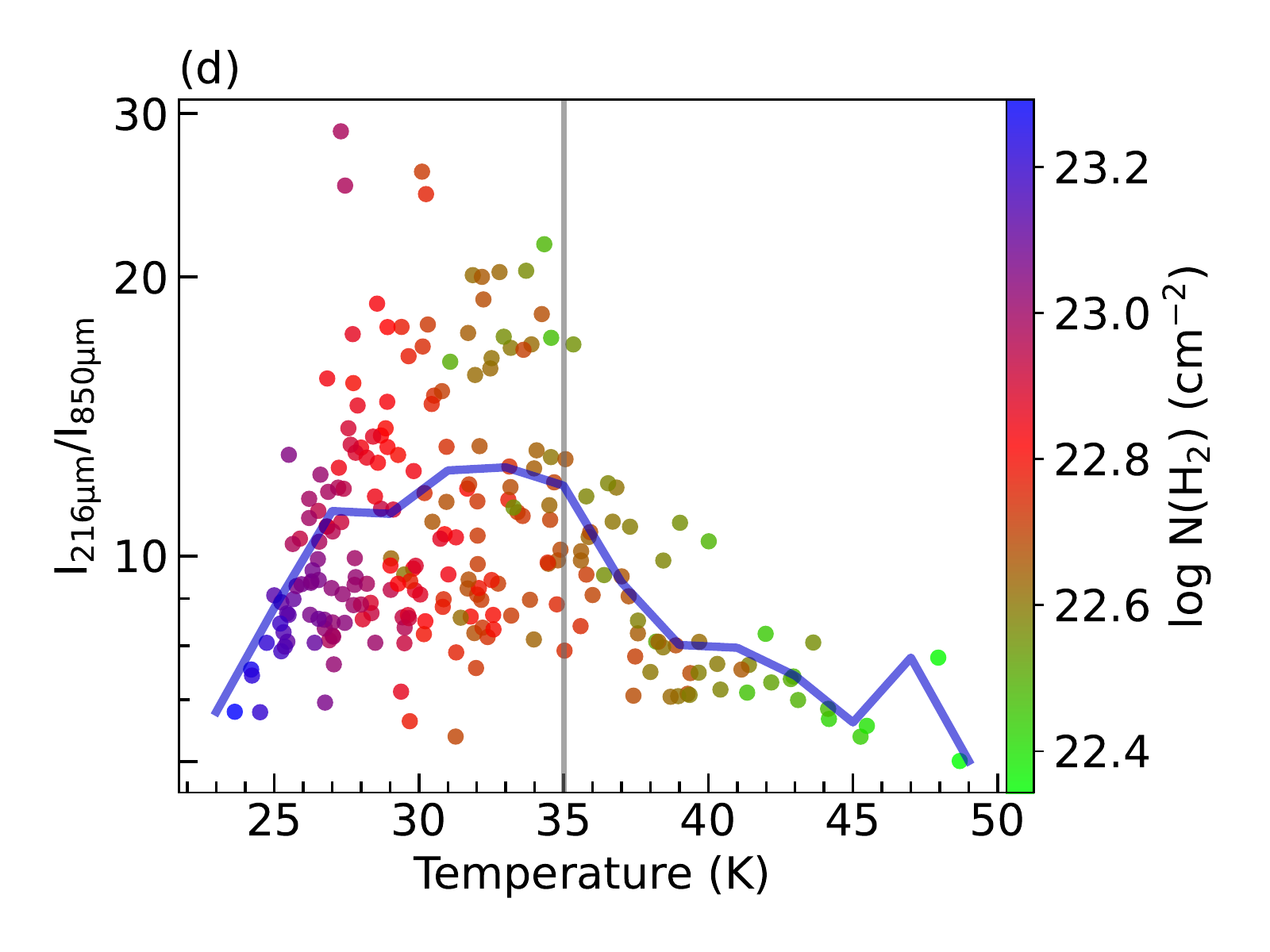}
    \caption{Variation of the ratio of polarization degree with temperature (a), column density (b), and intensity ratio (c). The relationship between the ratio of intensity with temperature and column density is shown in (d). The horizontal line represents $p_{216\mu m}/p_{850\mu m}=1$. The solid blue line represents the running mean in each figure. The vertical line in (a) indicates the transition from $p_{216\mu m}/p_{850\mu m}\leqslant1$ to $p_{216\mu m}/p_{850\mu m}>1$.}
    \label{fig:pRatio}
\end{figure*}

We used the S216 and J850 polarization data for this analysis as they have comparable beam sizes and cover a larger region of overlapping observations around Sgr A$^*$. S216 observation was smoothed and binned to the same beam size and resolution as the J850 data and the overlapping regions from both observations were used. The models of polarization spectrum from \citet{Lee2020} and \citet{Tram2021Ophi} predict a peak in the spectrum at about $100-200$ \micron\ for the observed temperature range for the separate dust model. In this case, the polarization degree at 216 \micron\ is larger than that at 850 \micron, resulting in $p_{216\mu m}/p_{850\mu m}>1$. However, for the composite dust model, the ratio $p_{216\mu m}/p_{850\mu m}<1$ due to the flat nature of the polarization spectrum in this environment. To probe the dust model in the Galactic centre, we have analyzed the relationship between this ratio of polarization degree with dust temperature, gas column density, and the ratio of total intensities ($I_{216\mu m}/I_{850\mu m}$) as shown in Figure \ref{fig:pRatio}. 

The $p_{216\mu m}/p_{850\mu m}-T_{\rm d}$ plot from Figure \ref{fig:pRatio}a shows a clear change in slope at about $T_{\rm d}\sim35K$, where the ratio changes from $p_{216\mu m}/p_{850\mu m}\lesssim1$ to $p_{216\mu m}/p_{850\mu m}>1$. This could be considered as another evidence for the RAT-D mechanism where the disruption results in a separate population of better-aligned silicate grains from a composite structure of silicate and carbon grains \citep[probably composite grain aggregates in the form of monomers;][]{Tatsuuma2019}. This change in slope is not observed in $p_{216\mu m}/p_{850\mu m}-N_{\rm H}$ relation shown in Figure \ref{fig:pRatio}b where the data points are more scattered with no evident trend. It is also interesting to note that the model predicted temperature at which RAT-D affects the polarization, $T_{\rm d}\sim30-34$ K \citep{Lee2020,Tram2021Ophi}, is in the same range where we see a drop in polarization in the J850 observation and also where we see a raise in the  $p_{216\mu m}/p_{850\mu m}$ ratio.

The ratio of polarization degree seems to drop linearly with the ratio of intensity as shown in Figure \ref{fig:pRatio}c. The highest ratio of the degree of polarization is observed in the region of high temperature and low column density, and this also corresponds to the emission only from silicate grains that are aligned. The ratio of intensity in this region is low may be due to the combined effect of grain disruption and low column density in this region, where a larger column of gas/dust is traced at a longer wavelength resulting in more intensity at 850 \micron. When the ratio of intensity increases, this is from the region where we have composite silicate and carbon grains which are not perfectly aligned. Hence we see that even though the ratio of intensity increases, the fraction of polarization reduces.

\subsection{Limitations}
The results presented here are biased by the projection effect in our derived values for $T_{\rm d}$, $N_{\rm H}$, and \mS.
The Galactic centre is a complex environment, one of the most extreme regions that is accessible for observations. Careful modeling of the matter distribution with multiple components along the line of sight is necessary to get an in-depth understanding of dust physics in this region. Earlier studies have proposed that the change in the polarization degree within the central parsec of the Galactic centre could be due the change in the direction of the magnetic field with respect to the plane of the sky \citep{Aitken1986,Roche2018}. This paper gives an overall picture of the dust physics at the Galactic centre and a detailed study of the strength and the morphology of the magnetic field will be presented in our follow up papers.  

\section{Summary} \label{Summary}
We have used the thermal dust polarization observations from SOFIA/HAWC+ and JCMT/SCUPOL at the wavelengths 53, 216, and 850 \micron\ to study the properties of dust and its alignment mechanisms in the region of about 30 pc around the Galactic centre. Our main results are as follows:
\begin{enumerate}[label=\arabic*.,leftmargin=*]
    \item The polarization observed at 53 \micron\ and 216 \micron\ can be explained using the RAT-A theory where we see a raise in the degree of polarization with temperature. However, the polarization at 850 \micron\ showed an inverse trend at higher temperatures which is contrary to what is expected from RAT-A theory.
    \item Tangling of the magnetic field was found to affect the polarization at 53 \micron, especially in the temperature range of $T_{\rm d}\sim30-38$ K where there was an observed drop in polarization.
    \item The highest polarization of $p\sim13\%$ was observed at 216 \micron\ from a region of low column density, high temperature, and low magnetic field tangling. 
    \item We estimated $a_{\rm align}$, $a^{\rm Lar}_{\rm max}$, $a^{\rm DG}_{\rm max\_JB}$, $a_{\rm disr}$, and $a_{\rm disr\_max}$ of grains which determine the level of the observed polarization. The minimum and maximum size of grains that can be aligned by RATs in this region was found to be $a_{\rm align}\sim0.1$ \micron\ and  $a^{\rm Lar}_{\rm max}\sim10$ mm.
    \item Magnetic relaxation was found to play a significant role in the grain alignment even when the grains were assumed to be PM in nature, due to the strong magnetic field.
    \item The region of highest polarization at 216 \micron\ was found to have the largest range of grains that can be aligned through MRAT mechanism, even with a low level of iron inclusion of $N_{\rm cl}=20$.
    \item The dust grains at the Galactic centre can achieve perfect alignment with $f_{\rm high-J}=1$ where $\delta_{\rm mag}>10$ if the grains are assumed to be SPM with $N_{\rm cl}\geqslant20$.
    \item We estimated the mass of the aligned dust grains using a parametric model to determine $f_{\rm high-J}$ and it has a weak correlation with the observed polarization degree. 
    \item The disruption size of grains was found to be around 1 \micron\ which could explain the drop in the polarization at higher temperatures observed in J850 data, as large grains contribute significantly to the observed emission at longer wavelengths.
    \item We used the ratio of polarization $p_{216\mu m}/p_{850\mu m}$ to predict the nature of the dust grains and found that the ratio changes from $p_{216\mu m}/p_{850\mu m}\leqslant1$ to $p_{216\mu m}/p_{850\mu m}>1$ beyond $T_{\rm d}\sim35$ K. This can be taken as further evidence of RAT-D where the composite mixed population of carbon and silicate grains can transform into separate silicate and carbon populations due to RAT-D, with the latter silicate grains having a higher degree of grain alignment.
\end{enumerate}

\section*{Acknowledgements}
We thank the anonymous referee for their useful comments and suggestions.  T.H. acknowledges the support of the National Research Foundation of Korea (NRF) grant funded by the Korea government (MSIT: 2019R1A2C1087045). This work was partly supported by a grant from the Simons Foundation to IFIRSE, ICISE (916424, N.H.). This research uses polarization data observed with the NASA/DLR Stratospheric Observatory for Infrared Astronomy (SOFIA). SOFIA is jointly operated by the Universities Space Research Association, Inc. (USRA), under NASA contract NNA17BF53C, and the Deutsches SOFIA Institut (DSI) under DLR contract 50 OK 2002 to the University of Stuttgart. This work made use of Astropy:\footnote{http://www.astropy.org} a community-developed core Python package and an ecosystem of tools and resources for astronomy \citep{astropy:2013, astropy:2018, astropy:2022}. This research made use of APLpy, an open-source plotting package for Python \citep{aplpy}.

\section*{Data Availability}
The data underlying this article will be shared on reasonable request to the corresponding author.


\bibliographystyle{mnras}
\bibliography{references} % if your bibtex file is called example.bib

%\begin{thebibliography}{99}
%\end{thebibliography}

%%%%%%%%%%%%%%%%%%%%%%%%%%%%%%%%%%%%%%%%%%%%%%%%%%

%%%%%%%%%%%%%%%%% APPENDICES %%%%%%%%%%%%%%%%%%%%%

\appendix

\section{Supplementary figures}
\begin{figure*}
    \centering
    \includegraphics[trim={2cm 0 0 0}, scale=0.5]{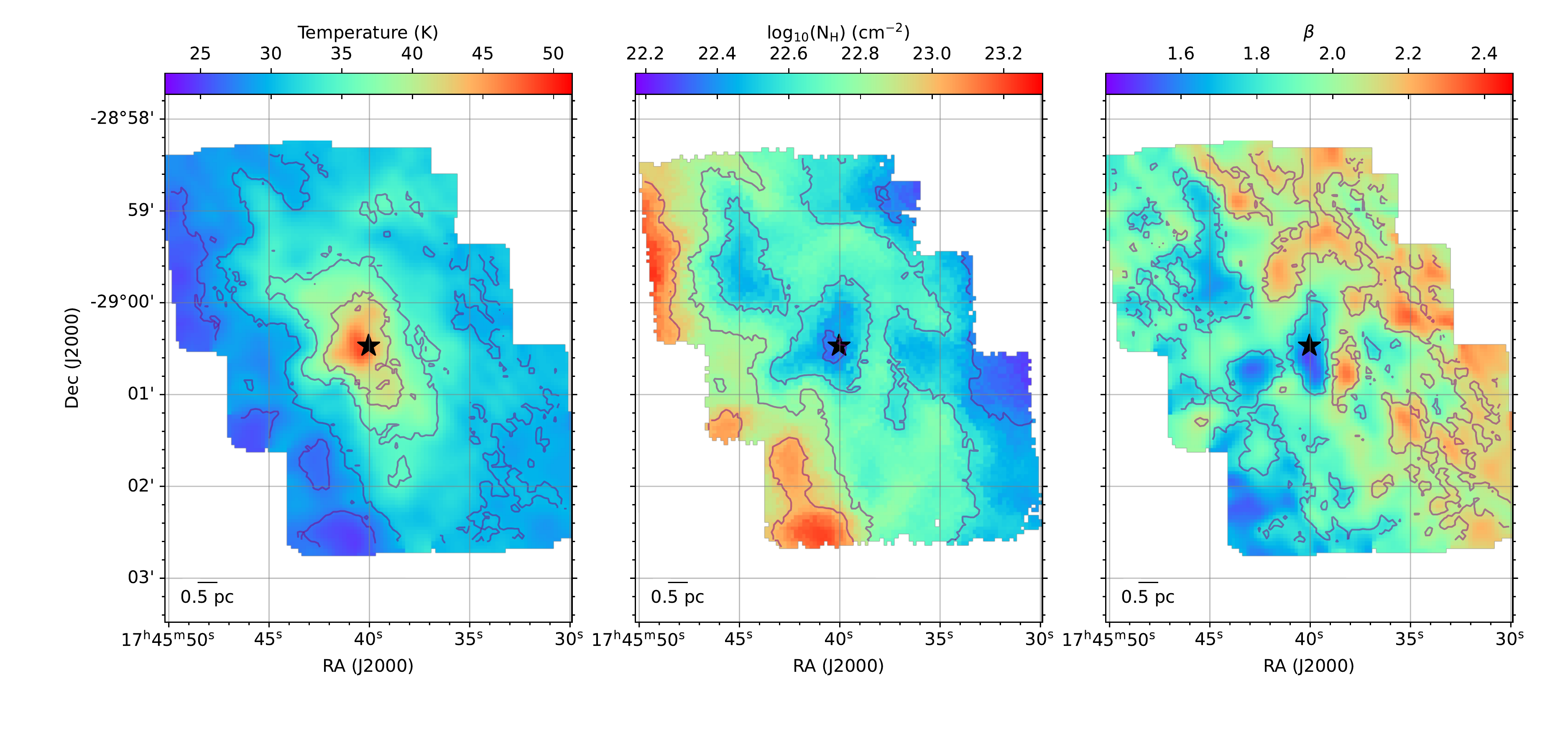}
    \includegraphics[scale=0.35]{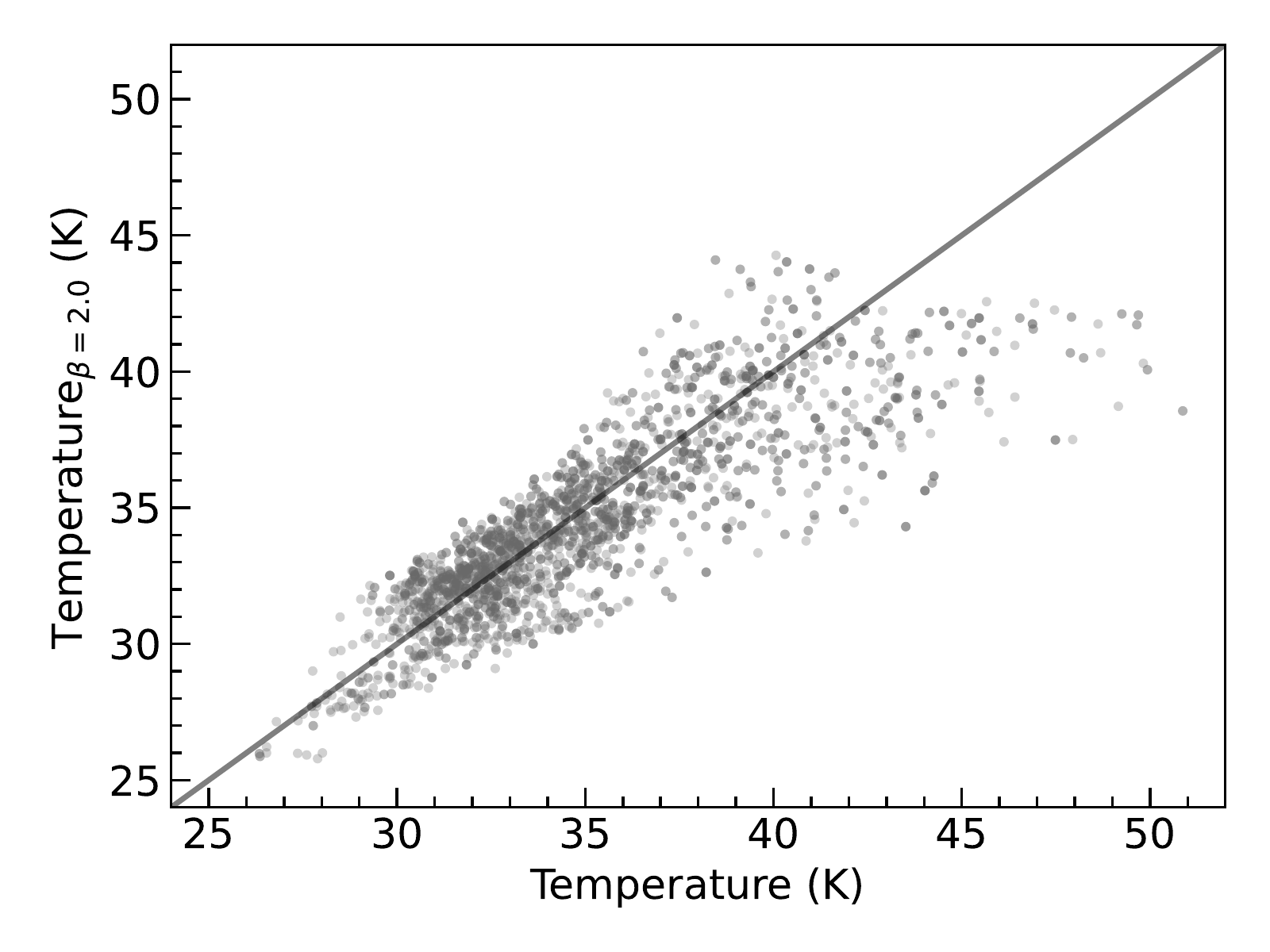}
    \includegraphics[scale=0.35]{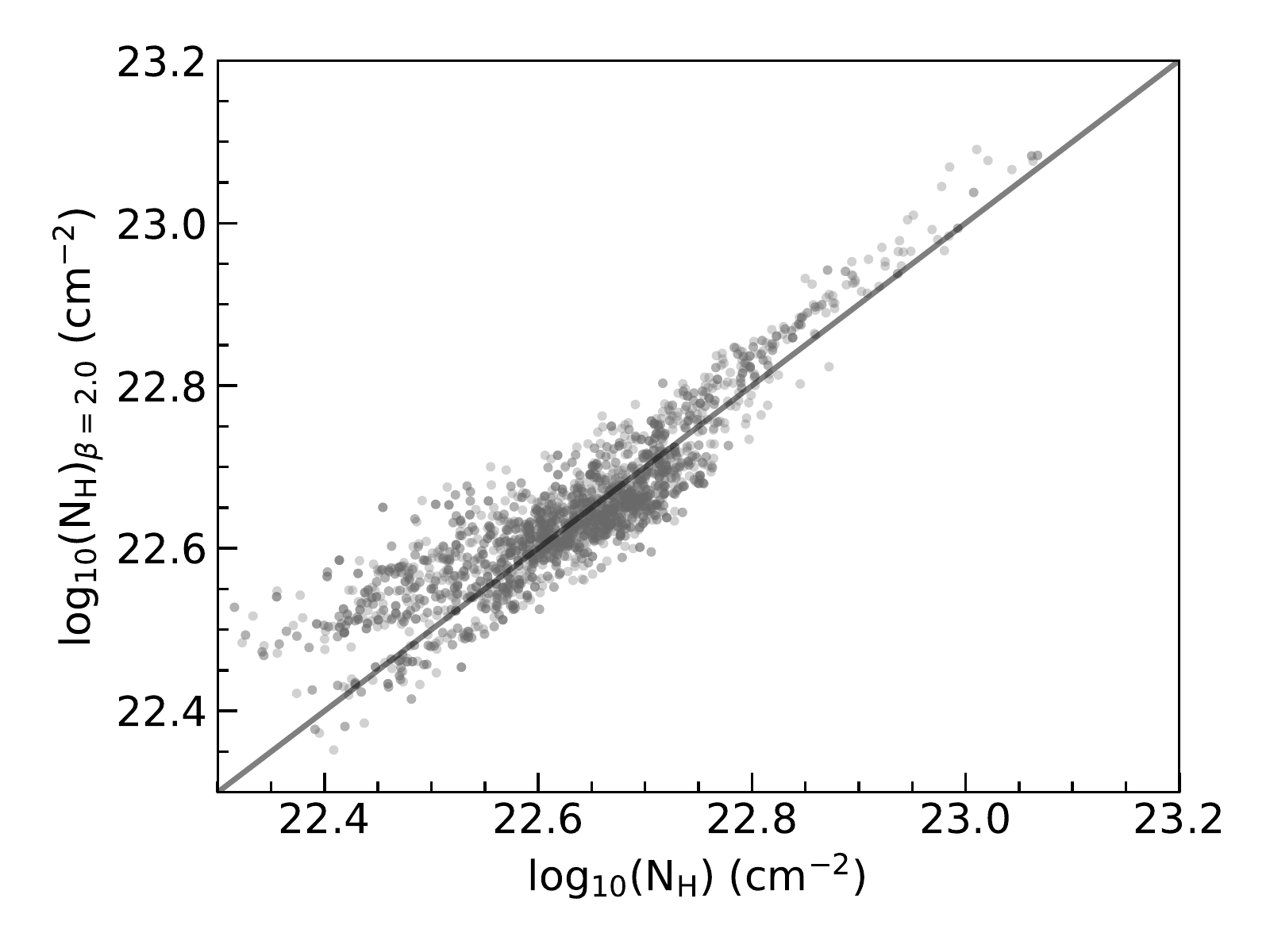}
    \includegraphics[scale=0.35]{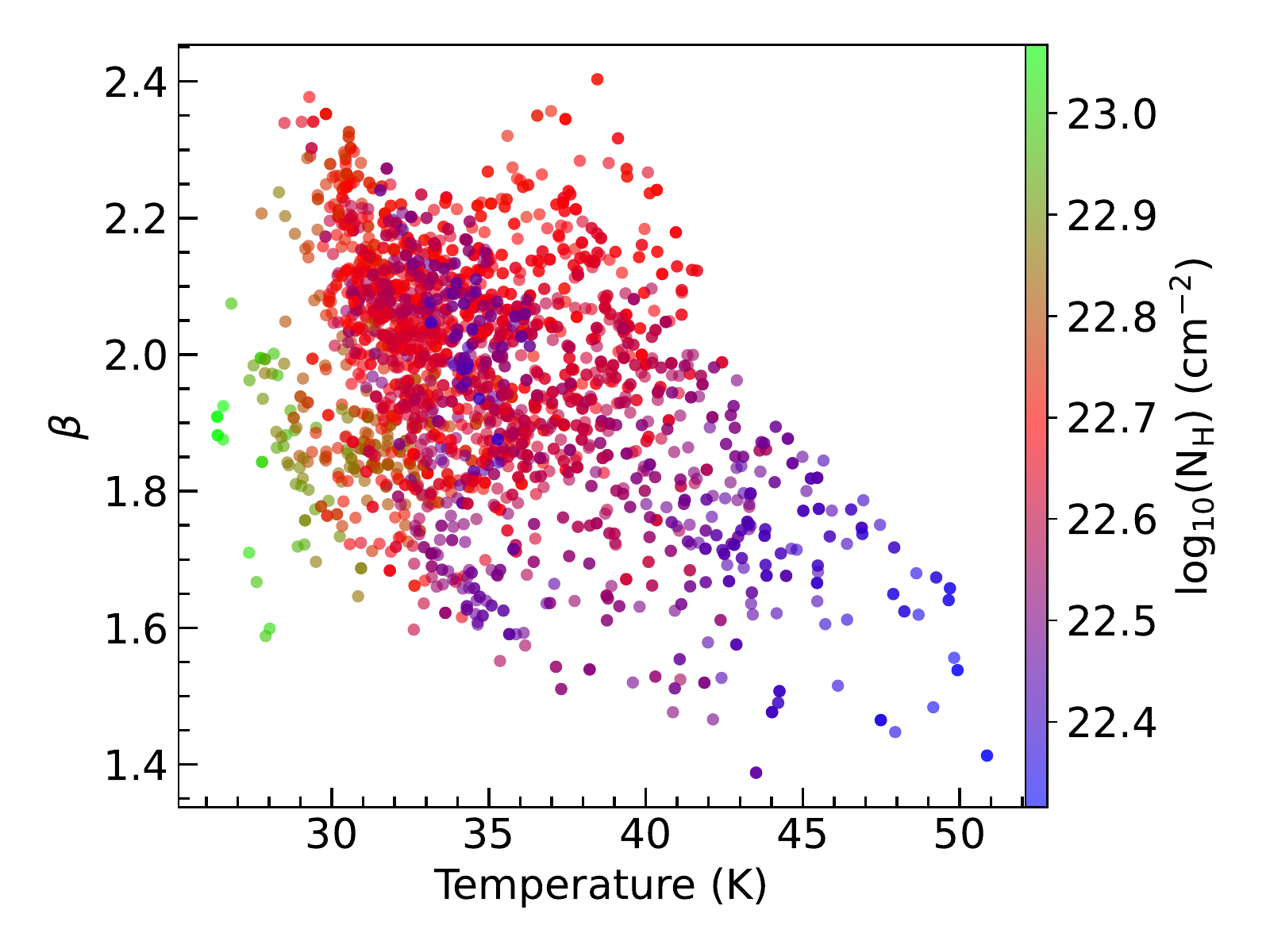}
    \caption{The top row shows the dust temperature ($T_{\rm d}$), gas column density ($N_{\rm H}$), and dust spectral index ($\beta$) derived using a modified black body fit to the \emph{Herschel} PACS and SPIRE data for the S53 map region shown in Figure \ref{fig:Sofia_5_pol_map}. The bottom row is the comparison between $T_{\rm d}$ and $N_{\rm H}$ derived by a varying $\beta$ and with $\beta$=2 as well as the relation between estimated value of $\beta$ with $T_{\rm d}$ and $N_{\rm H}$ for the S53 data region.}
    \label{fig:Sofia5_mbb_maps}
\end{figure*}

\begin{figure*}
    \centering
    \includegraphics[trim={2cm 0 0 0}, scale=0.5]{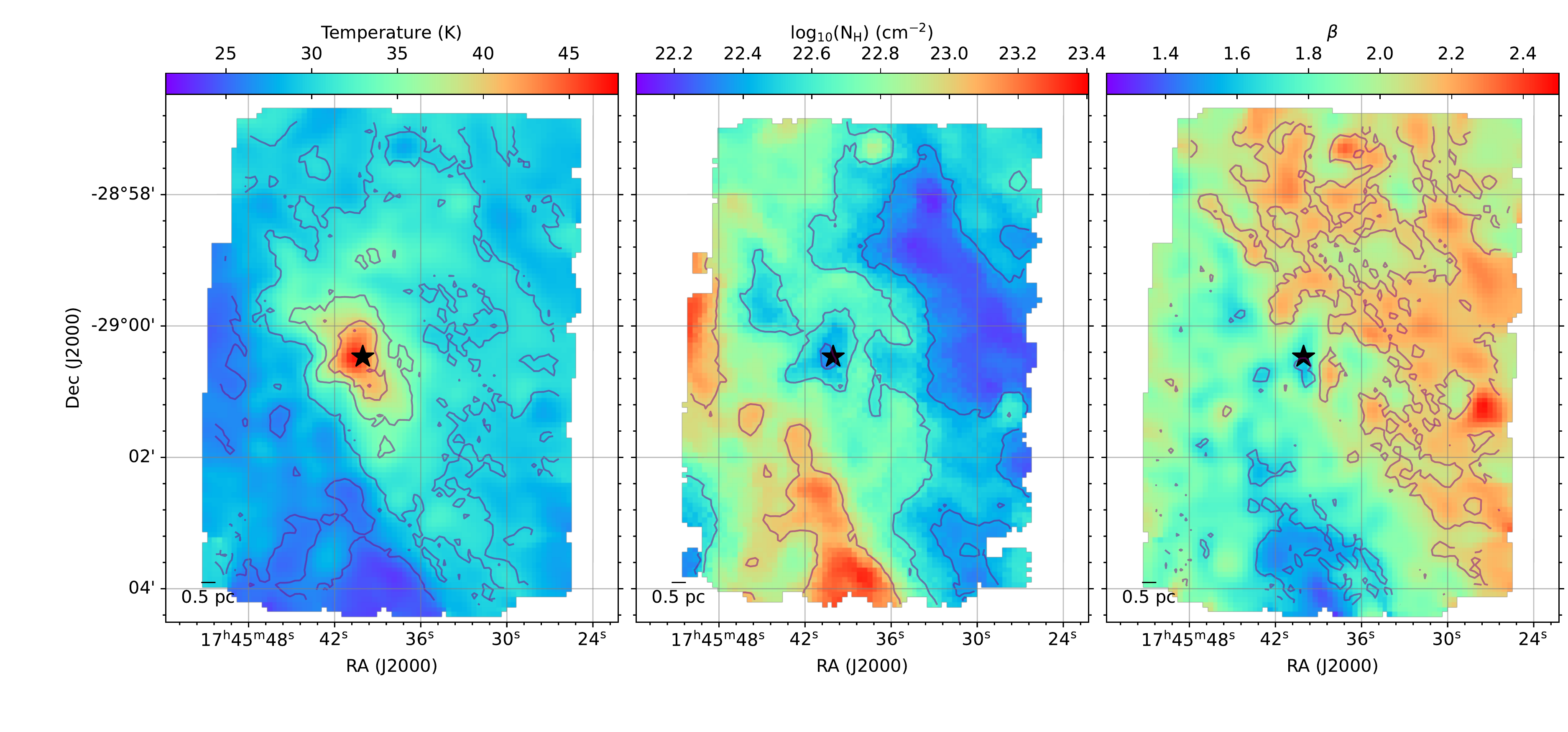}
    \includegraphics[scale=0.35]{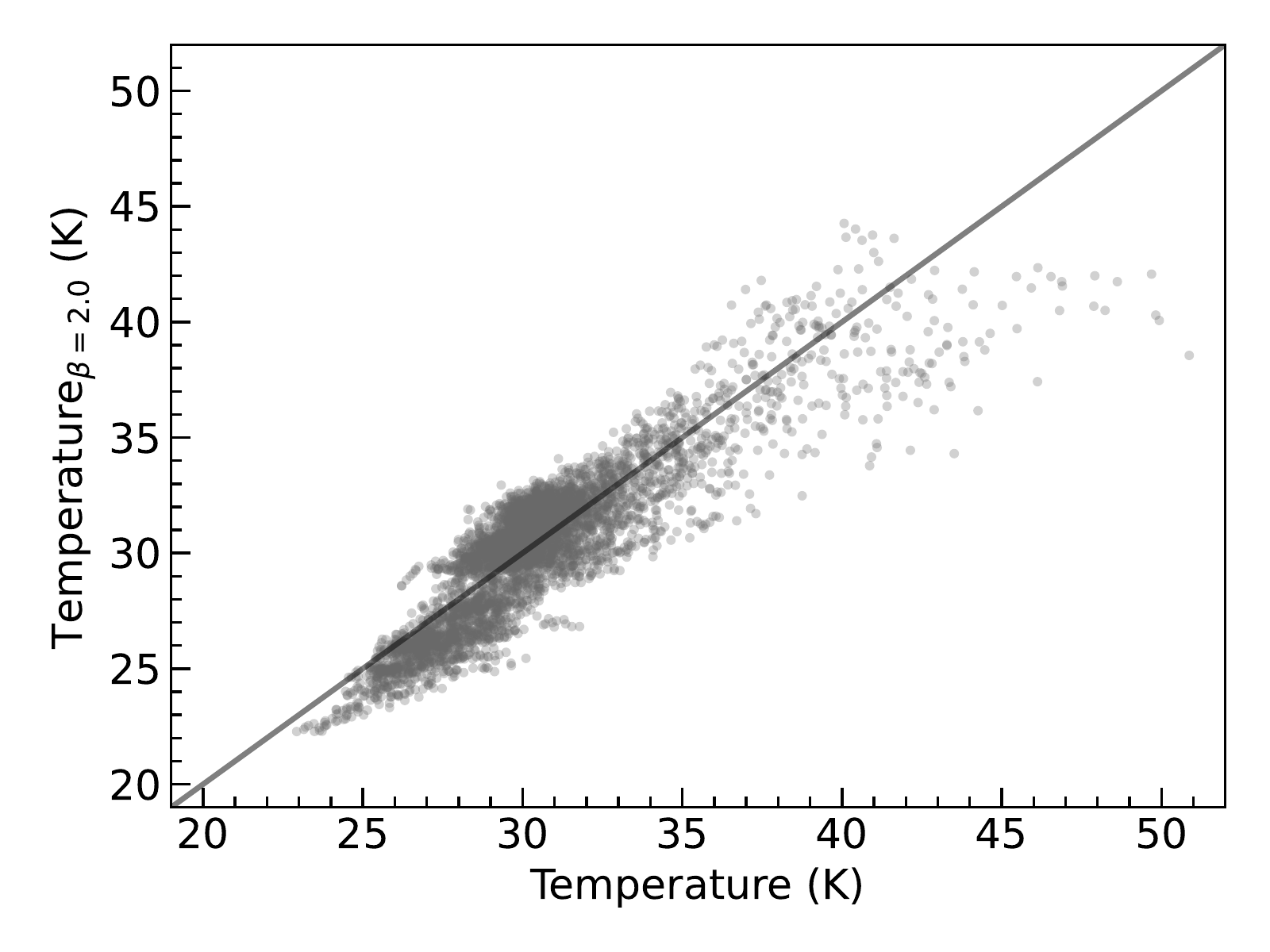}
    \includegraphics[scale=0.35]{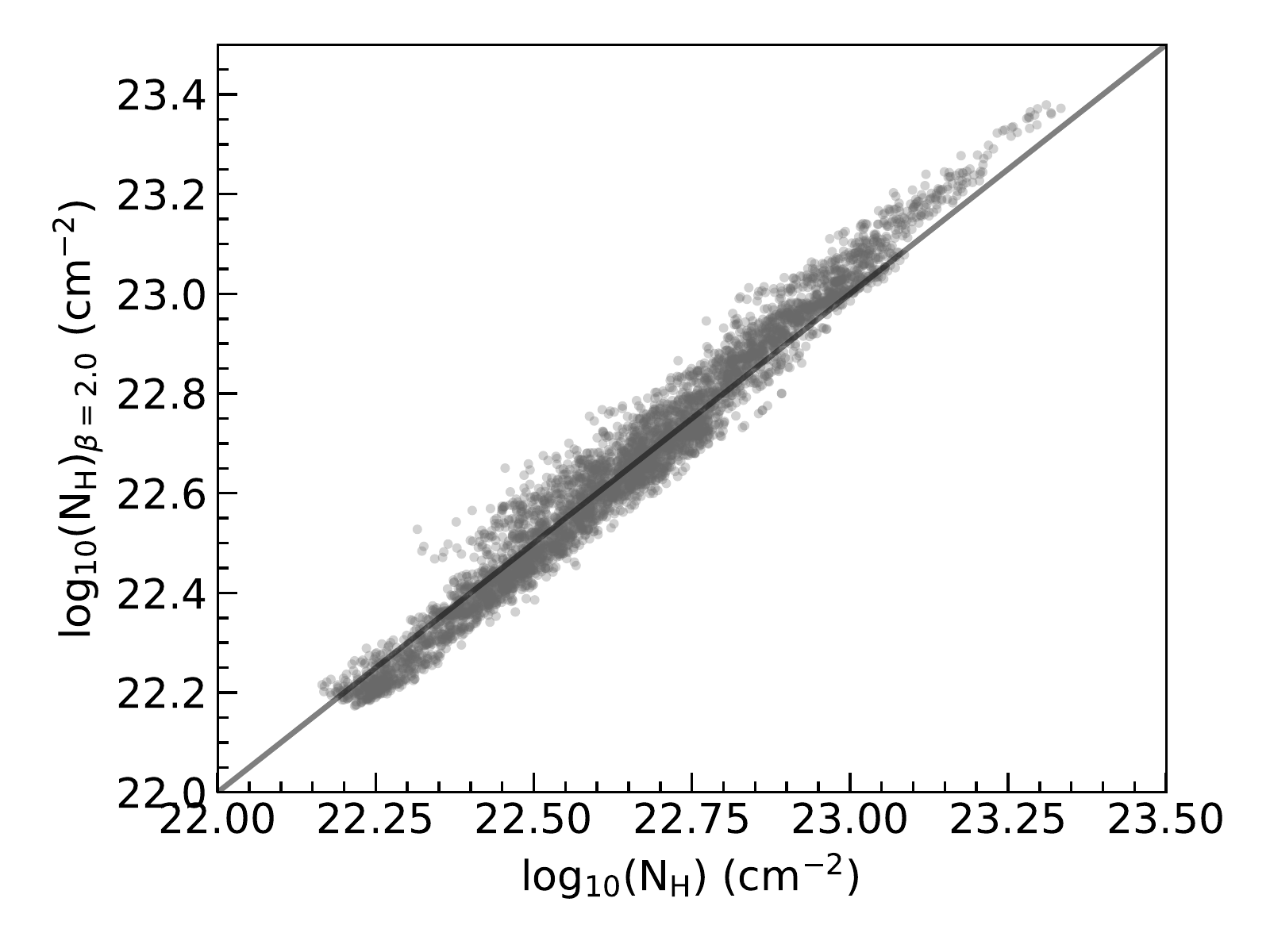}
    \includegraphics[scale=0.35]{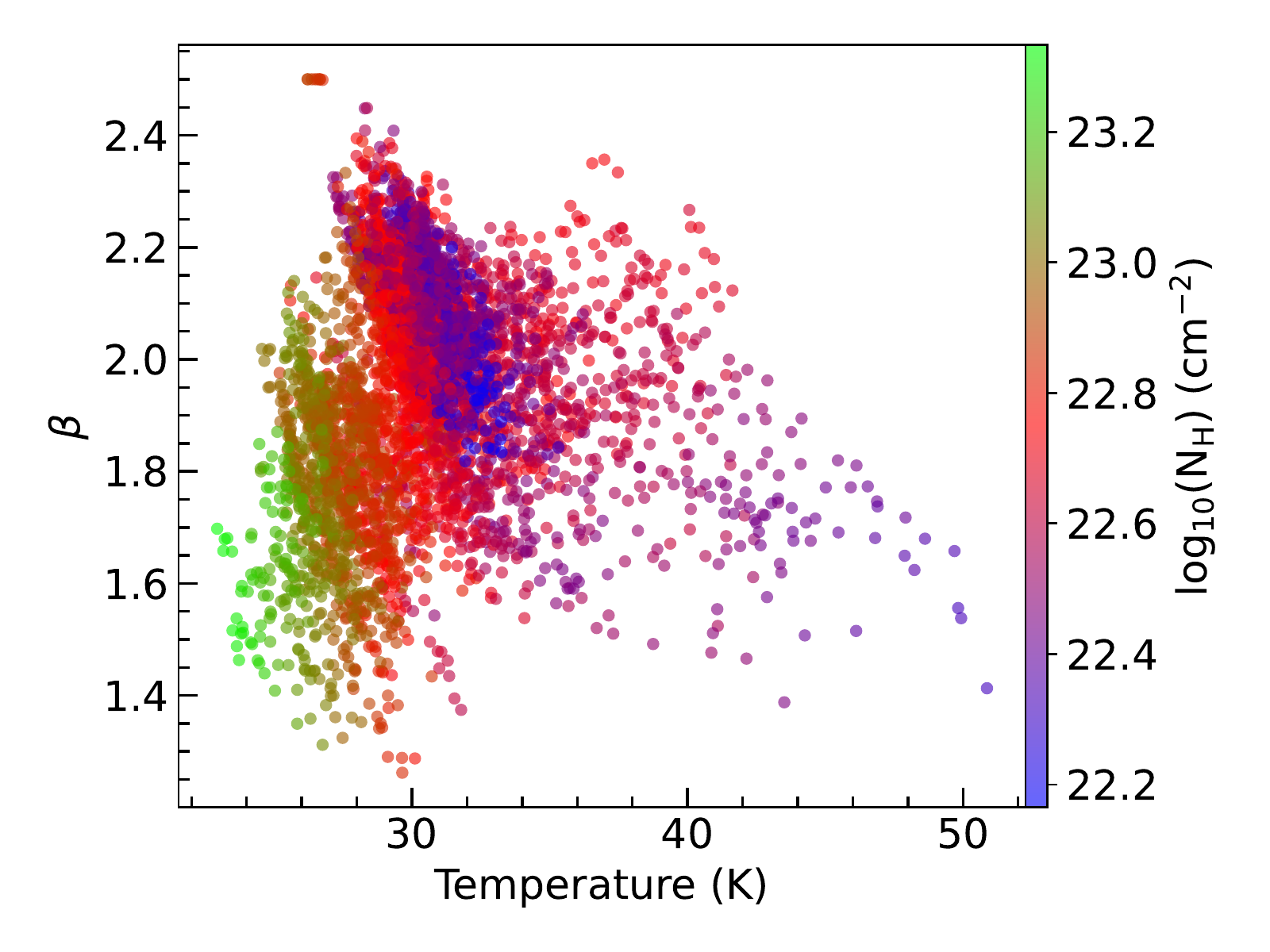}
    \caption{Same as Figure \ref{fig:Sofia5_mbb_maps} but for S216 observation shown in Figure \ref{fig:Sofia_10_pol_map}.}
    \label{fig:Sofia10_mbb_maps}
\end{figure*}

\begin{figure*}
    \centering
    \includegraphics[trim={2cm 0 0 0}, scale=0.5]{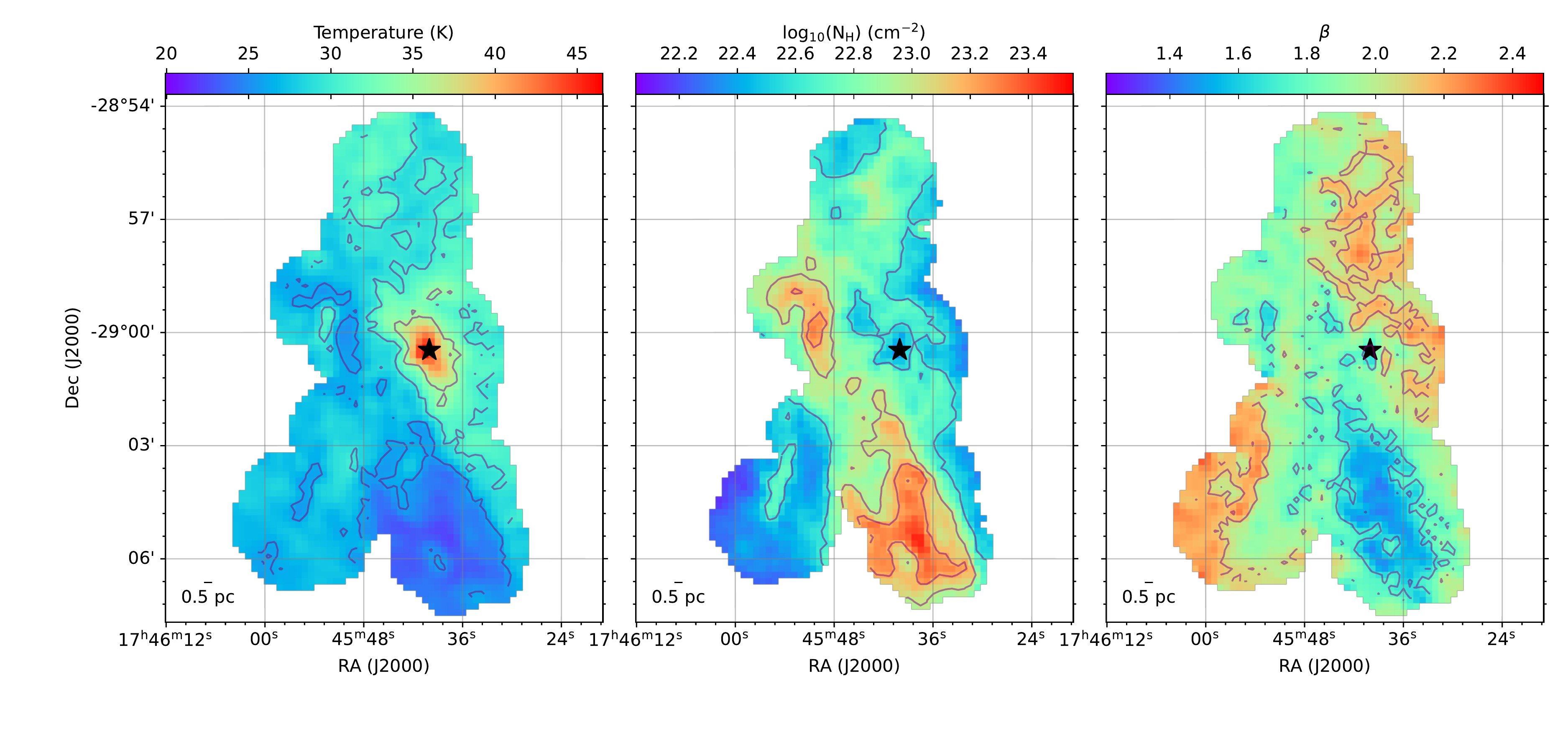}
    \includegraphics[scale=0.35]{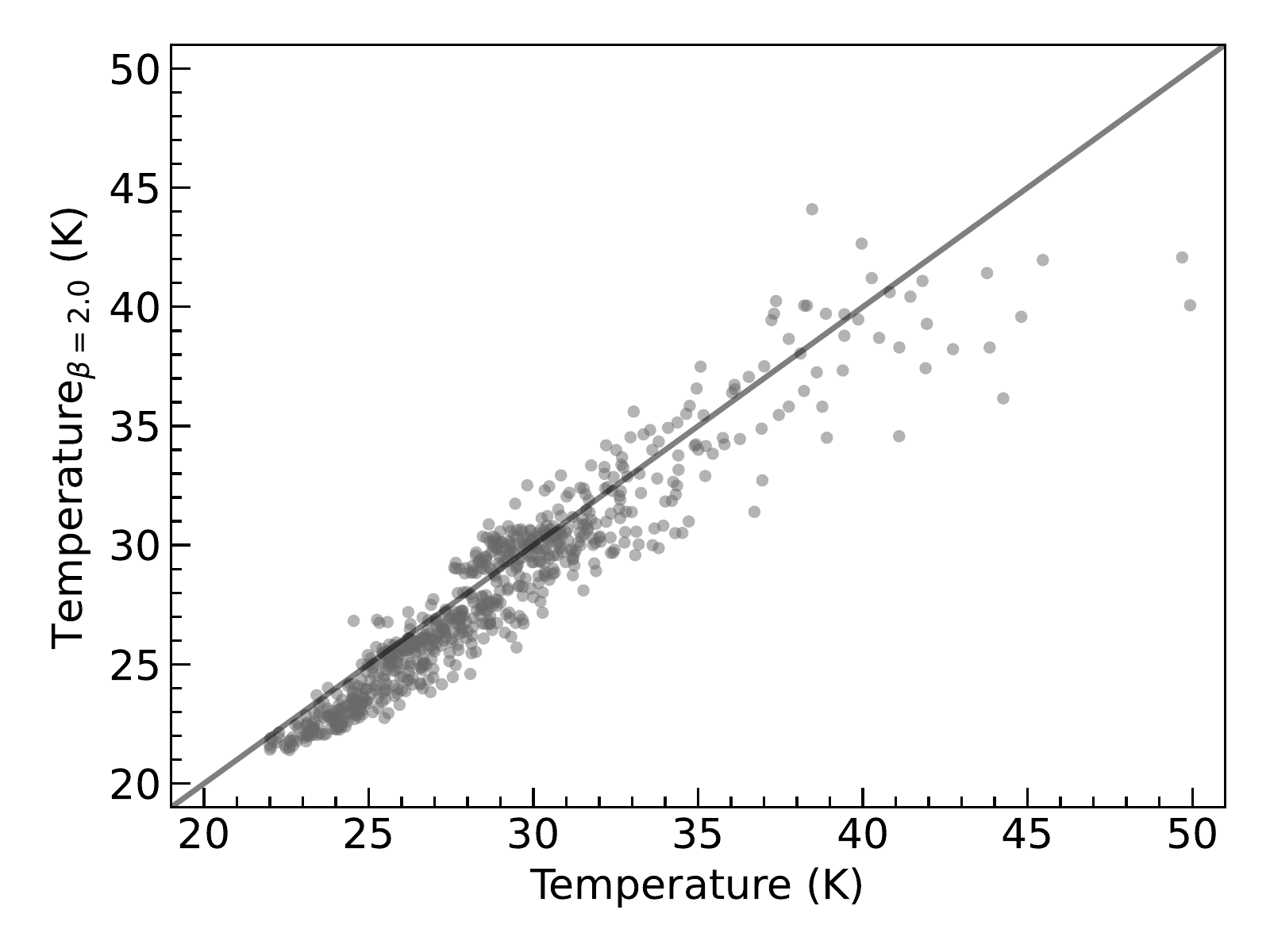}
    \includegraphics[scale=0.35]{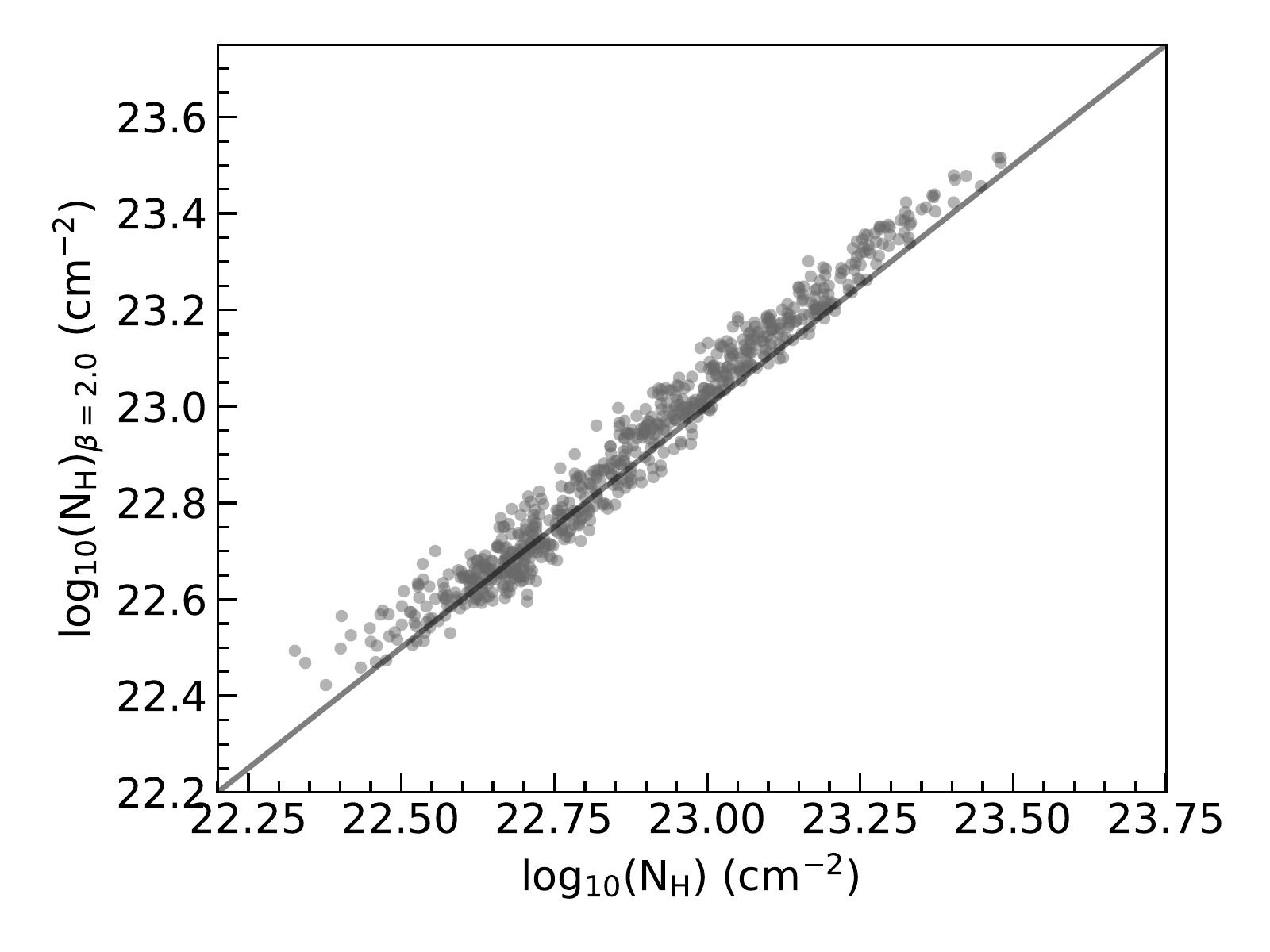}
    \includegraphics[scale=0.35]{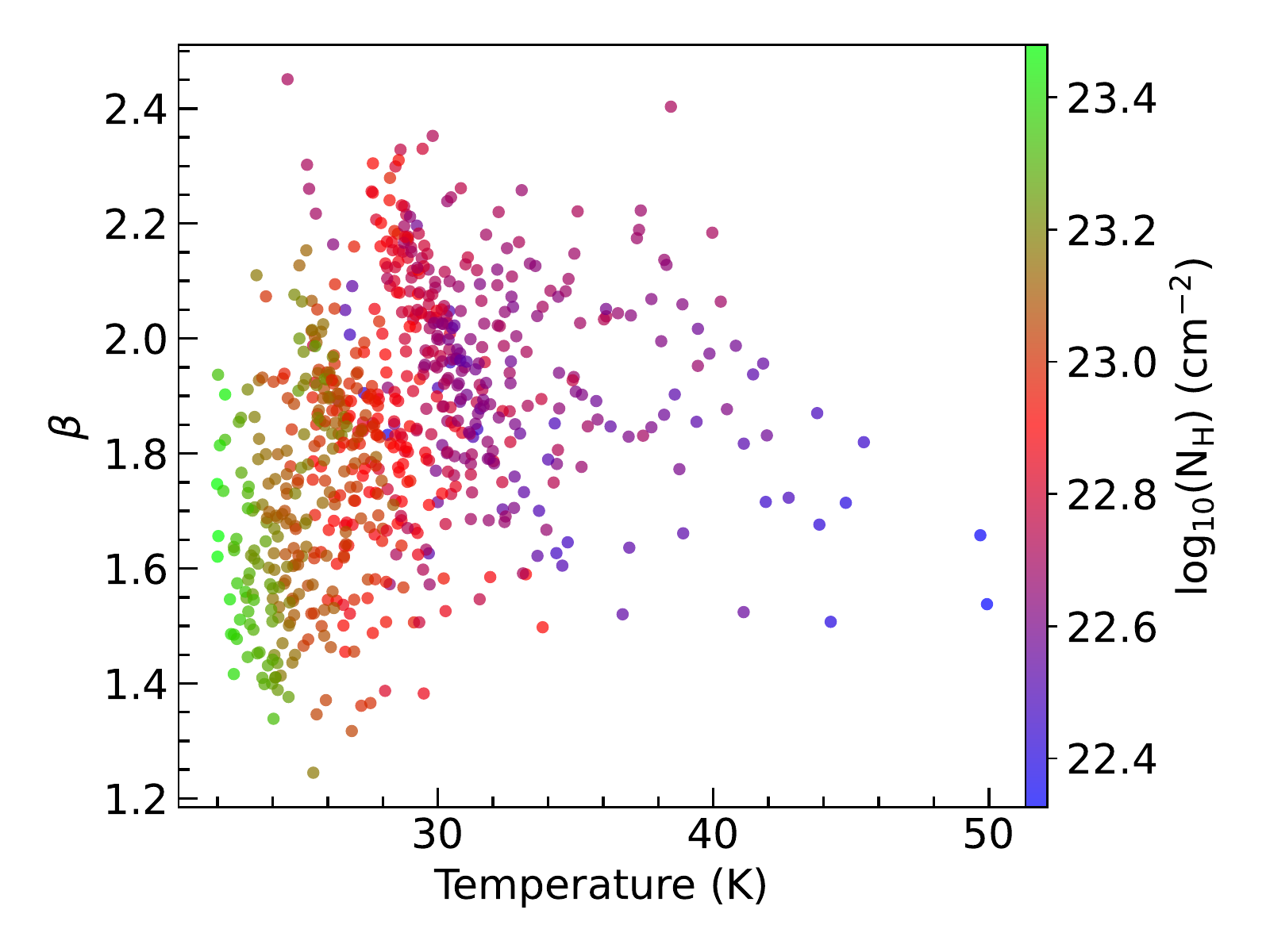}
    \caption{Same as Figure \ref{fig:Sofia5_mbb_maps} but for J850 observation shown in Figure \ref{fig:JCMT_pol_map}.}
    \label{fig:JCMT_mbb_maps}
\end{figure*}

\begin{figure}
    \centering
    \includegraphics[scale=0.5]{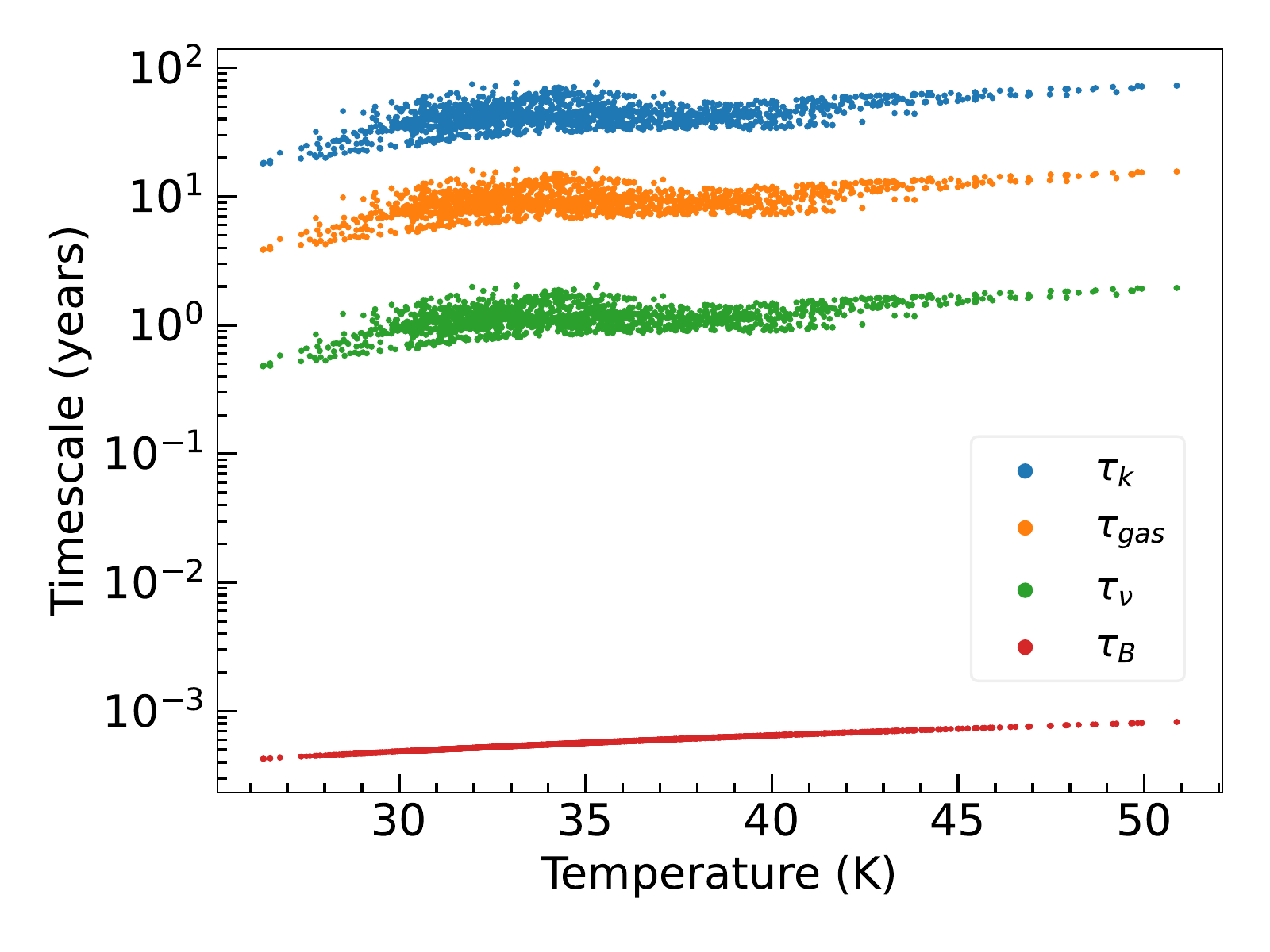}
    \caption{Variation of the timescales of external alignment of grains with dust temperature for S53 observation. The Larmor precession timescale ($\tau_{B}$) is the quickest compared to the radiative precession timescale ($\tau_k$), the gas damping timescale ($\tau_{\rm gas}$), and the mechanical precession timescale ($\tau_v$). The estimates of the timescales were made using the analytical formulae given by \citet{Hoang2022AJ}. Larmor precession leads to the alignment of grains along the magnetic field while the radiative precession leads to grain alignment along the direction of incident radiation and the mechanical precession leads to grain alignment along the direction of gas flow. In the diffuse ISM and in strong magnetic field environments, Larmor precession dominates over the other mechanisms and the magnetic field is the preferred direction of grain alignment.}
    \label{fig:time_scale}
\end{figure}

%%%%%%%%%%%%%%%%%%%%%%%%%%%%%%%%%%%%%%%%%%%%%%%%%%

% Don't change these lines
\bsp	% typesetting comment
\label{lastpage}
\end{document}